\documentclass[aps,prd,reprint,superscriptaddress,showpacs,floatfix,nofootinbib]{revtex4-2}  
\usepackage{graphicx}
\usepackage{amsmath,amssymb,graphics,setspace}
\usepackage{color}
\usepackage[english]{babel}
\usepackage{url,bm}
\usepackage{amsfonts}
\usepackage[separate-uncertainty = true]{siunitx}
\usepackage{dcolumn}
\usepackage{appendix}
\usepackage{lmodern}
\usepackage[draft=false]{microtype}
\usepackage{esdiff}
\usepackage[table,dvipsnames]{xcolor}
\usepackage[babel]{csquotes}
\usepackage{textcomp}
\usepackage{listings}
\usepackage{enumitem}
\usepackage{lipsum}
\usepackage{tabularx}
\usepackage{makecell}
\usepackage{soul}
\usepackage{enumitem}
\usepackage{afterpage}
\usepackage{float}
\usepackage{wasysym}

\usepackage[hidelinks,draft=false]{hyperref}
\usepackage{natbib}
\newcommand{\mathsym}[1]{{}}
\newcommand{\unicode}[1]{{}}
\newcommand{\rtHz}{\ensuremath{\sqrt{\text{Hz}}}}
\newcommand{\Pwat}{P_{\text{H}_2\text{O}}}

\newcommand{\ORCIDiD}[1]{\href{https://orcid.org/#1}{\includegraphics[width=2ex]{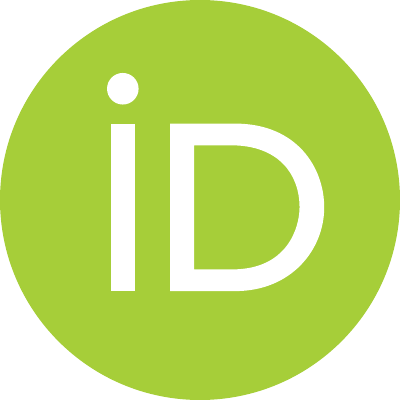}}}

\newcommand{\OMS}{\text{OMS}}
\newcommand{\GRS}{\text{GRS}}
\newcommand{\SNR}{\text{SNR}}

\DeclareSIUnit\year{yr}
\DeclareSIUnit\bar{bar}

\newcommand{\etr}{\mathrm{etr}}
\newcommand{\cw}{\mathcal{CW}}

\newcommand{\bSig}{\bm{\Sigma}}
\newcommand{\bW}{\bm{W}}
\newcommand{\bQ}{\bm{Q}}

\newcommand{\bPi}{\bm{\Pi}}

\newcommand{\bU}{\bm{U}}

\definecolor{HotPink}{RGB}{255 ,0,128}
\definecolor{cardinal}{rgb}{0.77, 0.12, 0.23}
\definecolor{Gorange}{RGB}{240,131,36}
\definecolor{lightgray}{gray}{0.85}

\makeatletter
\renewcommand*{\@fnsymbol}[1]{\ifcase#1\or a \or * \or b \or c \or d \or e \or f \or g \or h \or i \or ** \else\@ctrerr\fi}
\makeatother

\begin{document}
\renewcommand{\figurename}{FIG.} 
\renewcommand{\tablename}{TABLE} 

 \title{In-depth analysis of LISA Pathfinder performance results:\texorpdfstring{\\}{}
Time evolution, noise projection, physical models, and implications for LISA}

\author{M~Armano}\altaffiliation[Current address: ]{\addressh;~\addressgg}\affiliation{\addressa} 
\author{H~Audley}\affiliation{\addressb}
\author{J~Baird}\affiliation{\addressca}
\author{P~Binetruy}\thanks{Deceased.}\affiliation{\addressc} 
\author{M~Born}\affiliation{\addressb}
\author{D~Bortoluzzi}\affiliation{\addressf}
\author{E~Castelli}\altaffiliation[Current address: ]{\addressu}\affiliation{\addressi} 
\author{A~Cavalleri}\affiliation{\addressk}
\author{A~Cesarini}\affiliation{\addresso}
\author{V~Chiavegato}\affiliation{\addressi} 
\author{A\,M~Cruise}\affiliation{\addressj}
\author{D\,Dal~Bosco}\affiliation{\addressi} 
\author{K~Danzmann}\affiliation{\addressb}
\author{M~De Deus Silva}\affiliation{\addressa}
\author{I~Diepholz}\affiliation{\addressb}
\author{G~Dixon}\affiliation{\addressj}
\author{R~Dolesi}\affiliation{\addressi}
\author{L~Ferraioli}\affiliation{\addressl} 
\author{V~Ferroni}\affiliation{\addressi} 
\author{E\,D~Fitzsimons}\affiliation{\addressm}
\author{M~Freschi}\affiliation{\addressa}
\author{L~Gesa}\thanks{Deceased.}\affiliation{\addressn}\affiliation{\addressntwo} 
\author{D~Giardini}\affiliation{\addressl}
\author{F~Gibert}\altaffiliation[Current address: ]{\addressee}\affiliation{\addressi}
\author{R~Giusteri}\affiliation{\addressb}
\author{C~Grimani}\affiliation{\addresso} 
\author{J~Grzymisch}\affiliation{\addressh}
\author{I~Harrison}\affiliation{\addressp}
\author{M\,S~Hartig}\affiliation{\addressb}
\author{G~Heinzel}\affiliation{\addressb}
\author{M~Hewitson}\altaffiliation[Current address: ]{\addressh}\affiliation{\addressb}
\author{D~Hollington}\affiliation{\addressd}
\author{D~Hoyland}\affiliation{\addressj}
\author{M~Hueller}\affiliation{\addressi}
\author{H~Inchausp\'e}\altaffiliation[Current address: ]{\addresscb}\affiliation{\addressca}
\author{O~Jennrich}\affiliation{\addressh}
\author{P~Jetzer}\affiliation{\addressq}
\author{B~Johlander}\affiliation{\addressa}
\author{N~Karnesis}\altaffiliation[Current address: ]{\addressbb}\affiliation{\addressca}
\author{B~Kaune}\affiliation{\addressb}
\author{N~Korsakova}\affiliation{\addressca}
\author{C\,J~Killow}\altaffiliation[Current address: ]{\addressff}\affiliation{\addressr} 
\author{J\,A~Lobo}\thanks{Deceased.}\affiliation{\addressn}\affiliation{\addressntwo} 
\author{J\,P~L\'opez-Zaragoza}\affiliation{\addressn}\affiliation{\addressntwo}
\author{R~Maarschalkerweerd}\affiliation{\addressp}
\author{D~Mance}\affiliation{\addressl}
\author{V~Mart\'{i}n}\affiliation{\addressn}\affiliation{\addressntwo} 
\author{L~Martin-Polo}\affiliation{\addressa}
\author{F~Martin-Porqueras}\affiliation{\addressa}
\author{J~Martino}\affiliation{\addressca}
\author{P\,W~McNamara}\affiliation{\addressh}
\author{J~Mendes}\affiliation{\addressp}
\author{L~Mendes}\affiliation{\addressa}
\author{N~Meshksar}\affiliation{\addressl}
\author{M~Nofrarias}\affiliation{\addressn}\affiliation{\addressntwo} 
\author{S~Paczkowski}\affiliation{\addressb} 
\author{M~Perreur-Lloyd}\affiliation{\addressr} 
\author{A~Petiteau}\affiliation{\addressc}\affiliation{\addressca}
\author{E~Plagnol}\affiliation{\addressca}
\author{J~Ramos-Castro}\affiliation{\addresss}
\author{J~Reiche}\affiliation{\addressb}
\author{F~Rivas}\altaffiliation[Current address: ]{\addresscc}\affiliation{\addressi}
\author{D\,I~Robertson}\affiliation{\addressr} 
\author{G~Russano}\altaffiliation[Current address: ]{\addressx}\affiliation{\addressi} 
\author{L~Sala~\ORCIDiD{0000-0002-2682-8274}}\affiliation{\addressi}
\author{J~Slutsky}\affiliation{\addressu}
\author{C\,F~Sopuerta}\affiliation{\addressn}\affiliation{\addressntwo} 
\author{T~Sumner}\affiliation{\addressd}
\author{D~Texier}\affiliation{\addressa}
\author{J\,I~Thorpe}\affiliation{\addressu} 
\author{D~Vetrugno}\affiliation{\addressi} 
\author{S~Vitale~\ORCIDiD{0000-0002-2427-8918}}\affiliation{\addressi} 
\author{G~Wanner}\affiliation{\addressb}
\author{H~Ward}\affiliation{\addressr} 
\author{P~Wass}\affiliation{\addressd}\affiliation{\addressdd} 
\author{W\,J~Weber}\affiliation{\addressi} 
\author{L~Wissel}\affiliation{\addressb}
\author{A~Wittchen}\affiliation{\addressb}
\author{C~Zanoni}\affiliation{\addressf} 
\author{P~Zweifel}\affiliation{\addressl}

\collaboration{LISA Pathfinder Collaboration}\thanks{Corresponding authors:\\  \href{mailto:lorenzo.sala@unitn.it,stefano.vitale@unitn.it}{lorenzo.sala@unitn.it~\ORCIDiD{0000-0002-2682-8274}\\stefano.vitale@unitn.it~\ORCIDiD{0000-0002-2427-8918}}}

\def\addressa{European Space Astronomy Centre, European Space Agency, Villanueva de la Ca\~{n}ada, 28692 Madrid, Spain}
\def\addressb{Albert-Einstein-Institut, Max-Planck-Institut f\"ur Gravitationsphysik und Leibniz Universit\"at Hannover,
Callinstra{\ss}e 38, 30167 Hannover, Germany}
\def\addressc{IRFU, CEA, Universit\'e Paris-Saclay, F-91191 Gif-sur-Yvette, France}
\def\addressca{Universit\'e Paris Cit\'e, CNRS, Astroparticule et Cosmologie, F-75013 Paris, France}
\def\addresscb{Institut f\"ur Theoretische Physik, Universit\"at Heidelberg, Philosophenweg 16, 69120 Heidelberg, Germany}
\def\addressd{Physics Department, Blackett Laboratory, High Energy Physics Group, Imperial College London, Prince Consort Road, London SW7 2BW, United Kingdom}
\def\addresse{Dipartimento di Fisica, Universit\`a di Roma ``Tor Vergata'',  and INFN, sezione Roma Tor Vergata, I-00133 Roma, Italy}
\def\addressf{Department of Industrial Engineering, University of Trento, via Sommarive 9, 38123 Trento, 
and Trento Institute for Fundamental Physics and Application / INFN}
\def\addressg{Airbus Defence and Space, Claude-Dornier-Strasse, 88090 Immenstaad, Germany}
\def\addressh{European Space Technology Centre, European Space Agency, 
Keplerlaan 1, 2200 AG Noordwijk, The Netherlands}
\def\addressi{Dipartimento di Fisica, Universit\`a di Trento and Trento Institute for 
Fundamental Physics and Application / INFN, 38123 Povo, Trento, Italy}
\def\addressj{The School of Physics and Astronomy, University of
Birmingham, B15 2TT Birmingham, United Kingdom}
\def\addressk{Istituto di Fotonica e Nanotecnologie, CNR-Fondazione Bruno Kessler, 
    I-38123 Povo, Trento, Italy}
\def\addressl{Institut f\"ur Geophysik, ETH Z\"urich, Sonneggstrasse 5, CH-8092 Z\"urich, Switzerland}
\def\addressm{The UK Astronomy Technology Centre, Royal Observatory, Edinburgh, Blackford Hill, Edinburgh EH9 3HJ, United Kingdom}
\def\addressn{Institut de Ci\`encies de l'Espai (ICE,~CSIC), Campus UAB, Carrer de Can Magrans~s/n, Cerdanyola del Vall\`es~08193, Spain}
\def\addressntwo{Institut d'Estudis Espacials de Catalunya (IEEC), Edifici RDIT, C/ Esteve~Terradas, 1, desp.~212, Castelldefels~08860, Spain}
\def\addresso{DISPEA, Universit\`a di Urbino Carlo Bo, Via S. Chiara, 27 61029 Urbino/INFN, Italy}
\def\addressp{European Space Operations Centre, European Space Agency, 64293 Darmstadt, Germany }
\def\addressq{Physik Institut, 
Universit\"at Z\"urich, Winterthurerstrasse 190, CH-8057 Z\"urich, Switzerland}
\def\addressr{SUPA, Institute for Gravitational Research, School of Physics and Astronomy, University of Glasgow, Glasgow G12 8QQ, United Kingdom}
\def\addresss{Department d'Enginyeria Electr\`onica, Universitat Polit\`ecnica de Catalunya,  08034 Barcelona, Spain}
\def\addressu{Gravitational Astrophysics Lab, NASA Goddard Space Flight Center, 8800 Greenbelt Road, Greenbelt, MD 20771 USA}
\def\addressx{INAF Osservatorio Astronomico di Capodimonte, I-80131 Napoli, Italy}
\def\addressy{INFN - Sezione di Napoli, I-80126, Napoli, Italy}
\def\addressz{Dipartimento di Fisica ed Astronomia, Universit\`a degli Studi di Firenze and INFN - Sezione di Firenze, I-50019 Firenze, Italy}
\def\addressaa{OHB Italia S.p.A, Via Gallarate, 150 - 20151 Milano, Italy}
\def\addressbb{Department of Physics, Aristotle University of Thessaloniki, Thessaloniki 54124, Greece}
\def\addresscc{Universidad Loyola Andaluc\'ia, Department of Quantitative Methods, Avenida de las Universidades s/n, 41704, Dos Hermanas, Sevilla, Spain}
\def\addressdd{Department of Mechanical and Aerospace Engineering, MAE-A, P.O. Box 116250, University of Florida, Gainesville, Florida 32611, USA}
\def\addressee{isardSAT SL, Marie Curie 8-14, 08042 Barcelona, Catalonia, Spain}
\def\addressff{Qioptiq, Denbighshire, United Kingdom}
\def\addressgg{AGH University of Krakow, Al. Mickiewicza 30, 30-059 Krak\'{o}w, Poland.}
\date{August 21, 2024}

\begin{abstract}
We present an in-depth analysis of the  LISA Pathfinder differential acceleration performance over the entire course of its science operations, spanning approximately 500 days. We find: (1) The evolution of the   Brownian noise that dominates the acceleration amplitude spectral density (ASD),  for frequencies $f\gtrsim \SI{1}{mHz}$, is consistent with the decaying pressure due to the outgassing of a single gaseous species.
(2) Between $f=\SI{36}{\micro\hertz}$ and $\SI{1}{\milli\hertz}$,  the acceleration ASD  shows a  $1/f$ tail in excess of the Brownian noise of almost constant amplitude,  with $\simeq 20\%$  fluctuations over a period of a few days, with no particular time pattern over the course of the mission. (3) At the lowest considered frequency of $f=\SI{18}{\micro\hertz}$, the ASD significantly deviates from the $1/f$ behavior, because of temperature fluctuations that appear to modulate a quasistatic pressure gradient, sustained by the asymmetries of the outgassing pattern.
We also present the results of a projection of the observed acceleration noise on the potential sources for which we had either a direct correlation measurement or a quantitative estimate from dedicated experiments. These sources account for approximately 40\% of the noise power in the  $1/f$ tail. Finally, we analyze the possible sources of the remaining unexplained fraction and identify the possible measures that may be taken to keep those under control in LISA.
\end{abstract}


\pacs{04.80.Nn, 07.05.Kf, 95.55.-n}

\maketitle
~
\section{Introduction} 
\label{sec:introduction}

{LISA  \cite{Amaro-Seoane2017} is a gravitational wave observatory being developed by the European Space Agency (ESA) in collaboration with international partners, which has recently entered its final implementation phase. LISA targets a frequency range between $\SI{20}{\micro\hertz}$ and 1\,Hz, which is inaccessible to ground-based observatories due to terrestrial gravitational noise.

In this frequency range, LISA is expected to detect an extremely rich spectrum of new sources  \cite{Amaro-Seoane2017}, examples being: binaries of black holes with masses millions of times that of the Sun, formed in galaxy collisions, observable throughout the entire Universe; the plunge of compact objects into massive black holes, allowing the study of the gravitational field close to the event horizon of these; tens of thousands of detached stellar binaries in the Milky Way, including the inspiral stage of ultracompact binaries, with black holes and neutron stars, that will appear years later as merging sources in the audio band of terrestrial detectors.

To prepare for LISA, ESA launched and operated the LISA Pathfinder (LPF) mission \cite{PhysRevLett.120.061101,armano:subfemtog} between December 2015 and July 2017. The scientific goal of LPF was to demonstrate that parasitic forces on a test mass (TM), to be used as a geodesic reference in  LISA, may be suppressed below the required noise level. }

To that aim, the mission carried a miniature version of one of the LISA interferometric arms, that is, \SI{2}{kg}-size free-orbiting TMs, separated by a few tens of centimeters, and an interferometric readout measuring their relative acceleration along the line joining their respective centers of mass \cite{LPFPKS}.

The main results of the mission have been reported in \cite{PhysRevLett.120.061101,armano:subfemtog}, showing that the mission had surpassed its goals. In the interest of brevity, those papers focused on just two experimental acceleration measurement runs, and we presented a rather synthetic discussion on the possible physical origin of the observed residual acceleration noise.  

Throughout the 16 months of the mission, however, we performed many more acceleration measurements that allowed us to observe the time evolution of the performance, and its sensitivity to the mission operating conditions. 

In addition, during all those acceleration noise measurements, we have been measuring a set of other physical parameters,  like magnetic fields, temperatures, parasitic torques acting on the TMs, and others. Those measurements have allowed us to set limits on the role of some of the possible physical sources of the observed acceleration noise.

In this paper, we present the evolution of the acceleration performance of the mission, and present a quantitative analysis of  the correlation of acceleration data with those additional physical parameters. Based on that, we present a more in-depth analysis of the possible physical origin of the observed acceleration noise, and discuss the implication for the performance of the LISA observatory.

\vspace{2mm}
\noindent The paper is organized as follows:
\begin{enumerate}[label=(\roman*)]
\item In Sec.~\ref{sec:exp} we summarize the key features of the experiment, the operating conditions, and the measurement runs;
\item In Sec.~\ref{sec:accPSD} we describe, for all runs,  the measured acceleration power spectra and  their  key features;
\item Section~\ref{sec:brown} focuses on Brownian noise and its time evolution;
\item Section~\ref{sec:excess} describes the evolution of the noise in the excess of the Brownian contribution, including some long-term drift of the acceleration; 
\end{enumerate}
Section~\ref{sec:excess} concludes the part of the paper dedicated to experimental findings.
{To make the main text more readable, we have put significant information on these sections in the Appendixes. In particular, Appendix~\ref{app:CPSD} contains a detailed description of our spectral estimation methods, some aspects of which we could not trace in the standard literature on data processing.

The paper then proceeds to the discussion of the findings above. Such discussion is based on many extensive analyses, some of which could merit stand-alone papers. To ensure comprehensive coverage  of the relevant LPF experimental evidence within a single article, we have included these analyses here. To maintain readability, these analyses are presented in a series of Appendixes, and we specify in the title of each discussion section the relevant Appendixes where the details of the underlying analyses can be found.

As for the discussion within the main body of the paper, this is organized as follows:
\begin{enumerate}[label=(\roman*)]
\item In Sec.~\ref{sec:brdisc} we discuss the implication of the findings on the Brownian noise and  very-low-frequency behavior for the gas environment of the test masses;
\item In Secs.~\ref{sec:excdisc} and~\ref{sec:noiseprojection} 
we discuss several possible noise contributions for which we have quantitative estimates;
\item In Sec.~\ref{sec:nonmod} we discuss the possible sources for the part of the noise that we could not account for in the previous sections and identify possible measures to keep it under control in LISA; 
\item In Sec.~\ref{sec:con} we finally give some short concluding remarks.
\end{enumerate}
}

\begin{figure}[htbp]
  \centering
  \includegraphics[width=0.95\columnwidth]{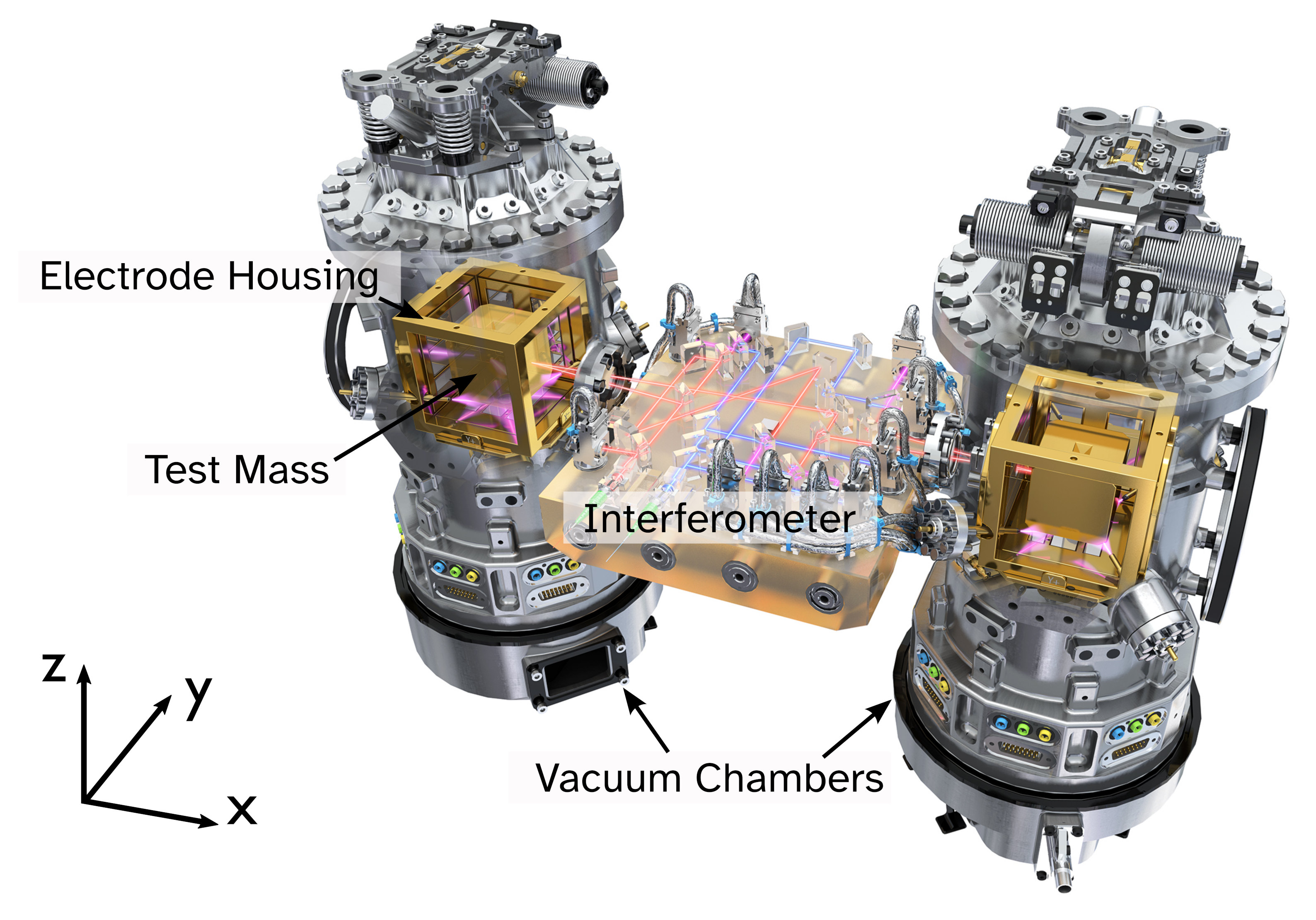}
  \caption{Rendering of the  LISA technology package. The rendering shows the two test masses hosted inside their respective electrode housings (some of the electrodes are not represented) and the vacuum chambers enclosing both test masses and electrode housings. The picture also shows the high stability optical bench hosting all interferometric readouts and many other features of the instrument, such as the launch lock, UV-light-based test mass neutralizer, etc. that are not relevant here. (Credits: ESA/ATG medialab)}
  \label{fig:LPFfigure}
\end{figure}

\section{\label{sec:exp}Summary description of the experiment} 
\subsection{The LISA technology package}
The instrument flown on LPF, the LISA technology package (LTP), has been described in detail in \cite{LPFPKS}. Here we summarize its essential features.


The LTP, depicted in Fig.~\ref{fig:LPFfigure}, carried two cubic Au-Pt TMs each with a mass of $M= \SI{1.928}{kg}$ and size $l=\SI{46}{mm}$. During operations, these TMs had no mechanical contact with their surroundings, and were nominally ``free falling'', each one at the center of a housing leaving 3 to \SI{4}{mm} clearance gaps to the faces of the TM. Each of these ``electrode housings'' (EHs) carried a series of electrodes facing all faces of the respective TM. These electrodes were used for two purposes. First, they were part of a capacitive sensor of the motion of the TM, relative to its housing, for all degrees of freedom (d.o.f). Second, they were used to apply electrostatic forces and torques to the TM, whenever needed.

Each EH, with its respective TM, was hosted inside a vacuum chamber, called a vacuum enclosure (VE), sealed by a dedicated vacuum valve. The valve connected the VE to a second outer volume that was connected by a vent duct to the outside of the spacecraft. 

The VE was needed both to allow for vacuum preparation on ground and to shield the TM from the spacecraft internal outgassing that had been predicted to be too large to achieve the desired vacuum level around a nonshielded TM. Further details about the vacuum preparation and handling are given a bit later.

In what follows we call the gravitational reference sensor (GRS) the system of the TM, its EH and VE, and all related accessories.

The main sensor for the relative motion of the TMs was a heterodyne laser interferometric system, called the optical metrology system (OMS)  \cite{PhysRevLett.126.131103}. For the purpose of this paper, it is important to recall that the interferometer measured six different d.o.f.: the relative displacement $\Delta x(t)$ between the TMs along the sensitive $x$-axis, joining their respective centers of mass; the relative displacement $x_1(t)$, along the $x$-axis, of one of the TM, called TM1, relative to the interferometer optical bench and, as a consequence, also relative to the spacecraft; the two angles of rotation $\eta (t)$ and  $\phi(t)$ for both TMs, around the $y$-axis and $z$-axis respectively. These six TM d.o.f., and also the remaining six, were also measured at all times by the capacitive sensors \cite{PRD_96_062004_capacitive}. However, the interferometric readout was approximately three orders of magnitude more sensitive than the capacitive one for all d.o.f. for which they were both available. All the data series analyzed in this paper have a sampling rate of \SI{10}{\hertz}.

In addition to the measurement of the TM motion, other physical quantities have been measured, by dedicated instruments, throughout the mission. In particular, we measured: the magnetic field vector at various locations, via a dedicated set of magnetometers \cite{magnetic-mnras}; the temperature at various critical locations,  via a dedicated set of thermistors  \cite{temperature}; the cosmic ray flux with a radiation monitor \cite{Canizares_2011,armano_characteristics_2018}.

Before closing this section, we want to specify some details of the vacuum handling that are useful for some of the discussions in the following sections. 

Once the  VE was evacuated on ground,  the vacuum valve was closed and the VE remained sealed for a few months until the valve was opened again once in orbit. Valve opening took place, within an hour for both VE, on February~3,~2016, i.e., \SI{62.4}{\day} after the launch. In the rest of the paper, we call $t_v$ this time of valve opening, while we call  $t_0$ the time of the launch, December~3,~2015 04:04:00~UTC. The outer volume, with its vent duct, was not evacuated on ground, and, having no seal, was exposed to space immediately after $t_0$.

In addition to the TM and EH, the VE contained various metal structural elements, two piezoelectric motors, used to center the TM within the EH and release it into free fall, and various thermistors and heaters. The VE also contained the $\sim 40$ cables needed for all these appliances, which crossed the VE wall through a set of vacuum feedthroughs. Finally, also included in the VE was a $\sim\SI{2}{kg}$ tungsten balance mass used to suppress the gravitational field at TM location \cite{gravity}. The outer chamber contained a high-output paraffin motor \cite{tibbitts_high-output_1992} used to activate the vacuum valve and the TM launch lock, with all the necessary cables.  

\subsection{\label{sec:dyn} Dynamical controls and data series formation}
LPF was a controlled dynamical system consisting of the spacecraft and the two TMs. More specifically, the spacecraft was forced to follow one of the TMs (TM1) along $x$ via an active control loop, using the spacecraft cold gas microthrusters as actuators \cite{PRDThrust}, known as drag-free control. 

Each TM rotation along $\phi$ and $\eta$ was kept fixed relative to the spacecraft by an active loop using electrostatic torques. These torques were applied via the above-mentioned electrodes.

No electrostatic actuation force was applied along $x$ on TM1, while a control loop (electrostatic suspension) kept the distance between the two TMs nominally fixed, by applying a suitable electrostatic force along $x$ on the other TM (TM2). All other degrees of freedom were also controlled, but the details are not relevant here.

{As the distance between the TMs was actively controlled using the $\Delta x$ signal the relative acceleration  $\Delta \ddot{x}$ could not be taken as a measurement of the differential disturbance force per unit mass that would act on the TMs in the absence of the control \cite{armano:subfemtog}. However, the applied feedback forces per unit mass $g_c(t)$ were known, so that they could be subtracted from $\Delta \ddot{x}$, which had been estimated numerically \cite{derivative},  to give an accurate estimate of  $\Delta g_\text{ext}$.}


In addition, acceleration data series were also corrected for the following effects:
\begin{enumerate}[label=(\roman*)]
    \item  Measured  inertial forces per unit mass  due to spacecraft rotation $g_i(t)$, which include the  centrifugal and the Euler force \cite{PhysRevLett.120.061101}. These effects will not be relevant for LISA  \cite{lisa-performance-model}.
    \item The  forces per unit mass generated by the motion of the TMs through  static force gradients in the spacecraft, as LISA data  can also be corrected for those.  Such force acting on TM$i$ is well approximated by the linear model $-\omega_i^2 x_i$, as described in \cite{PhysRevLett.120.061101}. 
    \item The interferometer spurious pickup $g_\text{CT}(t)$ of spacecraft motion along  d.o.f. different from $x$, due to crosstalk \cite{PhysRevD.108.102003_LPF_TTL_2023,PhysRevD.108.022008_TTL_LPF_Analytical}. This also includes the pick-up of the common mode motion of the TMs, described by a term $\delta_{x_1} \ddot{x}_1$. Some of these effects will also be present in LISA and will be analogously corrected \cite{lisa-performance-model}.
\end{enumerate}

Thus the  {corrected} differential force per unit mass data series used in the following analyses can be written as 
\begin{equation}
	\label{eq:deltag}
	\begin{split}
	    \Delta g(t)=& \Delta\ddot{x}_\OMS(t) +\omega_2^2 \Delta x_\OMS(t)+ (\omega_2^2-\omega_1^2) x_{1,\OMS}(t)\\& - g_c(t)-g_i(t) - g_\text{CT}(t).
	\end{split}
\end{equation}
Note that, in Eq.~\eqref{eq:deltag}, we have attached the suffix $\OMS$ to all coordinates to indicate that these have been measured by the relevant interferometers and not by the capacitive sensors. For these we will use the $\GRS$ suffix. Note also that  $\omega_1^2$ and $\omega_2^2$ above, as well as $\delta_{x_1}$, have been measured in dedicated calibration experiments \cite{PhysRevD.97.122002}. In particular $\omega_2^2 \approx \SI{-4.5e-7}{s^{-2}}$ is negative, while the differential stiffness $(\omega_2^2- \omega_1^2)$ is roughly 20 times smaller and thus neglected in this discussion here.

The feedback force  $g_c(t)$ in Eq.~\eqref{eq:deltag}, is the time series of the force that the onboard computer has commanded to the electrostatic force system. It was calibrated through an extensive series of  experiments \cite{PhysRevD.97.122002}, consisting of  the application to the TMs of a series of  forces oscillating at frequencies $\gtrsim \SI{1}{mHz}$ and of the measurement of the resulting acceleration, a quantity which is intrinsically calibrated in terms of the laser wavelength and of the onboard clock calibration.

This calibration campaign led us  to discover a systematic,  nonlinear error in the electronics, originating from an overlooked truncation of the digital voltage  commands \cite{ActNeda} that resulted in an amplitude-dependent fluctuation of the calibration. Therefore $g_c$ had to be recalculated  considering such an extra truncation. After this crucial correction, the calibration of $g_c$ was found to be stable at better than  $<1\%$,  independent of the amplitude of the applied forces and torques, throughout the entire mission \cite{PhysRevD.97.122002}.

$\Delta g (t)$ in Eq.~\eqref{eq:deltag} is our best estimate for the external differential force per unit mass $\Delta g_\text{ext}$. However the series is corrupted by the noise  $n_\OMS(t)$ in the differential interferometer readout $\Delta x_{\OMS}$. Such disturbance enters into $\Delta g(t)$ in Eq.~\eqref{eq:deltag}, both through $\Delta\ddot{x}_\OMS(t)$, and through $\omega_2^2 \Delta x_\OMS(t)$. Thus, the residual noise in $\Delta g$ can be evaluated as
 \begin{equation}
     \label{eq:deltag2}
     \Delta g = \Delta g_\text{ext}(t)+\ddot{n}_{\OMS}(t)+\omega_2^2 n_\OMS(t).
 \end{equation}

Finally, it is important to mention that occasional force transients were observed in the data \cite{lpf_glitch2022}. In ordinary runs these glitches occurred at an average rate of $\simeq 1\,\text{d}^{-1}$. These glitches have been removed from the data, as described in Ref.~\cite{lpf_glitch2022}, before any noise analysis.

\subsection{\label{sec:data runs}Data runs}
The mission scientific operations lasted from March~1, 2016 to July 18, 2017. During these more than 16 months, we performed  many uninterrupted ``noise runs'' during which the TMs and the satellite were in steady control conditions, with no purposely applied stimulus of any nature. 

We have performed many such noise runs, however we restrict the main analysis  to those with an overall duration of at least 2.5 days. Such a duration allows an estimate of acceleration power spectral density (PSD), down to about $\simeq \SI{18}{\micro\hertz}$ with reasonable accuracy  (see following sections).

We list in Table~\ref{tab:noise-only-runs}, for all these runs, the start and stop dates, the duration, and the instrument operating temperatures. Temperature values are the average of the eight thermometers placed on the two GRSs as described in \cite{temperature}. We note that 11 of these runs were performed at an operating temperature around 21--22~\si{\celsius}, while runs~10 and 11 were operated at around \SI{11}{\celsius}.



\renewcommand{\arraystretch}{1.2}
\begin{table}[htbp]
  \caption{List of the considered noise runs. {Run~2, of April 2016, overlaps the data published in \cite{armano:subfemtog}. The data in Ref.~\cite{armano:subfemtog} however, were presented without glitch removal, and as such four days at the beginning of the run were omitted from analysis due to a large glitch. In addition, for homogeneity with the other runs, we have removed  about one day at the end of   the time series of Ref.~\cite{armano:subfemtog} which contained a calibration signal.} Run~10 is the February 2017 run published in \cite{PhysRevLett.120.061101}.}
  \label{tab:noise-only-runs}
  \begin{tabular}{ccccc}
   No.& Start date & Duration\;& Mean time &\;Temperature\\
   &  & (d)& from launch (d) &(K)\\
    \hline
   1& ~\texttt{2016-03-21}~ &  5.3 &112&$295.37\pm 0.04$  \\
   2& \texttt{2016-04-04} &  9.3 &127&$295.30\pm 0.03$ \\
   3& \texttt{2016-05-16} &  3.2 &166&$294.97\pm 0.06$ \\
   4& \texttt{2016-06-19} &4.8 &202& $294.93\pm 0.01$\\
   5& \texttt{2016-07-17} &   2.8 &229& $296.62\pm 0.05$\\
   6& \texttt{2016-07-24} &  5.3 &237&$296.50\pm 0.02$ \\
   7& \texttt{2016-09-28} & 2.8 &302&$296.50\pm 0.02$ \\
   8& \texttt{2016-11-16} &  9.9  &354&$296.86\pm 0.04$\\
    9&\texttt{2016-12-26} & 18.5  &398&$295.38\pm 0.04$\\
   10& \texttt{2017-02-14} &  13.3 &446&$284.72\pm 0.03$\\
   11& \texttt{2017-05-18} &  4.3 &535&$284.2\;\pm 0.1$ \\
  12&  \texttt{2017-05-29} &  6.8 &547& $295.78\pm 0.01$\\
  13&  \texttt{2017-06-08} & 8.6  &558&$295.91\pm 0.03$\\
  \end{tabular}

\end{table}


The detailed operating conditions have been slightly different for different runs. We have listed in Table~\ref{tab:conf},  in  Appendix~\ref{app:runconf}, the few differences of settings that may bear some relevance for the noise performance.

For runs 1 to 11, the total charge of both TMs was maintained within the interval of $\pm\,3\times 10^7\,e$  \cite{PhysRevD.107.062007_Charging2023}, in order to keep the noise caused by voltage fluctuations negligible \cite{PhysRevLett.118.171101} (see Sec.~\ref{sec:decorrMCMC_noiseprojection}). For those runs, charge was measured just before and just after each run. For runs 12 and 13, the charge was measured just once before run 12. We could, however, extrapolate the missing final values from the measurement of cosmic ray flux \cite{PhysRevD.107.062007_Charging2023,ARMANO201828}. For run 12, such extrapolated final value is still in the aforementioned interval, while for run 13 it might have gone up to $4\times 10^7\,e$.

TM neutralization was performed using a noncontacting UV discharging system \cite{PhysRevD.98.062001} before the start of each run. As cosmic rays resulted in a steady increase of the charge, for most of the runs the charge was brought to a negative value within the $\pm\,3\times 10^7\,e$ interval at the start of the run, to have it drifting through zero during the run.

Two major events must be mentioned. First ESA operated LPF,  until April 6, 2017, on a Lissajous orbit near the L1 point of the Earth-Sun system. This orbit was unstable, and mission control had to make station-keeping maneuvers every few weeks to maintain it. On April 6, 2017 in preparation for the end of the mission,  mission control performed a final maneuver, called ``deorbiting burn'', to place the spacecraft on a trajectory that would not risk intercepting the Earth. The maneuver lasted a few days, and after that, until the end of the mission, no further station-keeping maneuver was performed.

In addition, between April 30, 2017 and  May 16, 2017, with the purpose of further reducing noise beyond the level reached in February 2017 \cite{PhysRevLett.120.061101}, we tried to operate the instrument near  \SI{0}{\celsius}, a value well outside its nominal operating range. The instrument entered into a rather unstable state with a very high rate of glitches \cite{lpf_glitch2022} and went back to its ordinary behavior only when the temperature was raised back to a value within its operating range.

During such \SI{0}{\celsius} cooling, we were able to perform a noise run lasting from 2017-05-03 to 2017-05-09 (for a total of \SI{5.8}{d}), not listed in Table~\ref{tab:noise-only-runs}. In this run, despite the low quality of the data, we were able to perform some noise measurements at frequencies around 1\,mHz, which we will mention later in the paper, in Sec.~\ref{sec:brown}.

Unfortunately, the temperature of this ``cold run'', has not been measured directly, as the electronics of the GRS thermometers clipped at $\simeq \SI{7}{\celsius}$. However we found that, upon a proper calibration, the average of a group of thermometers just outside the instrument bay was a good proxy of the average of the GRS thermometers, at all temperatures above $\simeq \SI{7}{\celsius}$. This proxy gives, for the cold run,  $T\simeq\SI{1.7}{\celsius}\simeq\SI{274.8}{K}$  with an overall uncertainty not larger than \SI{0.5}{K}. 

{As a final note we want to mention that the station-keeping maneuvers limited the maximum duration of noise runs to one week until May 2016,  then to two weeks through November 2016, and finally to three weeks until the deorbiting maneuver in April 2017. In addition, run duration planning had also to take into account the need to use operation time for a variety of other planned experiments \cite{vitale:judo}. Nevertheless, given that LPF science requirements \cite{LPFPKS} required to match acceleration requirements down to just 1\,mHz, and  to just measure acceleration  noise down to 0.1\,mHz,  run durations are all fulfilling such frequency requirements with significant margin.}

\section{Acceleration PSD}
\label{sec:accPSD}
During all noise runs we have estimated the PSD of $\Delta g(t)$, $S_{\Delta g}(f)$ [or, equivalently, its square root, the amplitude spectral density (ASD)], as a function of frequency $f$,  by using the periodogram estimation method explained in Appendix~\ref{app:PSDestimate}. The method gives a Bayesian estimate for the posterior distribution of $S_{\Delta g}(f_i)$  over a given set of frequencies ${f_i}$.  
The frequencies within the set have been chosen such that the PSD estimates at different frequencies have minimal statistical correlations (see Appendix~\ref{app:CPSD/freqs} and \cite{PhysRevLett.120.061101}). In addition, for practical reasons, we have adjusted the selection  such that the fourth frequency is 0.1~mHz, the lower bound of the official LISA band.

\begin{figure}[htbp]
  \centering
  \includegraphics[width=1\columnwidth]{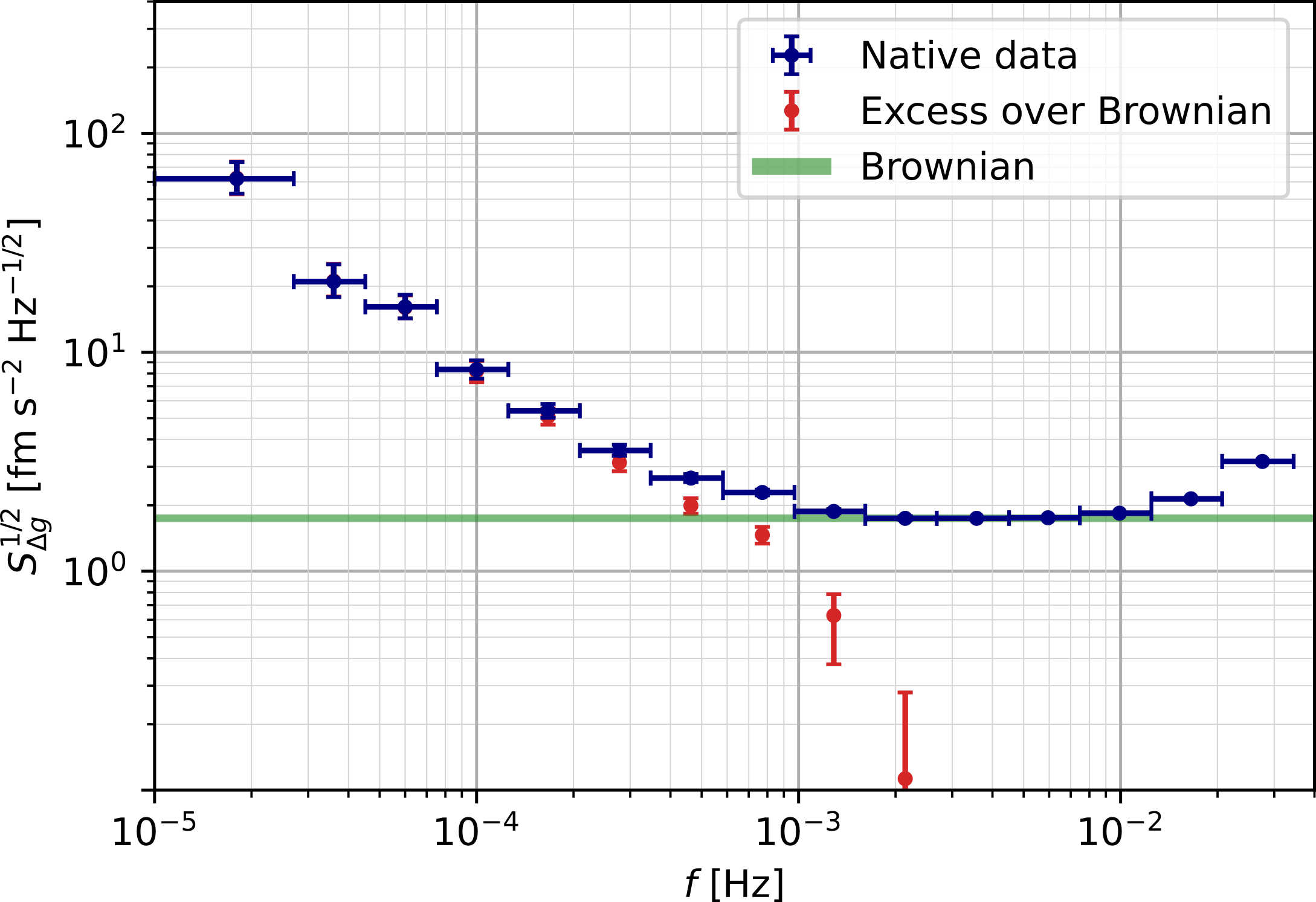}
  \caption{Blue points: ASD of  $\Delta g(t)$, $S_{\Delta g}^{1/2}$ for run~10 of Table~\ref{tab:noise-only-runs} (February 2017) as a function of frequency $f$.  The vertical error bars delimit the 1$\sigma$ interval (see Appendix~\ref{app:CPSD} for the definition). The horizontal bars indicate the effective width of the spectral window contributing to each of the points. Red points, estimated excess over Brownian noise; green thin band, uncertainty band for the estimate of the Brownian noise.}
  \label{fig:brownsubtr}
\end{figure}

The blue points in Fig.~\ref{fig:brownsubtr} illustrate the result of the procedure for run~10 of Table~\ref{tab:noise-only-runs}, which is the February 2017 run that we published in \cite{PhysRevLett.120.061101}. 

The figure shows that, as already noted in \cite{PhysRevLett.120.061101}, the ASD and the PSD have three different branches:
\begin{enumerate}
    \item a low-frequency branch with an approximate $f^{-1}$ behavior ($f^{-2}$ for the PSD);
    \item an approximately frequency-independent branch above about 1\,mHz;
    \item a rising branch above about 10\,mHz.
\end{enumerate}
As explained in \cite{armano:subfemtog}, the rising branch is due to the interferometer readout noise $n_\text{OMS}$. The details on the origin of  this branch, dominated by interferometer phase readout and frequency fluctuations,  may be found in \cite{PhysRevLett.126.131103}. We will not discuss it any further.

In \cite{armano:subfemtog} we have attributed the frequency-independent branch, with PSD value $S_\text{Brown}$, to Brownian noise due to gas damping.  To separate it from the other two branches,  we use, for each run, the following procedure.

\begin{figure*}[htbp] 
  \centering
  \includegraphics[width=1\textwidth]{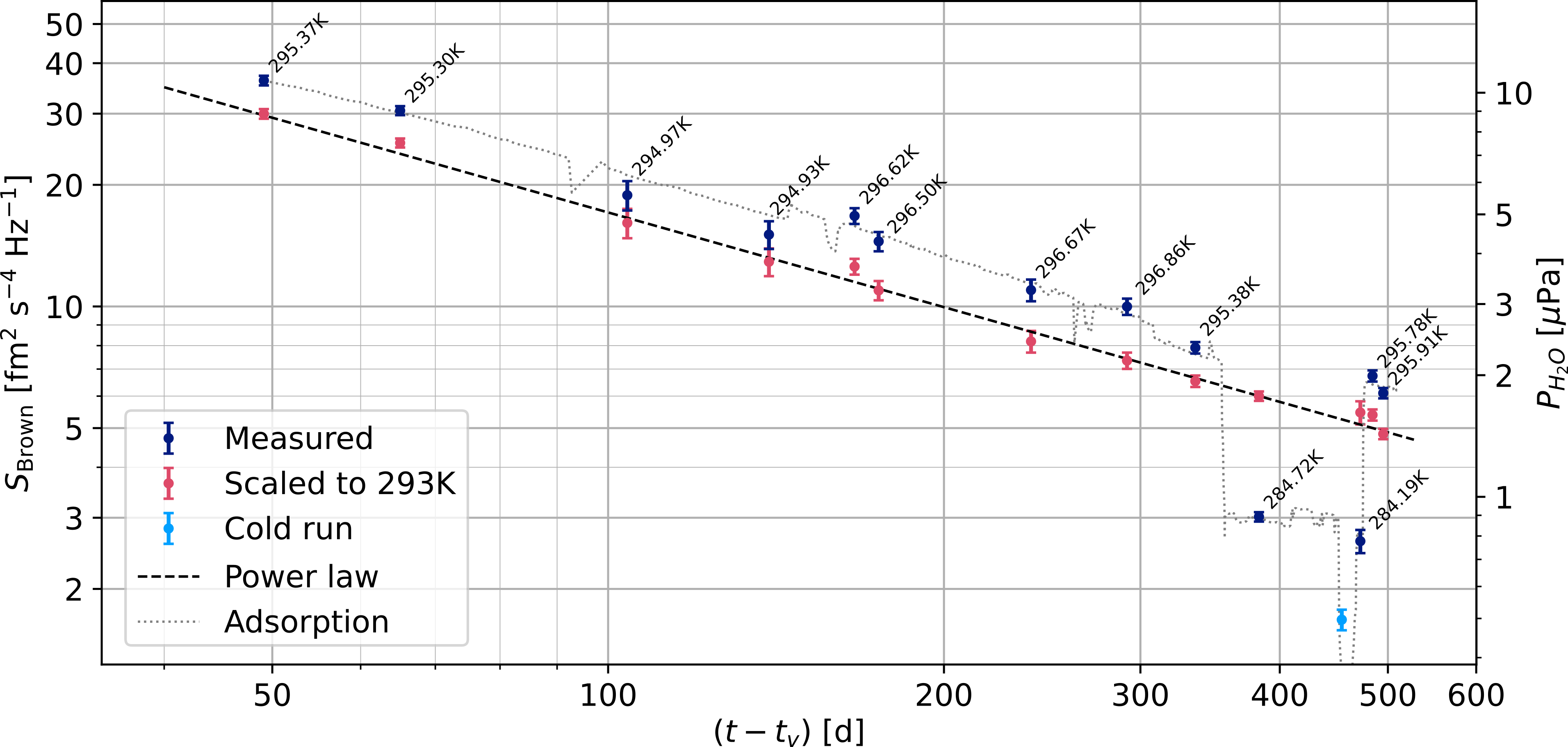}
  \caption{Dark blue dots, PSD of Brownian noise $S_\text{Brown}$ as a function of  $(t-t_v)$, the time  since venting of the vacuum enclosure. Error bars correspond to 1$\sigma$ intervals. For convenience, call outs repeat the mean temperatures $T$  of the runs from Table~\ref{tab:noise-only-runs}. Light blue dot, $S_\text{Brown}$ for the cold run ($T\sim\SI{275}{\kelvin}$).
  Data may be converted into the mean of the pressures $\Pwat$ in the two GRSs (right vertical scale) by using the conversion factor in \cite{PhysRevLett.103.140601} and by assuming that the residual gas consists of water molecules.
  Red dots, PSD of Brownian noise of ordinary runs, scaled as $S_\text{Scaled}=S_\text{Brown}\,e^{-T_a\left(\frac{1}{T}-\frac{1}{T_0}\right)}$, with  $T_0=\SI{293}{\kelvin}$ and $T_a=\SI{7.0}{\kilo\kelvin}$; dashed black line,  $S_\text{Scaled}(t)\propto(t-t_v)^{-0.80}$. See text for details; dotted gray line, best fit assuming adsorption quasiequilibrium. See Sec.~\ref{sec:brdisc} for details.}
  \label{fig:brown}
\end{figure*}

\begin{enumerate}[label=(\roman*)]


\item We fit the experimental PSD data $\Pi_{\Delta g,k}(f_i)$ of  run $k$ to the PSD model 
\begin{equation}
\label{eq:def_excessbrown}
S_{\Delta g,k}(f_i)=S_{\Delta g_{e},k}(f_i)+S_{\text{Brown},k},
\end{equation}
that is, with a frequency-independent term $S_{\text{Brown},k}$, plus the excess $S_{\Delta g_e,k}(f_i)$ that depends on the frequency $f_i$.
\item To avoid numerical instabilities, we limit the analysis to $f<\SI{3}{mHz}$, that is, to the lowest ten frequencies of Table~\ref{tab:freq}. However, to better constrain $S_{\text{Brown},k}$ we add one further data point consisting of the minimum value of  $\Pi_{\Delta g,k}(f_i)$ for $i>10$, $\Pi_{\Delta g,k,\text{min}}$. We fit this data point with the simpler model $S_{\Delta g,k,\text{min}}=S_{\text{Brown},k}$, that is, with no excess over the Brownian noise; this is a reasonable assumption, also supported by the results of the following analyses. This additional term also prevents the fit to put $S_{\text{Brown},k}$ to zero.
\item The fit is a  Monte Carlo Markov Chain (MCMC) estimate of the posterior 
\begin{equation}
    \begin{split}
        &p\left(S_{\Delta g_{e},k}(f_i),S_{\text{Brown},k}\vert \Pi_{\Delta g,k}(f_i),\Pi_{\Delta g,k,\text{min}}\right)_{i\in[1,10]}\propto\\
        &\propto \prod_{i=1}^{10}p\left(\Pi_{\Delta g,k}(f_i)\vert S_{\Delta g_{e},k}(f_i)+S_{\text{Brown},k}\right)\times\\
        &\times p\left(\Pi_{\Delta g,k,\text{min}}\vert S_{\text{Brown},k}\right)
    \end{split}
    \label{eq:post}
\end{equation}
Here the function $ p(\Pi \vert S)$, the probability distribution of the experimental PSD, is that given in  Appendix~\ref{app:PSDestimate}.
\item We scan the domain of the logarithm of all parameters $S_{\Delta g_{e},k}(f_i)$ and $S_{\text{Brown},k}$, and do not multiply the posterior in Eq.~\eqref{eq:post} by any explicit prior. This is equivalent to using the Jeffreys prior on all parameters \cite{Jeffreys}. 
However, again to avoid numerical instabilities, we constrain all PSD parameters, both Brownian and excess, to be larger than $1\,\text{am}^2\,\text{s}^{-4}/\text{Hz}$.
\end{enumerate}

The results of such procedure are again shown in Fig.~\ref{fig:brownsubtr} for run~10 of Table~\ref{tab:noise-only-runs} and for all runs in Figs.~\ref{fig:batch1} and~\ref{fig:batch2} in Appendix~\ref{app:allpsd}. 

Note that Figs.~\ref{fig:batch1} and~\ref{fig:batch2} show that the procedure gives consistent results for all runs, with the Brownian uncertainty band always very close to the average of native data in the range $\SI{2}{\milli\hertz}\le f \le \SI{5}{\milli\hertz} $.

\section{\label{sec:brown}Time evolution of Brownian noise}
From the procedure described in the previous section, we obtain an estimate of $S_\text{Brown}$ for all runs. Results are plotted in Fig.~\ref{fig:brown}, as blue dots, as a function of  $t-t_v$, the time since venting to space of the VE.

As discussed in \cite{armano:subfemtog,PhysRevLett.120.061101},  the PSD of Brownian noise due to gas damping is 
\begin{equation}
S_\text{Brown}=\frac{4 k_B T}{M^2} P l^2 \epsilon \left(\frac{32\,m}{\pi k_B T}\right)^{1/2}\equiv \kappa P
\label{eq:brownbase}
\end{equation}
where $P$ is the gas pressure, $l$, as already defined, is the side length of the test mass,  $m$ is the
mass of the gas molecules, and $\epsilon$ is a coefficient of proportionality that depends on the EH and TM  geometry.
Reference~\cite{PhysRevLett.103.140601} estimates $\kappa$ to be  $\kappa \simeq \SI{1.7}{\femto\meter^2\,\second^{-4}\,\hertz^{-1}/\micro\pascal}$, for a single LPF TM, for $T=293\,\text{K}$, and for a  gas consisting of water molecules. Thus, if indeed the frequency-independent branch represents Brownian noise, $\Pwat=S_\text{Brown}/(2\kappa)$, which can be read on the right vertical axis of Fig.~\ref{fig:brown}, gives a measurement of the mean of the pressures of the two GRSs. Note that, with the effective conductance from the interior of the  EH to the outer space estimated to be $\simeq \SI{19}{\liter/\second}$ for water, a pressure of $\SI{1}{\micro\pascal}$ corresponds to an outgassing rate of $\SI{1.9e-7}{\milli\bar\,\liter/\second}$.

The figure also shows that, remarkably, the data can be scaled to follow a straight line, by just multiplying them by a single ``activation'' factor $e^{-T_a\left(\frac{1}{T}-\frac{1}{T_0}\right)}$, with $T_a$ a properly chosen activation temperature, and $T_0$ an arbitrarily chosen, convenient common temperature for the scaled data. {Note that this is a much stronger temperature dependence, a factor 2.5 for a change in temperature of approximately 10\,K,  than that due to the  $ T^{3/2}$ factor in Eq.~\eqref{eq:brownbase}, $\simeq 3\%$ for the same 10\,K, and is dominated by  the  temperature dependence of the outgassing, and thus of $P$.}

{The scaled data may be fit to
\begin{equation}
\label{Eq: brown}
    S_\text{Brown}(t)= a \left(\frac{t_v}{t-t_v}\right)^\gamma
\end{equation}
where $a$ and $\gamma$ are two fitting parameters (see Fig.~\ref{fig:brown}). Actually, a simultaneous fit leaving also $T_a$ as a free parameter gives $T_a=(7.0\pm0.2)~\text{kK}$, $\gamma=(0.80\pm0.02)$, and finally $a=(27.0\pm0.7)~\si{\femto\meter^2\,\second^{-4}/\hertz}$.

Such power-law behavior is commonly observed, after an  initial rapid pressure decay phase, during the pump down of vacuum systems \footnote{{The presence of the initial decay phase implies that the law should not be extrapolated back to the start of the pump down, that is  when $t\to t_v$ in Eq.~\eqref{Eq: brown}.}} \cite{jousten_handbook_2016}. The exponent $\gamma$ is found to be  $\gamma \simeq 0.5$ in the case of diffusion-dominated outgassing, and  $\gamma\simeq 1$ in the case of isothermal quasiequilibrium  of water with the vacuum chamber metal walls, in both cases with good agreement with simple models for the underlying phenomena. Intermediate values for $\gamma$ are found for more complex systems \cite{chiggiato}.}




Given the uncertainty on the temperature and on the quality of the PSD data, we have left the cold run datum out of rescaling and  fitting. This is further discussed in Sec.~\ref{sec:brdisc}. Here we only note that the  cold run datum is lower than those for the lowest temperature ordinary runs. This shows that the TM acceleration noise in the millihertz bandwidth is truly Brownian down to a level well below $\SI{2}{\femto\meter\,\second^{-2}/\rtHz}$ per TM and not saturating to a significant level reflecting other important noise sources in this band.

For the sake of the discussion in Sec.~\ref{sec:brdisc}, we also report in Fig.~\ref{fig:brown}  the best fit for one common model for vacuum evolution under the hypothesis of quasiequilibrium between surface water readsorption and outgassing. The discussion on Brownian and vacuum environment continues in the mentioned Sec.~\ref{sec:brdisc}.

\section{\label{sec:excess} Stability and time evolution of  excess noise}
In this section we discuss the stability and evolution of $S_{\Delta g_{e}}$, the excess noise over Brownian, over the duration of the mission. The 
details of the variations of $S_{\Delta g_{e}}$ can be tracked in Figs.~\ref{fig:batch1} and~\ref{fig:batch2} in Appendix~\ref{app:allpsd}. Here we present a few summary analyses with the main aim of quantifying noise stationarity. 

We first discuss the compatibility of the observations at each frequency with a model considering a single, stationary, run-independent noise process.

Having identified significant variations from run to run of the lowest frequency datum, mostly in the initial phase of the mission,  we separate the discussion on the time evolution of the excess noise for $f>\SI{18}{\micro\hertz}$  from that at $f=\SI{18}{\micro\hertz}$, which we treat in Sec.~\ref{sec:lfbin} together with long-term drifts. 

\subsection{\label{sec:common}Fit to a common stationary Gaussian excess for all runs}
\begin{figure*}[htbp]
  \centering
  \includegraphics[width=0.85\textwidth]{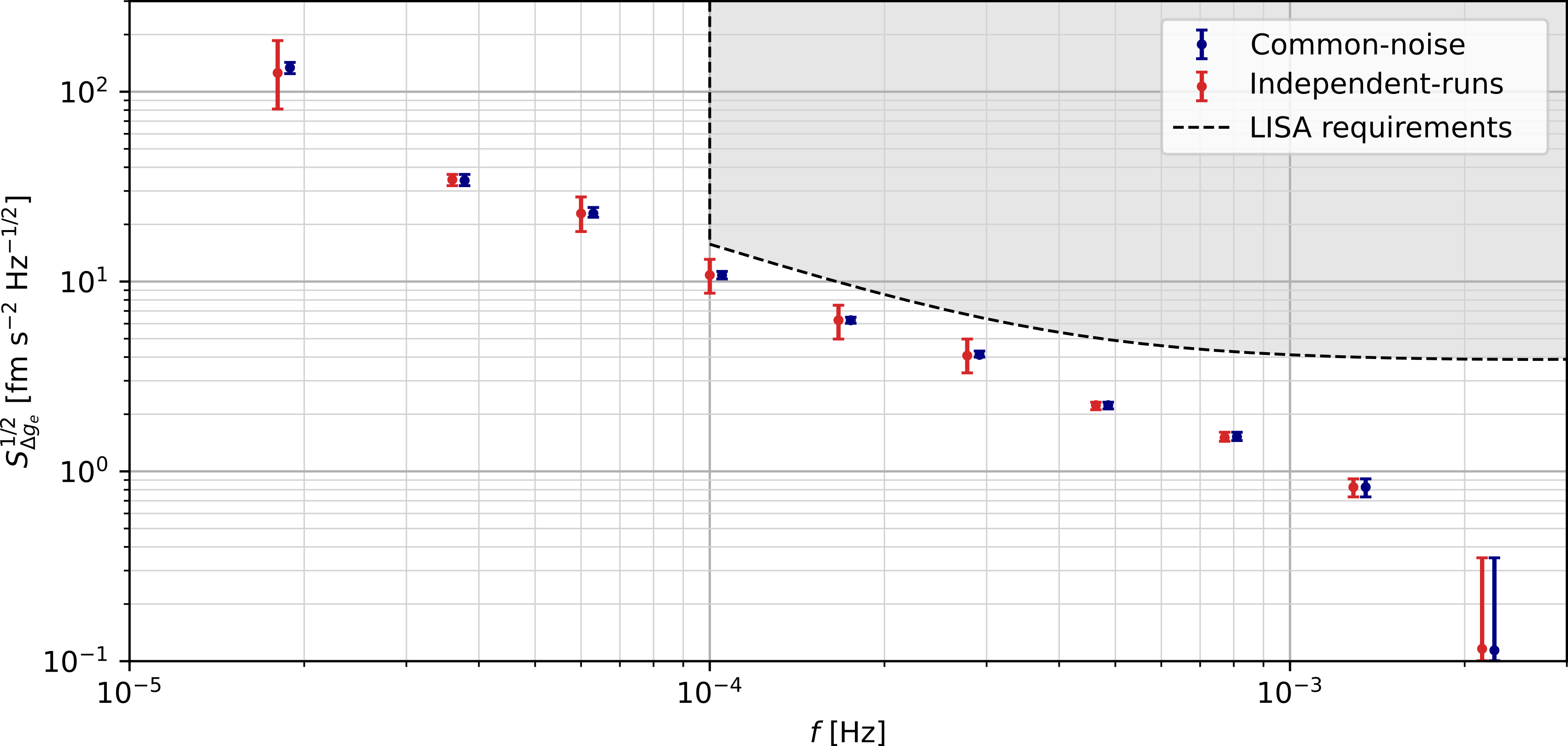}
  \caption{Summary of  ASD of $\Delta g$ excess noise, over all runs of Table~\ref{tab:noise-only-runs} and Figs.~\ref{fig:batch1} and \ref{fig:batch2}, as a function of the frequency.  Blue data, median and $\pm 1\sigma$ quantiles of the inferred posterior distribution of the excess for the entire set of runs, common-noise model (see text for details); red data, median and $\pm 1\sigma$ quantiles of the inferred posterior distribution, independent-runs model (see text for details).
  Red data have been slightly shifted in frequency for the sake of readability, but they all refer to the same frequency as that of blue data. Also reported are current LISA requirements adapted from \cite{SciRd}.}
  \label{fig:average}
\end{figure*}

In Fig.~\ref{fig:average} we report, at each frequency $f_i$, the inferred posterior distribution for $S_{\Delta g_e}(f_i)$ (blue data points) assuming that the excess is a common number to all runs, $S_{\Delta g_{e},k}(f_i)=S_{\Delta g_{e}}(f_i)$. We call this the ``common-noise model''. The inference is done by using a collective posterior consisting of the product of the posteriors in Eq.~\eqref{eq:post}, for all values of $k$, having dropped the dependence of the excess on $k$.

We have done a simple Akaike information criterion comparison \cite{Akaike} between the common-noise model and the model consisting of the separate Bayesian fits to the data of the 13 different runs discussed in Sec.~\ref{sec:accPSD}. The posterior for the latter is just the product of the posteriors used for the separated fits and thus depends on  143 parameters: 10 excess noise coefficients for each of the 13 runs, plus 13 Brownian noise coefficients. The common excess noise model depends instead on only 10 excess noise coefficients and 13 Brownian noise ones, for a total of 23 coefficients.\\
The Akaike test favors the common-noise model, with a relative likelihood ratio of $\simeq 10^{-20}$. This indicates that the hypothesis of completely independent excess noise values across the runs substantially overfits the data. However, this does not tell if the best of the two fits is a good fit. 

We have then done a posterior predictive check \cite{gelman2020bayesian} on the common-noise model described in the following. We first find the set of parameter values $\hat{\theta}_\text{best}=\{S_{\Delta g_{e}}(f_1),S_{\Delta g_{e}}(f_2),\dots, S_{\text{Brown},1},S_{\text{Brown},2},\dots\}$ that maximizes the posterior likelihood.  

Having $\hat{\theta}_\text{best}$,  for any set of data, either  observed or simulated from the posterior, we can then  calculate, as a measure of discrepancy at any given frequency $f_i$, the log-likelihood of the data at that frequency, conditional on $\hat{\theta}_\text{best}$
\begin{equation}
    \Lambda_i=\sum_{k=1}^{13}\log\,p\left(\Pi_{\Delta g,k}(f_i)\vert\hat{\theta}_\text{best} \right)
\end{equation}

Note that in an ordinary least square fit to independent Gaussian data, the log-likelihood would be, modulo an irrelevant additive constant,  just the sum of the square residuals, a quantity commonly used to measure the discrepancy between the data and the model. Thus the use of $\Lambda_i$ may be seen as just an extension of the method of square residuals to Wishart distributed data.

We have then generated simulated data by sampling the posterior for the parameters $\{S_{\Delta g_{e}}(f_2),S_{\Delta g_{e}}(f_3),\dots, S_{\text{Brown},1},S_{\text{Brown},2},\dots\}$, and by generating then simulated periodograms $\Pi_{\Delta g,k}(f_i)$ from the proper Wishart distribution. For each simulated periodogram, we have calculated  $\Lambda_i$, with $1\le i\le9$  obtaining then an expected posterior predictive distribution for each of these parameters. 

We have restricted the analysis to frequencies $ f_i \le f_9$, as,  at higher frequencies, the model breaks down anyway due to the dominance of Brownian noise, particularly for the earlier runs.

We have then calculated the probability $p$ that $\Lambda_i$ is less than value $\Lambda_{i,0}$ calculated for the real data, that is the cumulative distribution function for $\Lambda_i$ evaluated in $\Lambda_{0,i}$. For $i\in\{2,7,8,9\}$ we find that $p>0.2$, while for the other frequencies we find $p \le 0.002$. This shows that the common-noise model, despite being more informative than the separated fits, is not predictive of the observations.

To make it predictive, the common-noise model posterior must be widened by a (frequency-dependent) factor. To estimate how large this factor should be, we have repeated the simulation above, this time by generating the simulated periodogram for run $k$ at frequency $f_i$, by multiplying the sample for $S_{\Delta g_{e}}^{1/2}(f_i)$ from the original posterior, common to all runs, by a random variable $\gamma_{k,i}$. For each frequency $f_i$, all $\gamma_{k,i}$, were extracted from the same gamma distribution with unit mean value and standard deviation $\sigma_{\gamma_i}$. This model in essence allows $S_{\Delta g_{e}}^{1/2}(f_i)$ to fluctuate from run to run by a mean  relative amount $\sigma_{\gamma_i}$. This fluctuation adds up to the natural fluctuation of the periodogram for stationary Gaussian time series in generating the observed data. We call this model the ``independent-runs'' model.

The independent-runs model becomes predictive of the observation at all frequencies,  $p\ge 0.2$, if $\sigma_{\gamma_i}=0$ for $i\in\{2,7,8,9,10\}$, $\sigma_{\gamma_i}=0.4$ for $i=1$, and $\sigma_{\gamma_i}=0.2$ for $i\in\{3,4,5,6\}$. We report the data for the independent-runs posterior, in Fig.~\ref{fig:average}, red points.  

Note that the largest variation is for $i=1$, that is $f=\SI{18}{\micro\hertz}$. This is dominated by the large initial decrease of the ASD in the initial part of the mission that can be noticed by inspection in Fig.~\ref{fig:batch1} in Appendix~\ref{app:allpsd}. We will discuss this further in Sec.~\ref{sec:lfbin}.

The common-noise model is more predictive for subsets of data including a reduced number of runs. We find, for instance, that for the groups including runs 1--4  of Table~\ref{tab:noise-only-runs}, and for that including run~11--13, the model is reasonably predictive at all frequencies, while for the groups 5--8 and 9--10, the model is predictive at all frequencies but one. It must be said that other groups of runs, with no temporal continuity, may give similar results, and that this apparent improvement might just be a consequence of the reduced discrimination power of the posterior test if applied to a smaller number of data.

\subsection{\label{sec:oneoverf} \texorpdfstring{$1/f$}{1/f} model, and evolution of the excess noise for \texorpdfstring{\normalfont{$f>\SI{18}{\micro\hertz}$}}{f>18uHz}}
To study if the observed variations of excess noise level are correlated with operational conditions we have found a useful  ``figure of merit''  that summarizes the excess noise across the entire band $f>f_1$. We introduce it in the following. 

We take advantage of the fact that, as said, the PSD in this frequency band scales approximately as $1/f^2$. Thus the  coefficient $\widetilde{S}_{\Delta g}$ of the $1/f^2$ term in a fit with the model 
\begin{equation}
    S_{\Delta g}(f_i)=\widetilde{S}_{\Delta g}\left(\frac{\SI{1}{\milli\hertz}}{f_i}\right)^{2}+S_\text{Brown},
    \label{eq:dstilde}
\end{equation}
gives a reasonable measure of the average power, smoothing any frequency-to-frequency variation.  Consistent with the discussion so far, we will focus on the evolution of its square root $\widetilde{S}_{\Delta g}^{1/2}$ and will call this the ``$1/f$'' model\footnote{We stress that in this model it is the ASD that depends on the frequency as $1/f$. This should not be confused with the popular model for flicker noise, for example in electronics, where it is the PSD that depends on the frequency as $1/f$.}. 

To make the analysis consistent, and avoid mixing any effect of the different duration of the various runs, we have partitioned the data for all runs in nonoverlapping stretches, all of the same duration of about 2.75 d. This is the duration of the shortest of the runs (run 5 of Table~\ref{tab:noise-only-runs}). We have then fitted the PSD data  for the 27 resulting ``partitions'', to the model of Eq.~\eqref{eq:dstilde}.

\begin{figure*}[htbp]
  \centering
  \includegraphics[width=1\textwidth]{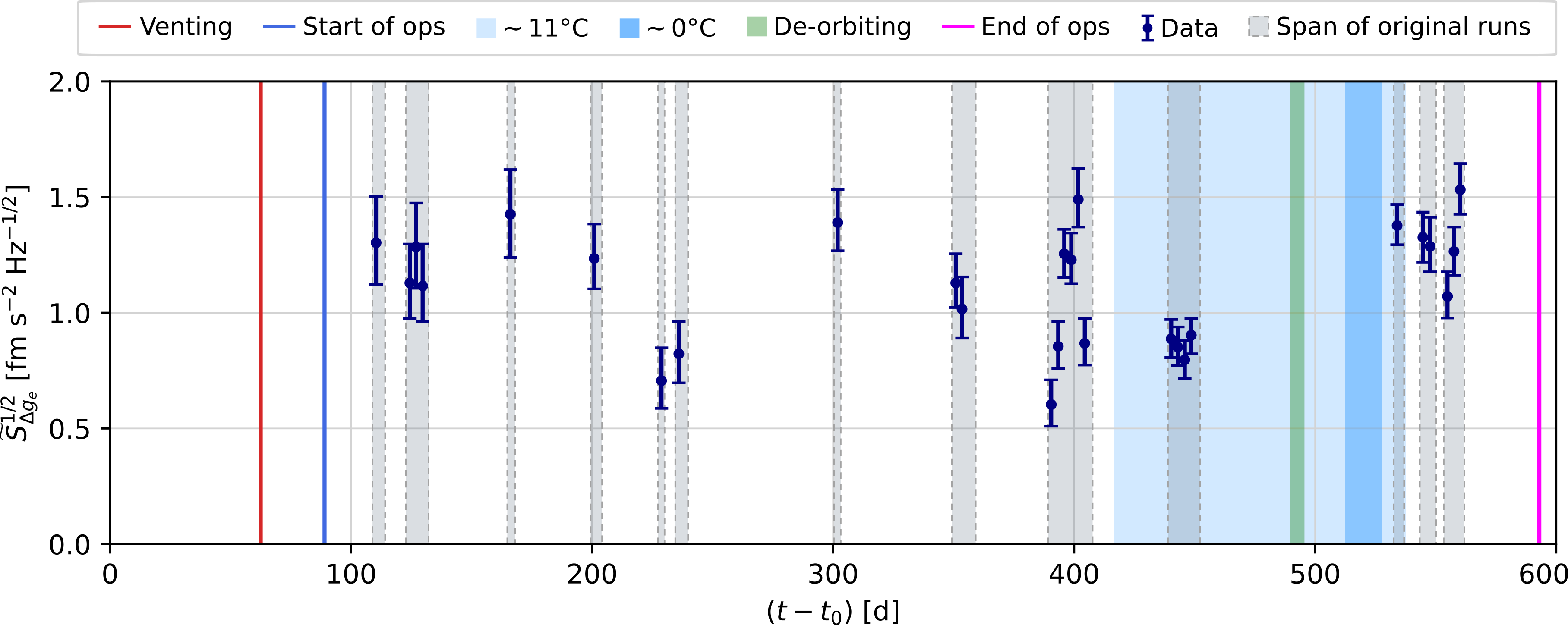}
  \caption{Time evolution of the $1/f$ tail amplitude of $S_{\Delta g}^{1/2}$, expressed as the corresponding  ASD at 1\,mHz $\widetilde{S}_{\Delta g_e}^{1/2}$, as a function of time since launch $(t-t_0)$, over the entire mission duration. Blue data, values for all the 2.75 d long, nonoverlapping stretches of data into which we partition the full set of the 13 run data series. The gray vertical bands with dashed edges represent the time span of the original runs to which the partition belongs. The picture also shows the date of the initial venting (red line), that of the beginning of the science operations (blue line), the epoch of the deorbiting burn (green band), and the date of the end of the science operations (magenta line). In cyan, the epochs during which the temperature was held at about \SI{11}{\celsius}; runs 10 and 11 of Table~\ref{tab:noise-only-runs} are both included in this span. Finally, in blue, the epoch of the \SI{0}{\celsius} cooling.}
  \label{fig:evolution}
\end{figure*}

The fit was again a Bayesian MCMC analysis,  supported by the posterior predictive test based on the likelihood, in close analogy with that used above to test the common-noise model. The only difference is that we now sum the log-likelihood on all frequencies considered. We find that all fits have $p\ge 0.1$ except for one for which $p=0.02$. Actually, for 2/3 of the 27 short runs $p>0.5$. The results for $\widetilde{S}_{\Delta g}^{1/2}$ are shown in Fig.~\ref{fig:evolution}.

Notice that the  LISA requirements for the ASD of $\Delta g$ in Fig.~\ref{fig:average}, include a $\propto 1/f$ branch, though just down to the lower corner at 0.1\,mHz. This branch would correspond to $\widetilde{S}_{\Delta g_e}^{1/2}=\SI{1.7}{\femto\meter\,\second^{-2}/\rtHz}$, significantly above all the points in Fig.~\ref{fig:evolution}. This is consistent with Fig.~\ref{fig:average}.

One feature that stands out in Fig.~\ref{fig:evolution} is that data fluctuate significantly more than the uncertainty on the single datum, even for the partitions of the same original run. Actually, the fluctuation of data from  partition of run~9 span a similar range as that of the remaining data. This large run-to-run fluctuation appears consistent with the poor fitting to the common-noise model. 

Just to confirm that these two facts are consistent, we have simulated the data assuming a common value for $\widetilde{S}_{\Delta g_e}^{1/2}$ for all 27 partitions,   and  Gaussian, stationary noise. This common value is extracted from the posterior for $\widetilde{S}_{\Delta g_e}^{1/2}$ of a cumulative fit to all 27  partitions with a common $1/f$ noise. The model is found, as expected, as nonpredictive as the common-noise model for the general case. 

To get a useful quantitative statistics to compare the simulated data to the actual ones, we form the sample of the 27 maximum likelihood values for $\widetilde{S}_{\Delta g_e}^{1/2}$, in essence, for the real sample, the dots of Fig.~\ref{fig:evolution}, and then calculate the sample standard deviation divided by the sample mean $\rho\equiv\sigma_{\tilde{S}_{\Delta g_e}}/\langle{\widetilde{S}_{\Delta g_e}}\rangle$.  For the sample of the actual data, we find  $\rho\simeq 0.23$.

The simulation gives instead a mean value for $\rho$ of $\langle\rho\rangle\simeq 0.11$ and projects a probability $< 10^{-3}$ of observing a value as large as $\rho=0.23$. 

Coherent with findings for the common-noise model, to give such a large value some reasonable probability, we had to widen the posterior, allowing for a 0.2 relative fluctuation of $\widetilde{S}_{\Delta g_e}^{1/2}$ from partition to partition. 

This is consistent with the observation that, joining the posterior for all 27 partitions, we get that $\widetilde{S}_{\Delta g_e}^{1/2}=(1.1\pm0.3)\,\si{\femto\meter\,\second^{-2}/\rtHz}$, while the width of the posterior for each run fluctuates of about $\simeq 0.1\,\si{\femto\meter\,\second^{-2}/\rtHz}$, implying a true fluctuation about $\simeq 0.25\,\si{\femto\meter\,\second^{-2}/\rtHz}$ from partition to partition, beyond the statistical uncertainty.

While the data of Fig.~\ref{fig:evolution} do not display any clear systematic long-term trend, or correlation with the experimental configuration parameters of Table~\ref{tab:conf}, they show a comparatively large difference, $\simeq 0.45\,\si{\femto\meter\,\second^{-2}/\rtHz}$, between the mean value of the four partitions of run~10 (438--451 days after launch in Fig.~\ref{fig:evolution}) and that of the six last partitions of runs 11--13 (531--561 days after launch in Fig.~\ref{fig:evolution}). 

Actually the difference is not much less, $\simeq 0.32\,\si{\femto\meter\,\second^{-2}/\rtHz}$, if, instead of just taking the mean of the partitions of run~10, one also adds the six partitions belonging to run~9, as runs~9 and 10  could be reasonably fit together with the common-noise model, as could also runs 11--13. The observation is particularly intriguing, as the $\SI{0}{\celsius}$ cooling, with its rather dramatic consequences \cite{lpf_glitch2022}, and the deorbiting burn took place just between these two epochs.

In the attempt to assess if the observed variation could still be due to a random fluctuation, given the comparatively large spread of the values in Fig.~\ref{fig:evolution}, we have resorted to a classical permutation test. We have done random permutations of the time coordinates of the data of Fig.~\ref{fig:evolution} and counted the number of times a step of any sign,  larger than the observed one, could be found between two adjoining sets of data, one six-sample long, like that of runs~11--13, and the other either four- or ten-samples long.

We find that the fraction of permutations in which we find a step $\le 0.45\,\si{\femto\meter\,\second^{-2}/\rtHz}$, between the six-sample and the four-sample sets, is $p=0.09$, while for the ten- and six-sample case, the chances of a step $\le 0.32\,\si{\femto\meter\,\second^{-2}/\rtHz}$ is $p=0.12$. These fractions are both too large to allow us to rule out the hypothesis that the observed increase is due to a random fluctuation.

\subsection{\label{sec:gauss} Gaussianity}

In addition to the lack of stationarity discussed so far, non-Gaussian distributed data may also contribute to the discrepancy between a simple common-noise model and the data. We have performed a coarse test on the Gaussianity of the periodograms used to calculate all experimental PSDs. 

We have first taken  the real and the imaginary parts of all periodograms used to produce the PSD at a given  frequency and for  a given run, and merged them into a single sample. This to increase the statistics for those frequencies and those runs for which the number of averaged periodograms was very small. We have done this merging taking advantage of the fact that for  Gaussian data the real and the imaginary part are independent and  have the same distribution.

We have then standardized each sample, by subtracting its mean and by dividing by its standard deviation. Limiting the analysis to the lowest nine frequencies we then have $9\times13=117$ of these samples, as the number of runs is 13. The number of periodograms in each sample may vary quite significantly, ranging from 2, for the lowest frequency bin and the shortest runs, up to 512
for $f_9$ and run~9. We exclude cases with just one periodogram, as the normalization procedure would be not significant.

\begin{figure}[htbp]
  \centering
  \includegraphics[width=1\columnwidth]{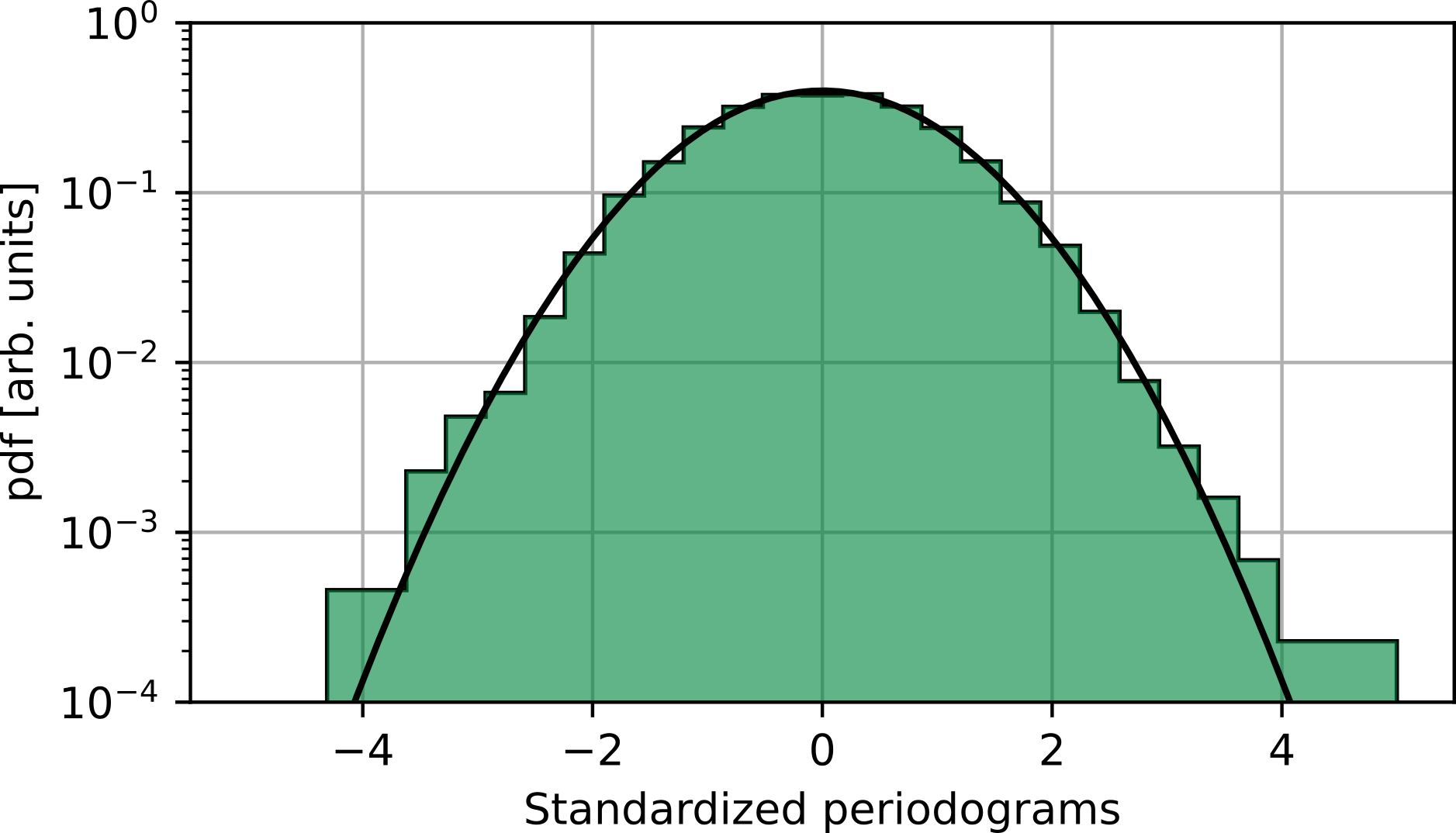}
  \caption{Gaussianity test of periodograms. Green bars, histogram of the periodogram amplitudes standardized as explained in the text; solid black line: probability density function of the zero-mean unit variance Gaussian distribution.}
  \label{fig:Gauss}
\end{figure}

As all elements of these 117 samples are normalized, if data are Gaussian, they should all come from  the same  zero-mean, unit variance Gaussian distribution. We have then merged them all in a single sample containing 12684 standardized periodograms. In Fig.~\ref{fig:Gauss} we report the histogram of this sample and, for comparison,  the probability density function of the zero-mean unit variance Gaussian distribution. 
The agreement between the two is supported by an Anderson-Darling test for Gaussianity, that gives $p>0.15$. 
We note that the test is rather crude, the statistic being dominated by the longest runs and the higher frequencies. However, the absence of any outlier significantly above $5\sigma$ excludes the presence of significant long tails.

\subsection{\label{sec:gamma} Associated angular acceleration noise}
For the purpose of characterizing the properties of the excess acceleration noise described above, we have investigated if there is any torque associated with it.  To this purpose, we have measured the torque acting on the test masses around $z$ and $y$.

As the rotation of the test masses around all axes is controlled, and as the angular acceleration of the spacecraft, which is common mode for both TMs, is rather large, in total analogy with the $\Delta g$ and the motion along $x$, we form the differential open-loop torque (per unit moment of inertia) acting on the test mass,
\begin{equation}
\label{eq:delta_gamma}
\begin{split}
 &\Delta\gamma_{\phi} = \ddot{\phi}_2-\ddot{\phi}_1-\frac{N_{z_2}-N_{z_1}}{I_{zz}} + \omega_{\phi_2}^2\phi_2-\omega_{\phi_1}^2\phi_1\\
  &\Delta\gamma_{\eta} = \ddot{\eta}_2-\ddot{\eta}_1- \frac{N_{y_2}-N_{y_1}}{I_{yy}}+ \omega_{\eta_2}^2\eta_2-\omega_{\eta_1}^2\eta_1,
\end{split}
\end{equation}
Here ${\phi}_i$ and ${\eta}_i$ are the angular displacements of TM$i$ along $z$ and $y$ respectively, $N_{z_i}$, $N_{y_i}$ are the control torques applied to TM$i$ along $z$ and $y$ respectively, $I_{zz}$ and $I_{yy}$ are the TM moments of inertia around $z$ and $y$ respectively, and finally the $\omega$'s are the angular stiffnesses. All signals had to be properly calibrated to ensure that the large, common mode angular acceleration of the spacecraft was duly suppressed.

\begin{figure}[htbp]
  \centering
  \includegraphics[width=1\columnwidth]{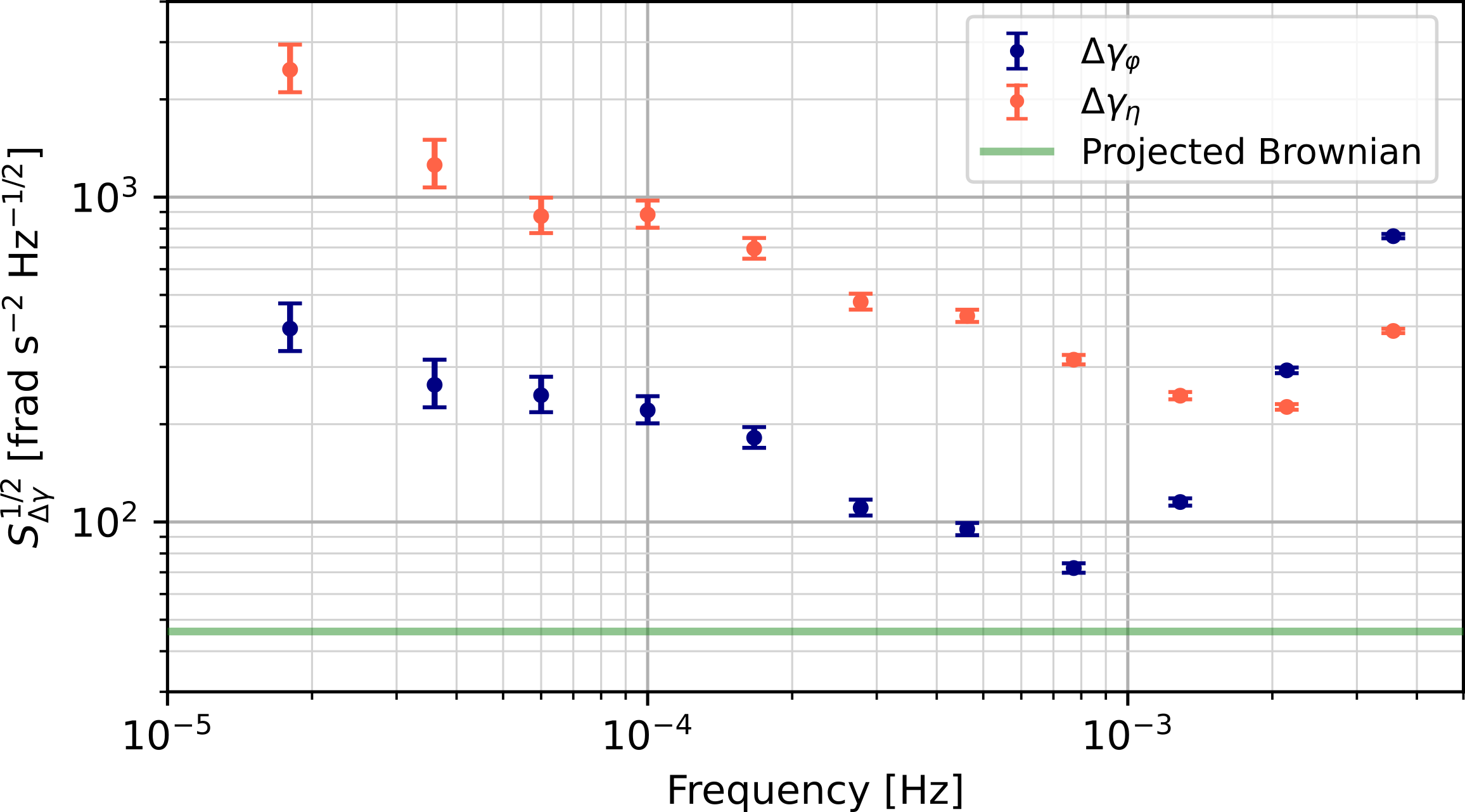}
  \caption{PSD of $\Delta\gamma_{\phi}(f)$ and $\Delta\gamma_{\eta}(f)$, as a function of frequency $f$, for run~10 of Table \ref{tab:noise-only-runs} (February 2017). The green band represents the uncertainty range for the projection of the PSD of the rotational Brownian noise around $\phi$, from the corresponding Brownian noise for $\Delta g$ (see text for details). The reported error on the projected value accounts just for the propagation of statistical errors. Additional inaccuracy due to various sources may amount to some additional 10\%.}
  \label{fig:gamma}
\end{figure}

The PSDs $S_{\Delta\gamma_{\phi}}(f)$ and $S_{\Delta\gamma_{\eta}}(f)$ are reported in Fig.~\ref{fig:gamma} for run~10 of Table~\ref{tab:noise-only-runs}.

The figure also shows the PSD of rotational Brownian noise along $z$, projected from the measured value of the PSD of the Brownian noise in  $\Delta g$ for the same run. The projection is based on calculating the rotational viscous damping coefficient from the linear one, as explained in  \cite{PhysRevLett.103.140601}. 

Both  $S_{\Delta\gamma_{\phi}}(f)$ and $S_{\Delta\gamma_{\eta}}(f)$ show a minimum much larger than the projected Brownian noise. This minimum is set by the crossover between the torque noise and the readout noise due to the angular interferometer, which is comparatively more noisy than  the linear one.

To assess if the excess $\Delta g$ noise carries any torque, we have measured, together with the  PSDs, also  the cross-spectral densities among  $\Delta\gamma_{\phi}$, $\Delta\gamma_{\eta}$, and $\Delta g$.

To get a simple parametrization of the associated torque we use the simple model of a vector pointlike force ${f}(t)$, applied to one of the test masses at the point $(x_0,y_0,z_0)$, relative to the test mass center, and with component $(f_x(t),f_y(t),f_z(t))$. Such force would also apply a torque $N(t)$ with $z$ and $y$ components given, respectively, by $N_z(t)=x_0 f_y(t)-y_0 f_x(t)$ and $N_y(t)=z_0 f_x(t)-x_0f_z(t)$. For such a force,
\begin{equation}
\label{eq:gammamodel}
\begin{cases}
 &S_{\Delta\gamma_{\phi}}(f)=S_{\Delta\gamma_{\phi},0}(f)+y_0^2\left(\frac{M}{I_{zz}}\right)^2 S_{\Delta g}(f)\\
  &S_{\Delta\gamma_{\phi}\Delta g}(f)=-y_0 \left(\frac{M}{I_{zz}}\right) S_{\Delta g}(f)\\
\end{cases}
\end{equation}
and similarly for $y$, with $-y_0\rightarrow z_0$. Here, $S_{\Delta\gamma_{\phi},0}(f)$ [$S_{\Delta\gamma_{\eta},0}(f)$] is the spectral density of any part of $\Delta\gamma_{\phi}(t)$ [$\Delta\gamma_{\eta}(t)$]  that is not correlated to $\Delta g(t)$.

The parametrization holds for an arbitrary system of forces, but while for a real pointlike force $\lvert y_0\rvert,\lvert z_0\rvert \le l=\SI{46}{mm}$, for an arbitrary system both parameters can take any value.  One important example is that of the force due to a voltage on only one of the electrodes facing the $x$-faces of the TM, for which  $|y_0|\simeq\SI{11}{mm}$  while $z_0=0$.

We have estimated the Bayesian posterior for the parameters $S_{\Delta\gamma_{\phi},0}(f)$, $S_{\Delta\gamma_{\eta},0}(f)$, $y_0$, and $z_0$ with an MCMC calculation, assuming Gaussian data, a Jeffreys prior for $S_{\Delta\gamma_{\phi},0}(f)$, $S_{\Delta\gamma_{\eta},0}(f)$, and a uniform prior for  $y_0$ and $z_0$. We performed this analysis including frequencies from the second to the seventh frequency bins (\SI{36}{\micro\hertz}\,--\,\SI{0.60}{\milli\hertz}), as in some of the runs, some eighth-frequency data already include the interferometer rising branch. 

\begin{figure}[tbp]
  \centering
  \includegraphics[width=1\columnwidth]{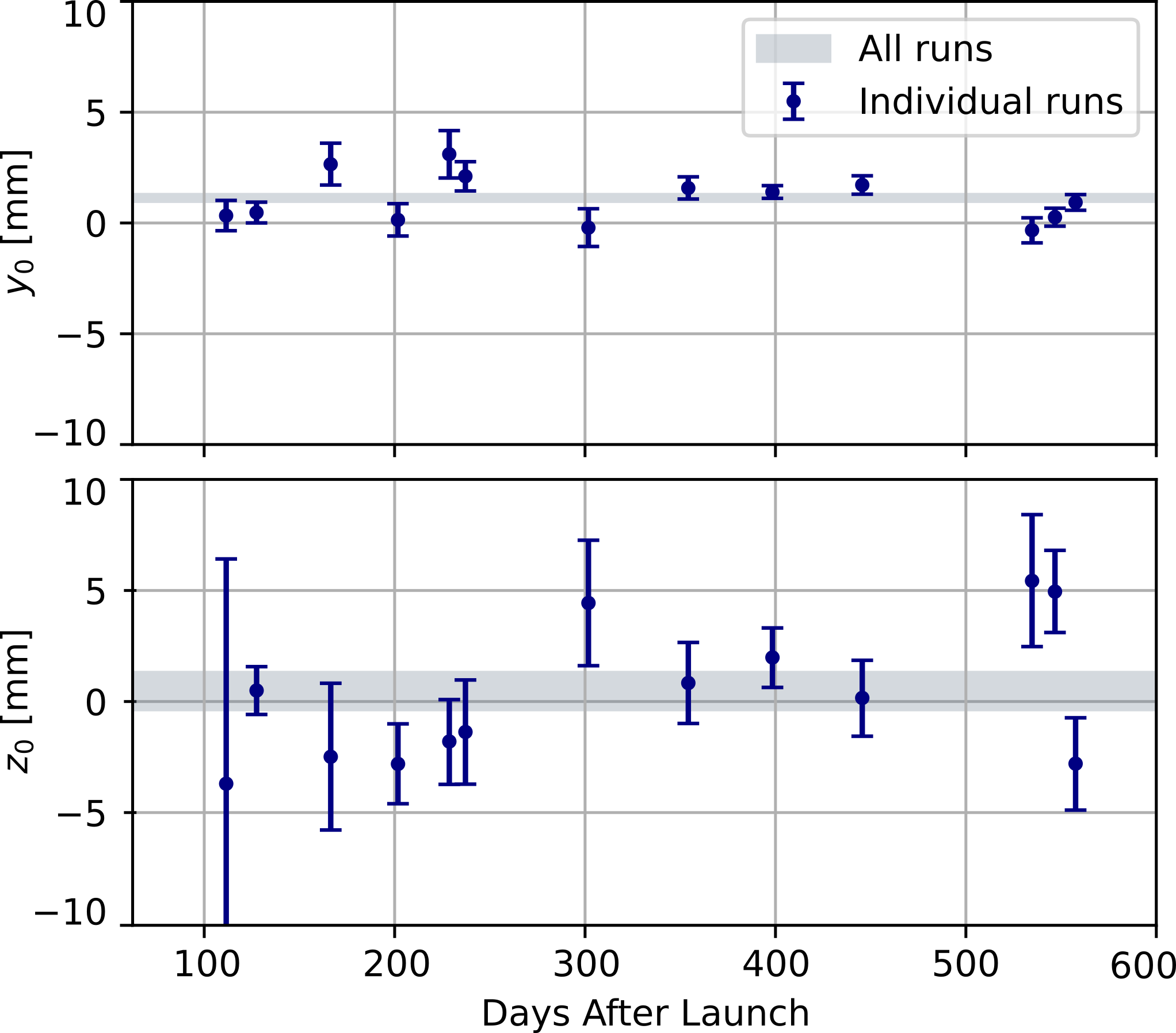}
  \caption{Top data points, effective lever arm  $y_0$ of force noise around the $z$-axis as a function of the epoch $t$ of the run. Error bars are $1\sigma$ width of the posterior. Light gray band $y_0=(1.13\pm0.15)\,\text{mm}$, the estimated value of the lever arm assuming a common value for all runs.
  Bottom: same as the top panel, but for the effective lever arm  $z_0$ of force noise around the $y$-axis. The light gray band is for $z_0=(0.5\pm0.8)\,\text{mm}$.} 
  \label{fig:leverarm}
\end{figure}

We have performed the calculation for both the individual runs, and also by assuming that the same value of $y_0$ or $z_0$ may fit all runs. The results are reported in Fig.~\ref{fig:leverarm}. 

We note that  for $y_0$ a systematic pattern appears, with the global fit giving  $y_0=(1.13\pm0.15)\,\text{mm}$, excluding then $y_0=\SI{0}{mm}$ and $y_0=\SI{11}{mm}$ with very large significance.  On the contrary, for $z_0$ data scatter on both sides of zero and the global fit gives $z_0=(0.5\pm0.8)\,\text{mm}$, i.e. a lever arm not significantly different from zero, though within an error which is significantly larger than that for $y_0$.

Reference~\cite{actuation-paper-bill} analyzes the force and torque noise due to gain fluctuations in the LPF TM actuation systems (see also Sec.~\ref{sec:decorrMCMC_noiseprojection}). That study has found that the $\phi$ angular acceleration noise excess at low frequencies is largely explained by actuation gain fluctuations in the circuitry used to produce both $x$ forces and $\phi$ torques.  Additionally, that study observes a slight correlation between $\Delta g$ and $\Delta \gamma_\phi$ arising from the gain fluctuations in these shared $x$-$\phi$ actuators, a correlation that is consistent with the $y_0$ arm length in the top panel of Fig.~\ref{fig:leverarm}. The slightly positive arm length comes from a slight asymmetry in the electrode voltage noises, as better explained in Ref.~\cite{actuation-paper-bill}. 


We have compared the results of our analysis, in particular for run~10  of Table~\ref{tab:noise-only-runs}, to those in \cite{actuation-paper-bill} and have found full compatibility between them, indicating that the detected torque noise correlated with $\Delta g$ is entirely explained by the shared gain fluctuations. We will further discuss the implications of these findings later in the paper, in Sec.~\ref{sec:nonmod}.

\subsection{\label{sec:sumexc} Summary note on the \texorpdfstring{$1/f$}{1/f} tail}
In conclusion, the overall picture is that, for frequencies in the range [\SI{36}{\micro\hertz}\,--\,\SI{0.60}{\milli\hertz}] the following is noted for noise ASD for $\Delta g_e$: 
\begin{enumerate}[label=(\roman*)]
    \item it is basically compatible with the $1/f$ model within a single run of $\simeq 2.5\,\text{d}$ duration;
    \item the  ``true'' amplitude of the $1/f$ branch may fluctuate on average by $\simeq \pm 20\%$ from one of these runs to another, this fluctuation being in addition to that expected from Gaussian stationary noise;
    \item the amplitude of such a nonstationary  extra fluctuation does not seem to follow any long-term pattern, neither decaying nor increasing over the course of the mission;
    \item there is no proven correlation between these extra fluctuations and any identifiable operational condition, though we cannot exclude that the \SI{0}{\celsius} cooling, or the deorbiting burst may have had some  effect on the noise after day 460 (runs 11--13);
    \item there is a small correlated torque associated with such $1/f$ force noise. The millimeter-size effective radius points to forces acting toward the center of the TM, or to forces that are almost uniform over the TM faces, ruling out simple possibilities like the force due to a noisy voltage on one of the $x$-electrodes;
    \item finally, it is worth noting that the approximate stability of the $1/f$ tail’s amplitude persisted despite numerous changes to the operational environment including station-keeping maneuvers, planned experiments with the LTP and DRS payloads, and unplanned spacecraft anomalies.  By comparison, LISA operations will be simpler with no orbital station keeping or planned experiments postcommissioning.
\end{enumerate}

\subsection{\label{sec:lfbin}Evolution of the lowest frequency datum and of long-term drifts}

 In Sec.~\ref{sec:excess}, we noted that the first-frequency point ($f_1=\SI{18}{\micro\hertz}$) deviates from the $1/f$ ASD behavior, 
consistently showing a noise level above that predicted by the $1/f$  fit model. To quantify the deviation from the model, we define $\widetilde{S}_{\Delta g_e}(f_1)$, i.e., the excess noise at frequency $f_1$, above $\widetilde{S}_{\Delta g}$, 
\begin{equation}
\widetilde{S}_{\Delta g_e}(f_1)\equiv (f_1/\SI{1}{\milli\hertz})^2S_{\Delta g_e}(f_1) -\widetilde{S}_{\Delta g}
\end{equation}
We have multiplied the data by the frequency factor $(f_1/\SI{1}{\milli\hertz})^2$ to make the comparison with the $1/f$ branch more immediate.

\begin{figure}[htbp]
  \centering
  \includegraphics[width=1\columnwidth]{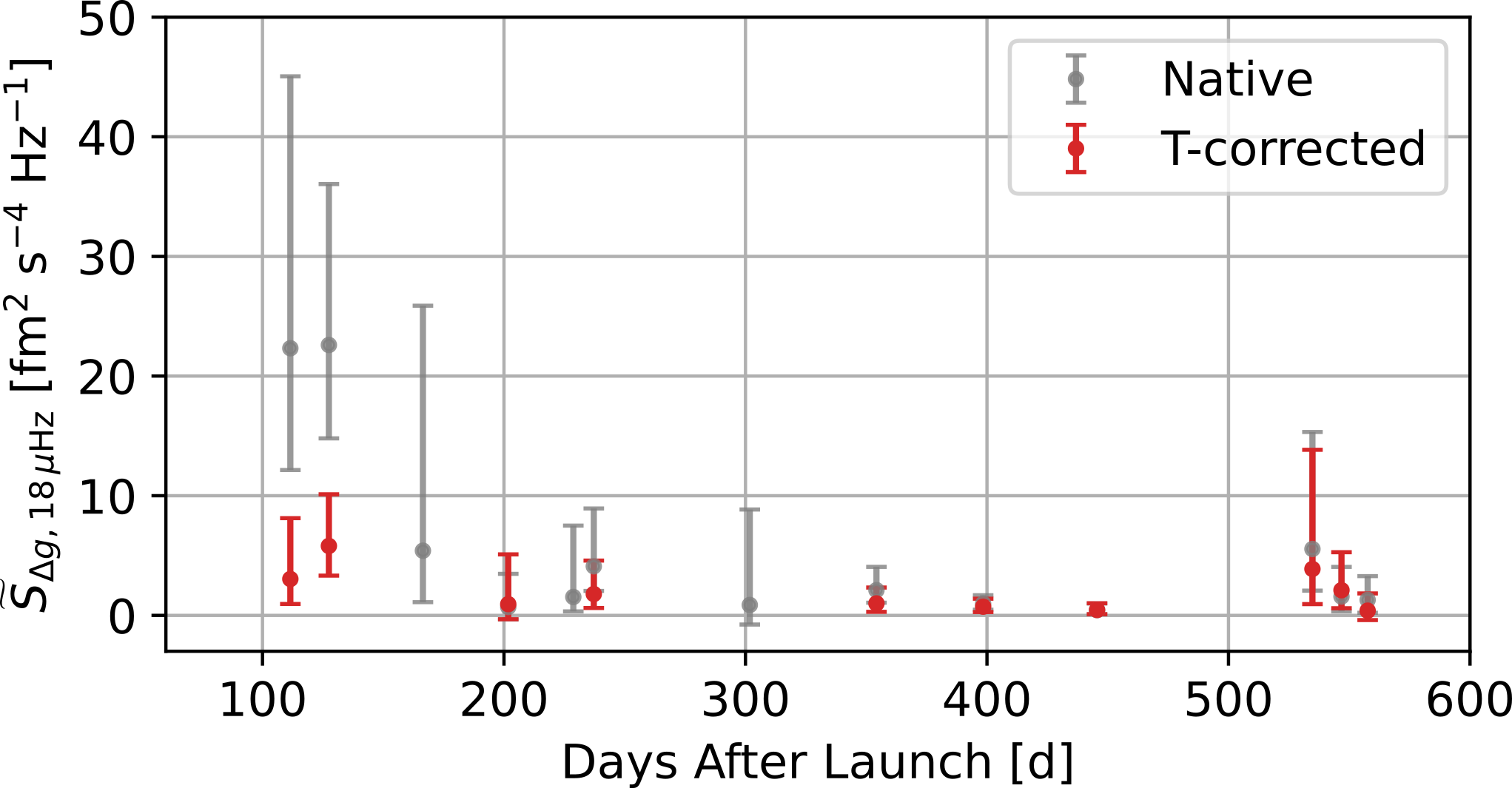}
  \caption{Evolution of the excess noise at the lowest frequency  $f_1=\SI{18}{\micro\hertz}$ over the course of the mission. Gray points: $\widetilde{S}_{\Delta g_e,1}\equiv \left(f_1/\SI{1}{mHz}\right)^2 S_{\Delta g_e}(f_1)$, for the 13 runs of Table~\ref{tab:noise-only-runs}, as a function of the mean time of the run since launch $(t-t_0)$; red points, residual after correcting for the effect of the temperature (see text at the end of the section for details). Only data for runs with more than one periodogram at $\SI{18}{\micro\hertz}$ could be corrected, hence in some cases there is no red point (see. Appendix~\ref{app:CPSD/decorrelation}).}
  \label{fig:firstbin} 
\end{figure}

In Fig.~\ref{fig:firstbin} (gray points) we show the time evolution of $\widetilde{S}_{\Delta g_e}(f_1)$ at the lowest frequency\footnote{Note that some of the PSD values in Fig.~\ref{fig:firstbin} take negative values, which would make the ASD imaginary. This is the reason why we prefer to use PSD.}.
Despite the large errors, the picture shows a significant initial decay pattern. In addition, all values of $\widetilde{S}_{\Delta g_e}(f_1)$ are significantly above zero, i.e. there is a noise excess above the $1/f$ model at \SI{18}{\micro\hertz} at all times during the mission.


Before discussing this behavior in further detail and before introducing the temperature correction  shown by the red points in the figure, we have to discuss the long-term drift, over scales of several hours and days,  that affects  $\Delta g$ data for all runs. We will show that such drifts share many of the features shown by $\widetilde{S}_{\Delta g,e}(f_1)$. 

This is not surprising, as the relative width of the spectral window at $f_1$ is wider, by construction,  than that for all other frequencies (See  Appendix~\ref{app:PSDestimate}). Thus a significant spectral leakage from the frequency band $f \ll f_1$, i.e. from the long-term drifts may be naturally expected.

\begin{figure}[htbp]
  \centering
  \includegraphics[width=1\columnwidth]{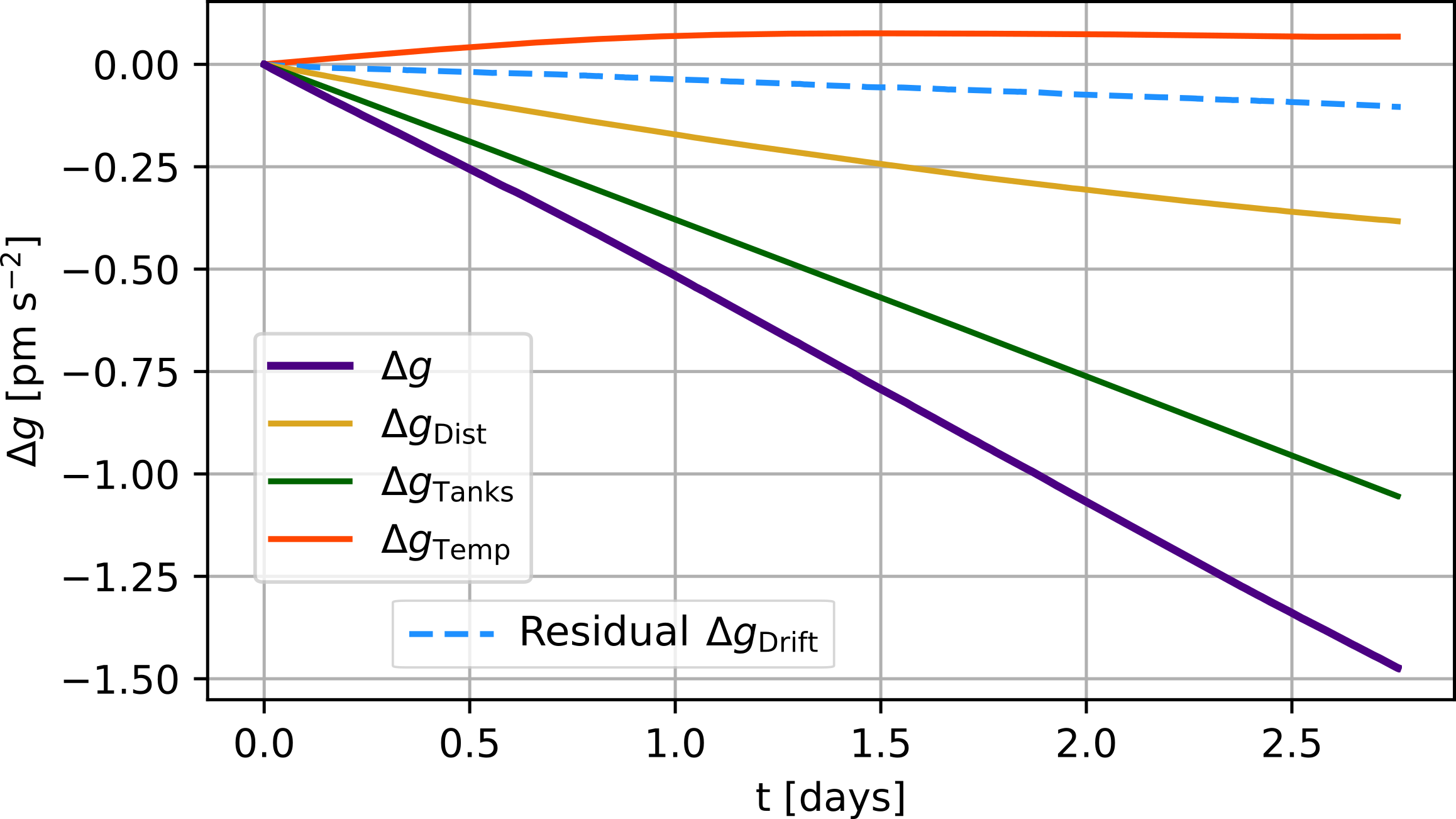}
  \caption{Long-term evolution of $\Delta g$ for run~7 of Table~\ref{tab:noise-only-runs} (purple line) and of various disturbances that may contribute to it. Green line, gravitational signal from propellant tank depletion; yellow line, mechanical distortion calculated assuming the nominal value of effective stiffness $\omega_d^2$; orange line, temperature contribution from best fit to data; dashed cyan line, residual drift. See text for details.} 
  \label{fig:drift}
\end{figure}

An example of the drift that, in addition to quasistationary noise, affects $\Delta g$ data for all runs is shown in Fig.~\ref{fig:drift}. 

The sources for such a drift may be many. There are two predictable ones:
\begin{itemize}
        \item {motion of massive parts of the instrument relative to the TMs, due to any kind of mechanical distortion, also produces a time-varying gravitational field and then a time-varying $\Delta g_\text{Dist}$. One important mechanical distortion is the rigid translation, along $x$, of each GRS relative to its own TM due to expansion and contraction of the optical bench to GRS separation. By GRS  we mean here vacuum chamber, electrode housing, etc., but not the TM itself. 
        
    Such distortion is rather likely, as the GRS has a rather complex interface, on one hand to the satellite, via a set of tens of centimeters long ceramics struts, and, on the other, to the optical metrology, through a sophisticated metal-glass interface. This mounting method is rather ``soft'' and prone to strain, while the GRS itself is a much more rigid assembly.
    
    Fortunately, the capacitive motion sensor gives a measurement of the relative motion of the GRS relative to its own TM, $\Delta X\equiv\left(\Delta x_\text{OMS}-\Delta x_\text{GRS}\right)$, and allows to predict $\Delta g_\text{Dist}$, as $\Delta g_\text{Dist}=\omega_d^2\Delta X$ (see  Fig.~\ref{fig:drift}) with $\omega_d^2$ a stiffness factor which is known to within a sufficient  accuracy. This is again explained in Appendix~\ref{app:driftcalc}.
    
    We note that this contribution to the overall drift is rather small, if not negligible, for all runs, except that for runs 7, 11, and 12 of Table~\ref{tab:noise-only-runs}.}
    
    \item {The use of propellant for the drag-free control slowly depletes the propellant tanks \cite{PRDThrust}. The propellant  in the tanks produces a gravitational field at the TMs location that results in a  differential acceleration  of the TMs with a significant component $\Delta g_\text{Tanks}$ along $x$. Due to depletion, this acceleration  drifts in time (see Fig.~\ref{fig:drift}). In Appendix~\ref{app:driftcalc} we show that, for a run in which the propellant tank $i$ and the thruster branch $j$ are used, $\Delta g_\text{Tanks}(t)=\kappa_{t,i} \kappa_{b,j} \Delta g_{\text{Tank},0,i,j}(t)+c$, where $\Delta g_{\text{Tank},0,i,j}(t)$ is a signal that can be entirely calculated,  $\kappa_{t,i}$ and $\kappa_{b,j}$ (where $t$ stands for tanks and $b$ stands for branch) are two factors  that reflect calibration uncertainty, and that are both 1 for the nominal case;  $c$ is a constant, depending on $i$ and $j$, with no real relevance for the following discussion.}
    \end{itemize}
    
Once the two sources above have been subtracted from the  $\Delta g$ data series, the residual  shows an evident correlation with the instrument temperature $T$, defined in Sec.~\ref{sec:data runs}, and its long-term variations (Fig.~\ref{fig:drift}). Once also such correlation has been suppressed, by  linear least squares fitting  $T$ to $\Delta g$ and by subtracting the best fit, some residual drift still remains (Fig.~\ref{fig:drift}).

We note from Fig.~\ref{fig:drift} that the raw drift (purple line) is  of order $\SI{0.5}{\pico\meter\,\second^{-2}/\day}$, while the residual unmodeled drift (dashed blue line) is of order $\SI{60}{\femto\meter\,\second^{-2}/\day}$.

To get some quantitative estimate of this residual drift, we have preliminarily corrected the data for the two predictable effects, $\Delta g_\text{Tanks}$ and $\Delta g_\text{Dist}$, and then fitted the corrected data to a linear combination of temperature and time as described in Appendix~\ref{app:driftcalc}. As explained therein, such fitting gives, for each run, a value for the mean partial derivative of $\Delta g$ relative to temperature $\partial \Delta g/\partial T$ and one for that relative to time $\partial \Delta g/\partial t$. 

\begin{figure}[htbp]
  \centering
  \includegraphics[width=1\columnwidth]{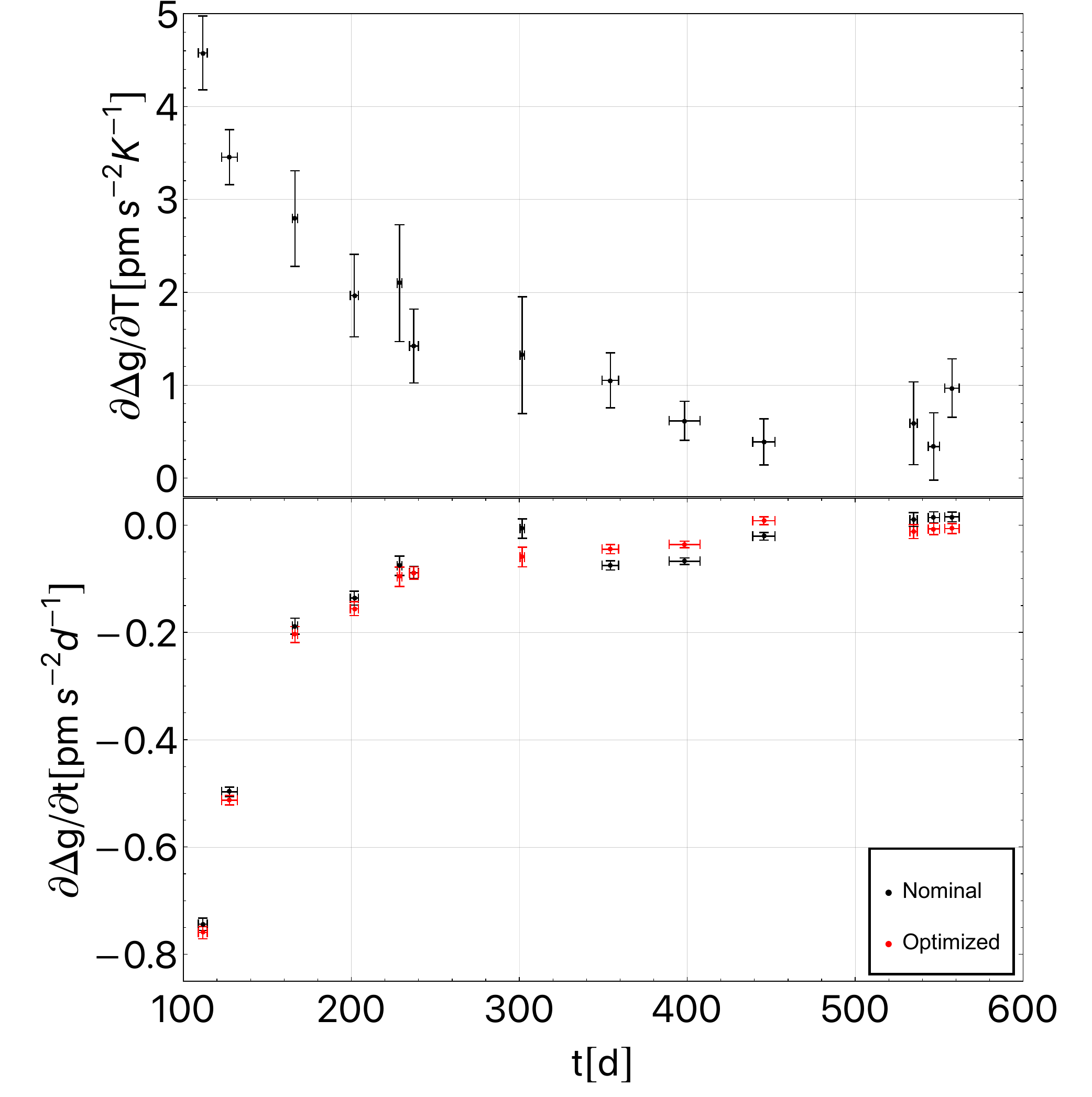}
  \caption{Upper: partial derivative of $\Delta g$ relative to  temperature $T$, $\partial \Delta g/\partial T$, as a function of the time of the run. Data refer to the nominal configuration, but those for the optimized configuration (see text for the definition) are numerically indistinguishable from them. Vertical errors are derived as explained in Appendix~\ref{app:driftcalc}. Horizontal bars represent the time span of the run.
  Lower: partial derivative of $\Delta g$ relative time $t$, $\partial \Delta g/\partial t$, as a function of the time of the run.  Black dots refer to the nominal configuration, while the red ones refer to the optimized configuration. The meaning of the error bars is the same as that for the upper panel.} 
  \label{fig:dgdtT}
\end{figure}

The above-mentioned preliminary correction requires some assumptions on the value of $\omega_d^2$, and on those for  $\kappa_{t,i}$ and $\kappa_{b,j}$. The results for the nominal calibration $\kappa_{t,i}=\kappa_{b,j}=1$ and $\omega_d^2=-3.32\times 10^{-7}\text{\,s}^{-2}$ are shown, as black points, in Fig.~\ref{fig:dgdtT} as a function of the time of the run.
The relative rapid decay of both quantities in the early phase of the mission led us to investigate the existence of any possible correlation with the similar decay of Brownian noise. 

\begin{figure}[htbp]
  \centering
  \includegraphics[width=1\columnwidth]{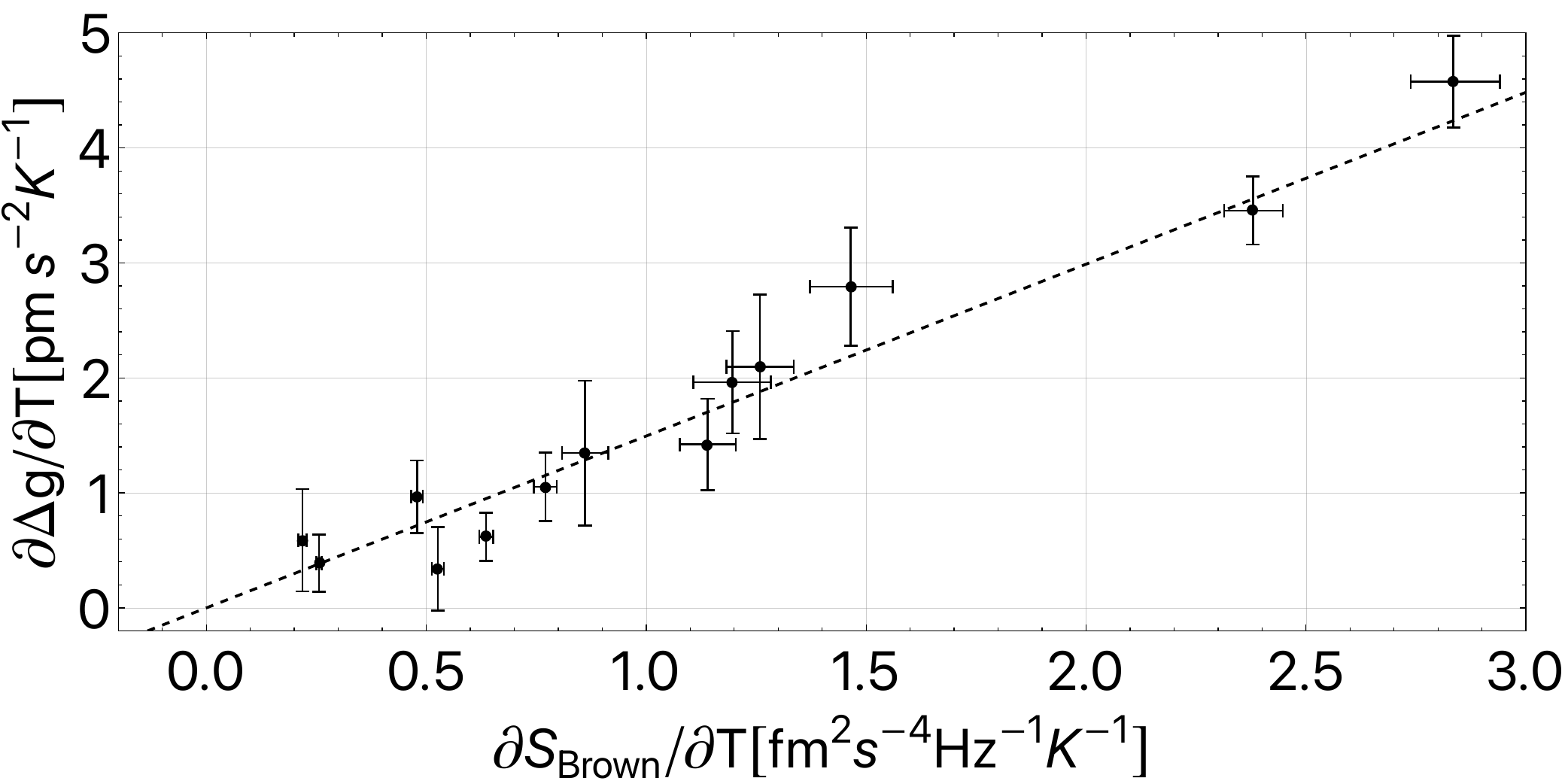}
  \caption{$\partial \Delta g/\partial T$ vs $\partial S_\text{Brown}/\partial T$ for the nominal calibration. Points are the data while the dashed line is the linear, least square best fit  $\partial \Delta g/\partial T= \alpha_T\; \partial S_\text{Brown}/\partial T$, with $\alpha_T=(1.49\pm0.07)\times 10^{18}\,\text{s/m}$.} 
  \label{fig:dgdTdSdT}
\end{figure}

We have  found a quite significant linear relation between $\partial \Delta g/\partial T$ and $\partial S_\text{Brown}/\partial T$; we calculate the latter, from Sec.~\ref{sec:brown}, as $\partial S_\text{Brown}/\partial T=(T_a/T^2)S_\text{Brown}$ (see Fig.~\ref{fig:dgdTdSdT}). 
A fit with the simple proportionality relation $\partial \Delta g/\partial T=\alpha_T \partial S_\text{Brown}/\partial T$ gives $\alpha_T=(1.49\pm0.07)\times 10^{18}\,\text{s/m}$ (see Fig.~\ref{fig:dgdTdSdT}). The $\chi$-squared test for the  goodness of such fit gives $p=0.8$.

Note, however, that the factor $(T_a/T^2)$  changes at most by some 6\% from run to run, thus the correlation between $\partial \Delta g/\partial T$ and $S_\text{Brown}$ itself is as strong as the other. Nevertheless, we will continue to discuss the case for $\partial S_\text{Brown}/\partial T$ for reasons that will be clear in the following.

\begin{figure}[htbp]
  \centering
  \includegraphics[width=1\columnwidth]{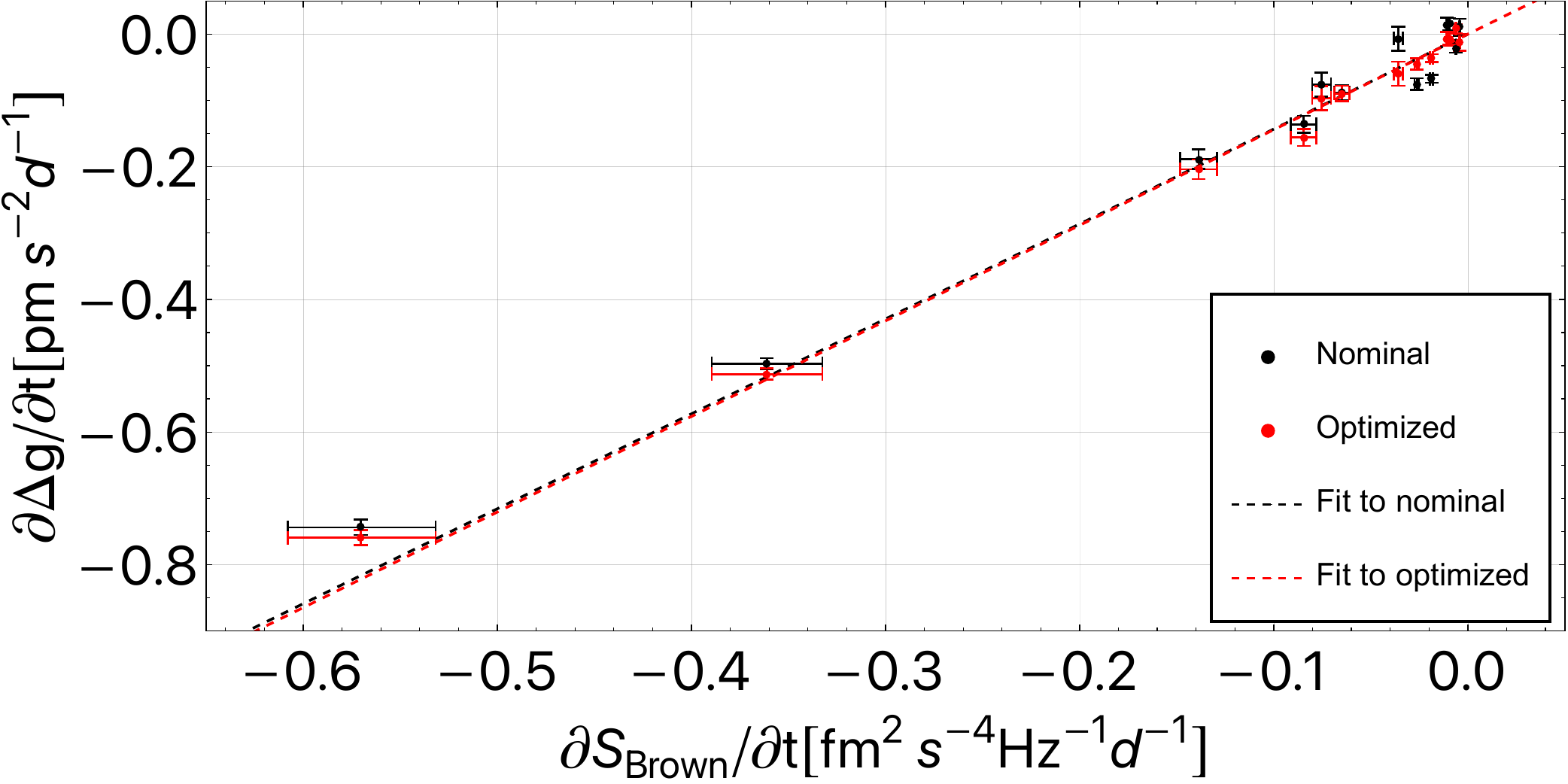}
  \caption{ $\partial \Delta g/\partial t$ vs $\partial S_\text{Brown}/\partial t$. Black points are the data for the nominal calibration, and  the black dashed line is the linear, least square best fit to the black points with $\partial \Delta g/\partial t= \alpha_t \partial S_\text{Brown}/\partial t$ and $\alpha_t=(1.4\pm0.2)\times 10^{18}~\text{s/m}$. Red data and line have the same meaning, but for the optimized calibration. In this case $\alpha_t=(1.44\pm0.07)\times 10^{18}~\text{s/m}$.} 
  \label{fig:dgdtdSdt}
\end{figure}

A plot of $\partial \Delta g/\partial t$ as a function of $S_\text{Brown}$ shows a rather low level of linear correlation. The linear correlation is instead rather significant between $\partial \Delta g/\partial t$ and $\partial S_\text{Brown}/\partial t=(-\gamma/(t-t_v))S_\text{Brown}$ [following from Eq.~\eqref{Eq: brown}]. A plot is shown in Fig.~\ref{fig:dgdtdSdt} together with a best fit to the data with $\partial \Delta g/\partial t=\alpha_t \partial S_\text{Brown}/\partial t$ that gives $\alpha_t=(1.4\pm0.2)\times 10^{18}\,\text{s/m}$.
The goodness of the fit to nominal data is rather poor, with negligible $p$-value. As the gravitational drift is large, even a limited change to the values of $\omega_d^2$, $\kappa_{t,i}$, and $\kappa_{b,j}$, used for its subtraction, may change the value of the residual drift and then of $\partial \Delta g/\partial t$. Thus the uncertainty on those  coefficients projects a large uncertainty on the true value of  $\partial \Delta g/\partial t$, uncertainty that is not taken into account while performing the goodness of fit test. 

To show that this is the case, we have searched if values of $\omega_d^2$, $\kappa_{t,i}$ and $\kappa_{b,j}$, other than the nominal ones, but still within the range of their uncertainties, may improve the quality of the fit. Actually, we have searched for the values that just give the maximum $p$-value in the goodness of fit test. 

We find that the maximum is attained when $\kappa_{t,1}=1.00$, $\kappa_{t,2}=1.05$, $\kappa_{t,3}=0.95$, $\kappa_{b,A}=1.00$, $\kappa_{b,B}=0.92$, and $\omega_d^2=-3.31\times 10^{-7}\,\text{s}^{-2}$. With this ``optimized'' calibration, $\alpha_t=(1.44\pm0.07)\times 10^{18}\,\text{s/m}$, and the $p$-value is $p=0.18$ (see Fig.~\ref{fig:dgdtdSdt}). Remarkably, moving to this optimized calibration, within errors, does not change the slope of the line.

The most important observation though, is that $\alpha_T=\alpha_t\equiv \alpha$ within the errors. This supports, with all the limitations and caveats that come from the empirical approach we had to use to process the data, that, for run $k$, the long-term evolution of $\Delta g$ obeys
\begin{equation}
    \Delta g_k(t)=\alpha S_\text{Brown}(t)+\Delta g_{0,k}= 2 \alpha \kappa \Pwat+\Delta g_{0,k}
    \label{eq:pgrad}
\end{equation}
with $\Delta g_{0,k}$ a constant that depends on the run, and is affected by many operational factors that may be different in different runs. 

In Eq.~\eqref{eq:pgrad} we have used the conversion from Brownian noise PSD to the mean pressure around the TM that we have discussed in Sec.~\ref{sec:brown}. It is interesting to note that also $\Delta g$ may be converted into an equivalent difference of pressure between the $x$-faces of one of the TMs, as $\Delta \Pwat=M\Delta g/L^2$, with $M$ the mass of the TM and $L$ the length of one of its edges. Then $\Delta \Pwat=((2 M \alpha \kappa)/L^2)\Pwat=(4.5\pm0.2)\times 10^{-3}\,\Pwat$. If the pressure difference was similar and opposite on both TMs, not unlikely given the mirror symmetry of the instrument, then all figures should be divided by 2.

The temperature dependence of $\Delta g$ at these very low frequencies is consistent with the transient behavior of the lowest frequency datum reported in Fig.~\ref{fig:firstbin}. Indeed a Bayesian decorrelation of temperature, following the method of Appendix~\ref{app:CPSD/decorrelation}, gives the  red points in Fig.~\ref{fig:firstbin}, which show a  suppressed  initial transient and a significantly reduced discrepancy from the $1/f$ tail.

\begin{figure}[htbp]
  \centering
  \includegraphics[width=1\columnwidth]{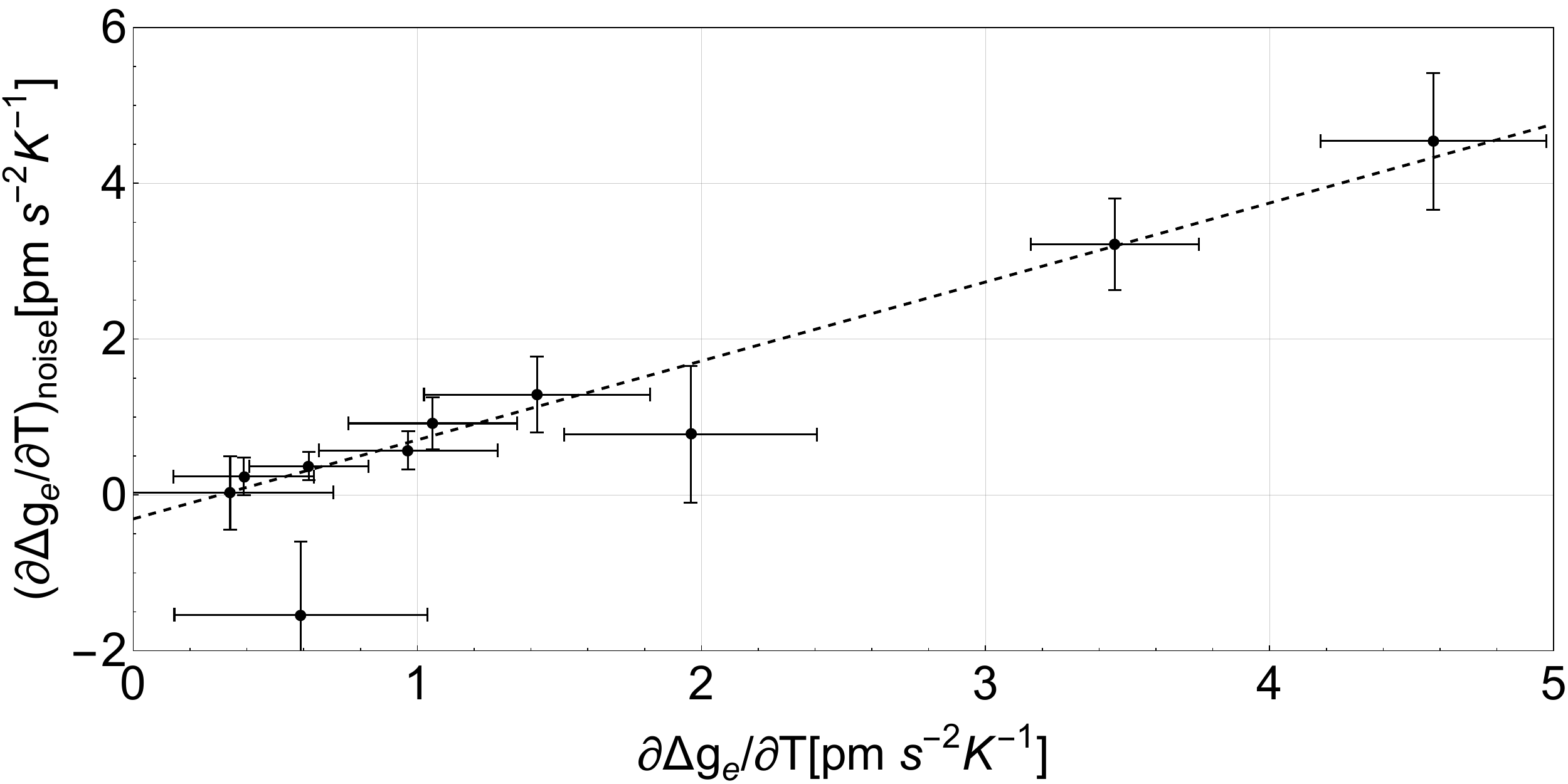}
  \caption{Temperature coefficient $\left(\partial \Delta g_e/\partial T\right)_\text{noise}$ from the Bayesian temperature decorrelation of the lowest frequency datum vs the temperature coefficient $\partial \Delta g_e/\partial T$ of Fig.~\ref{fig:dgdTdSdT}. The dashed line is the result of a linear least square fit, which gives $\left(\partial \Delta g_e/\partial T\right)_\text{noise}=\left(1.0\pm0.1\right)\partial \Delta g_e/\partial T+(-0.3\pm0.2)\,\text{pm\,s}^{-2}\,\text{K}^{-1}$, with a reduced $\chi$-square  $\chi^2 \simeq0.6$.} 
  \label{fig:noisevsdrift}
\end{figure}

Remarkably, the coefficient obtained from the decorrelation, $\left(\partial \Delta g_e/\partial T\right)_\text{noise}$, is in quantitative agreement with the values of  $\partial \Delta g_e/\partial T$ reported in Fig.~\ref{fig:dgdTdSdT} (see Fig.~\ref{fig:noisevsdrift}). This confirms our modelization.

We will discuss in Sec.~\ref{sec:brdisc} the implication of such findings for the nature of the vacuum environment of the TMs.

\section{\label{sec:discussion}Discussion}

In this  section we discuss the physical information we are able to gather from the observations described  so far. {Due to the extensive analyses required to gather this information, we will report only the main conclusions here. The detailed analyses are provided in the Appendixes.}

\subsection{\label{sec:brdisc}Brownian noise, long-term drift, and the TM vacuum environment (details in Appendix~\ref{app:BLTD})}

The conclusion of the analysis reported in Appendix~\ref{app:browniannoise} is that the observed temperature and time dependence of the Brownian noise is not consistent with a model in which the vacuum is dominated by water vapor in quasiequilibrium between thermal outgassing from the metal walls and readsorption onto them. This model describes well the behavior of clean, essentially metallic vacuum systems \cite{chiggiato}. A  fit to the data of Fig.~\ref{fig:brown} with a standard isotherm one would use for such a model, shown in  Fig.~\ref{fig:brown} as the dotted noisy line, is of relatively poor quality and can only be obtained with unphysical values for the fitting parameters.

The data are better explained by diffusion-limited outgassing from the polymer-rich, complex environment surrounding the TM. This would naturally lead to the observed single activation energy and to the fractional exponent in the power-law dependence of the data over time.

In addition, the analysis reported in Appendix~\ref{app:discT} indicates that such complex  TM surroundings, which create a rather asymmetric molecular flow impedance pattern, may also naturally explain the observed slowly varying pressure difference discussed in Sec.~\ref{sec:lfbin}. 

Finally, the observed tight link between the total residual pressure, as measured by the Brownian noise level, and the decaying drift in $\Delta g$, with both time and temperature, implies, as a minimum, the same desorption properties for the gas setting the overall residual pressure and the pressure gradients.  

The most simple explanation for this is that the dominant source of outgassing in the GRS is diffusion out of sources close to the TM, that is, located inside the EH or just outside its $x$-walls.
These sources may include the electrodes, the insulators and the structure of the EH, the tungsten balance mass, the cables placed close to the outer $x$-walls of the EH, and a few other elements.

These observations lead to the recommendation to perform, in preparation for LISA, in-depth qualification studies on the outgassing properties of these elements.

\subsection{\label{sec:excdisc} Excess noise and possible observational artifacts (details in Appendix~\ref{app:obsart})}

Here we begin discussing the possible sources of the observed $1/f$ tail. We first consider the role of some possible observational artifacts, while  a detailed projection of the noise onto the possible modeled sources is discussed further down.

\subsubsection{\label{sec:ifonoise} Role of interferometer noise.}

The contribution of interferometer noise $n_{\OMS}$, with ASD $S_{n_{\OMS}}^{1/2}$,  to $S_{\Delta g}^{1/2}$ is  $S_{\Delta g,n}^{1/2}=S_{n_{\OMS}}^{1/2}\left(4 \pi^2 f^2+\lvert \omega_2^2\rvert\right)$. 

Thus, for $f\le \lvert \omega_2\rvert/2\pi$,  a  branch of $S_{n_{\OMS}}^{1/2}$,  raising rapidly enough upon  decreasing frequency, may have contributed to $S_{\Delta g,e}^{1/2}$. 

In Appendix~\ref{app:ifonoise}  we use two independent  methods to put an upper limit on such a possible contribution. The first uses data taken with the test masses  held fixed by the blocking mechanism, the second exploits the independent capacitive measurement of the relative motion of the TMs $\Delta X_{\GRS}$.

The conclusion of the analysis is  a rather conservative  upper limit that places  this contribution  at most at  some 1\%--2\% in power of the total excess noise.

\subsubsection{\label{sec:glitches} Excess noise as a flow  of undetected glitches}

As soon as transient events, known as glitches, were observed in the data \cite{lpf_glitch2022} and removed from them, the question arose if the excess noise might be due to an undetected and nonremoved fraction of glitches.  Not only is the question a legitimate one, but it is also made particularly relevant by the observation that both the glitch properties and the excess noise appeared rather stable throughout the mission, despite the changes in operation conditions \cite{lpf_glitch2022}.

We have addressed the question by performing extensive simulations. The detail of such simulation work is reported in Appendix~\ref{app:glitchsimul}.

The basic conclusion is that the excess noise might be due to a Poisson flow of undetected glitches, but those glitches would belong to a distribution quite well separated in properties from that of the detected ones. In addition, the flow rate should be high enough that the resulting noise would be stationary and Gaussian, bearing no detectable feature proving its Poisson nature.

Thus if the $1/f$ tail is made of Poisson noise this is not related to the observed glitches, and its Poisson nature does not show up in the data.

\subsection{\label{sec:noiseprojection}Projection of  excess noise on modeled noise sources (details in Appendix~\ref{app:decor})}

In this section, we estimate the contribution of modeled sources of force noise to the observed excess over the Brownian noise.

Our approach is to give a quantitative estimate of those contributions whenever they appear to be significantly different from zero, while we try to establish an upper limit whenever the resolution of our methods limits the estimate.

We focus our analysis on the run performed in February 2017, run~10 of Table~\ref{tab:noise-only-runs}, which is the lowest noise one. We will also discuss, whenever relevant, the case for the other runs. 

We consider two categories of sources: the first includes the effect of physical quantities that have been measured during noise runs, synchronously with $\Delta g$; the second includes effects for which we have an estimate from different experiments, performed at different times from those of the noise runs.

For sources of the first category, we infer their contribution to the ASD of $\Delta g$ via a ``decorrelation'' method explained in Appendix~\ref{app:decor}. The method also returns the ``susceptibilities'' of  $\Delta g$ to these disturbances.
For those in the second category, we just give the best estimate of the contributed noise ASD.

\subsubsection{Decorrelation analysis of synchronous time series}

During noise runs we have measured, synchronously with $\Delta g$:
\begin{enumerate}[label=\textit{\alph*}.]
    \item {\label{item:tseries:a} The gravitational force loss due to fuel depletion;}
    \item{\label{item:tseries:b}The relative motion of the two GRS;}
    \item {\label{item:tseries:c}The temperature};
    \item the two temperature differences across the two electrode housings, in the $x$-direction;
    \item the three magnetic field components at four different locations;
    \item a series of spurious low-frequency voltages that have unintentionally been applied to the electrodes via the actuation circuitry nonlinearity.
\end{enumerate}

{A more precise definition of the above time series is contained in Appendix~\ref{app:decorrMCMC}. The decorrelation technique is described in Appendix~\ref{app:decor}. The three series listed in points \ref{item:tseries:a}, \ref{item:tseries:b} and \ref{item:tseries:c} above are separated in this analysis, as these time series are likely to be affected by significant and unknown readout noise (see Appendix~\ref{app:CPSDnoisy}), and the remaining ones, for which the readout noise is known to be negligible (see Appendix~\ref{app:CPSD/decorrelation}).

The results of the decorrelation analyses for the contribution of these sources to $\Delta g$ (see Appendix~\ref{sec:decnonoise},~\ref{app:decres}) are summarized in Fig.~\ref{fig:dec}.

\begin{figure}[htbp]
 \centering
 \includegraphics[width=1\columnwidth]{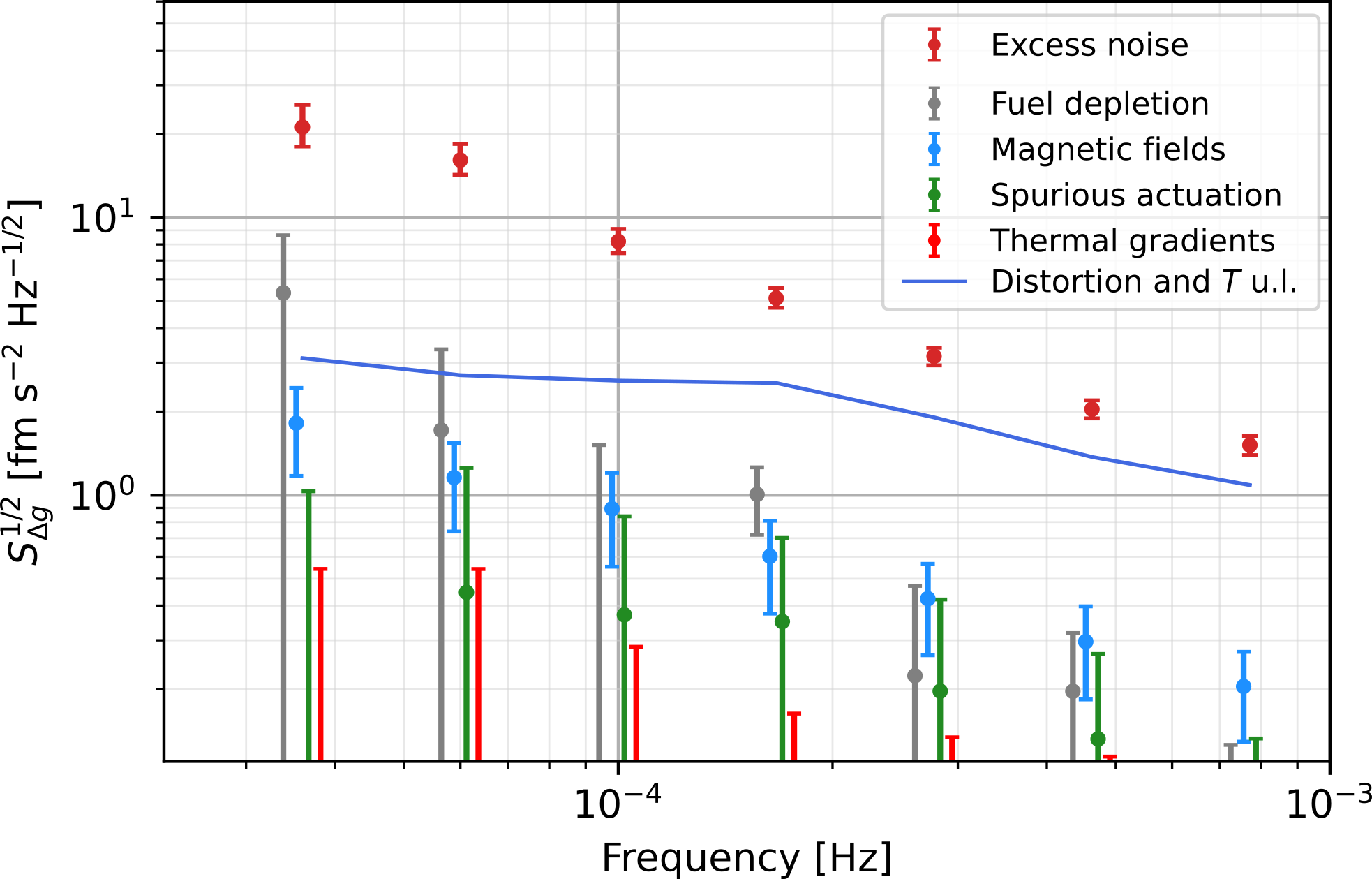}
 \caption{Decorrelation of synchronous time series for run~10, over  the [\SI{36}{\micro\hertz},~\SI{0.77}{mHz}] frequency band.  Frequencies are slightly shifted for clarity.
  Red points, ASD of total excess noise over Brownian, $S_{\Delta g_e}^{1/2}$, as in Fig.~\ref{fig:brownsubtr}; all other points, Bayesian posterior $\pm 1\sigma$ interval for the residual ASD after decorrelation for the time series indicated in the legend; solid blue line, $1\sigma$ upper bound for the contribution of LTP distortion and of pressure-mediated temperature effect to $S_{\Delta g_e}^{1/2}$. This figure summarizes Figs.~\ref{fig:decorrelation q} to~\ref{fig:XT} in Appendix~\ref{app:decor}.}
\label{fig:dec}
\end{figure}

In addition, whenever the analyses give a posterior estimate of susceptibility values, these are found to be compatible with our prior knowledge of the physical system (see Appendix~\ref{sec:decnonoise})}.

\subsection{Modeled contributions from different experiments}
\label{sec:decorrMCMC_noiseprojection}

As said, in addition to the disturbances discussed in the previous section, there is another set of effects that we could estimate from dedicated measurements separated from the noise measurements. The results of these experiments have been published in dedicated publications \cite{PhysRevLett.118.171101,actuation-paper-bill,BrigittePhD}. We list these disturbances in the following and use the results from those experiments.

\begin{figure}[htbp]
  \centering
  \includegraphics[width=1\columnwidth]{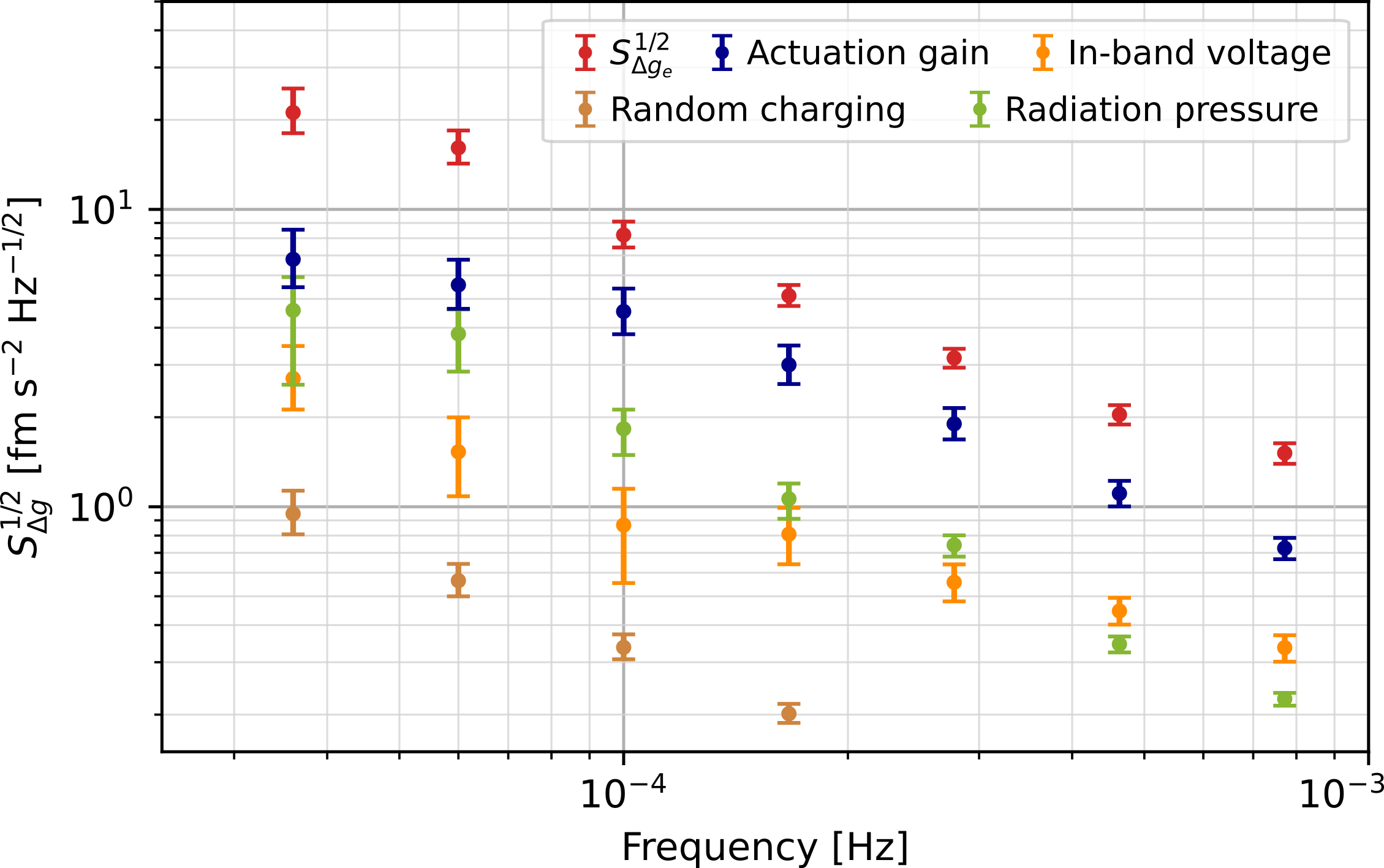}
\caption{\label{fig:noiseprojection} Estimated contributions to acceleration noise, for run~10 and within the [\SI{36}{\micro\hertz},~\SI{0.77}{mHz}] frequency band, of: actuation gain fluctuations (blue points, data adapted from \cite{actuation-paper-bill}), in-band voltage fluctuations (orange points, data adapted from \cite{actuation-paper-bill}), cosmic ray charging fluctuations (brown points, estimate taken from \cite{PhysRevLett.118.171101}), and laser radiation pressure fluctuations (green points, data adapted from \cite{BrigittePhD}). For reference, we also report the ASD of the excess noise over Brownian $S_{\Delta g_e}^{1/2}$ as in Fig.~\ref{fig:brownsubtr}, red points.}
\end{figure}

\begin{enumerate}[label=(\roman*)]
    \item Actuation gain noise. This is the largest known contribution to the LPF noise in $\Delta g_e$. It consists of the effect of the gain fluctuations of the amplifiers that apply the audio-frequency actuation voltages to the TMs. During LPF operations we performed an extensive experimental campaign to model this force. The details of the measurements  and of their results are the subject of a dedicated paper \cite{actuation-paper-bill}. In Fig.~\ref{fig:noiseprojection}, blue points, we report from \cite{actuation-paper-bill} the $\pm1\sigma$-credible interval of the posterior for its contribution to $S_{\Delta g}$. Note that gain fluctuations also affect rotational actuation \cite{actuation-paper-bill} and induce a correlation between $\Delta g$ and $\Delta \gamma_\phi$. The figures in Fig.~\ref{fig:noiseprojection} are in agreement with the effective crosstalk arm discussed in Sec.~\ref{sec:gamma}.
    \item In-band voltage noise. Actuation voltage fluctuations within the measurement frequency band induce noisy forces on the test masses by coupling to their {dc} counterparts. Results of the measurements on this effect performed during LPF operations can also be found in \cite{actuation-paper-bill} and are reported again in Fig.~\ref{fig:noiseprojection} as orange points.
    \item Random charging. The effect of noisy charging due to cosmic rays was estimated in \cite{PhysRevLett.118.171101} to be in quantitative agreement with an equivalent  Poisson flow of  single elementary charges arriving at a rate $\lambda_\text{eff}\sim 1 \times 10^{3} \text{ s}^{-1}$. 
    This Poisson charge flow converts into a force noise through the effective {dc} voltage across the TM electrode capacitor system $\Delta_{x}$, as described in Ref.~\cite{PhysRevLett.118.171101}. Considering residual {dc} voltages --- after compensation as in Table~\ref{tab:vcomp} --- of $|\Delta_{x,1}|\sim|\Delta_{x,2}|\sim\SI{5}{mV}$, the effect of random charging noise is shown as brown points in Fig.~\ref{fig:noiseprojection}.
    \item Laser radiation pressure. Fluctuation of the radiation pressure of the measurement laser beam induces a differential force on the TMs. The measurement beam reflects off TM1 and TM2, with nominal power  $P_{1}=\SI{2.4}{mW}$ and $P_{2}=\SI{1.2}{mW}$, respectively. Relative fluctuations  $\delta P_1/P_1$ and $\delta P_2/P_2$ induce a force $\Delta g\simeq \SI{8}{\pico\meter\,\second^{-2}} \,(\delta P_1/P_1)+\SI{4}{\pico\meter\,\second^{-2}}\,(\delta P_2/P_2)$. Unfortunately, on LPF there was no direct measurement of the instantaneous total optical power reflected off the test masses. However, \cite{BrigittePhD} proposes a thorough analysis of the range of values of the contribution of this phenomenon to $S_{\Delta g}$. The analysis gets two possible estimates depending on the (unknown) sign of the correlation between different light polarizations. In Fig.~\ref{fig:noiseprojection} (green points) we report the overall range spanned by the $\pm1\sigma$ uncertainty of both options. 
\end{enumerate}

\subsection{Summary of modeled contributions to excess noise}
\begin{figure}[htbp]
  \centering
  \includegraphics[width=1\columnwidth]{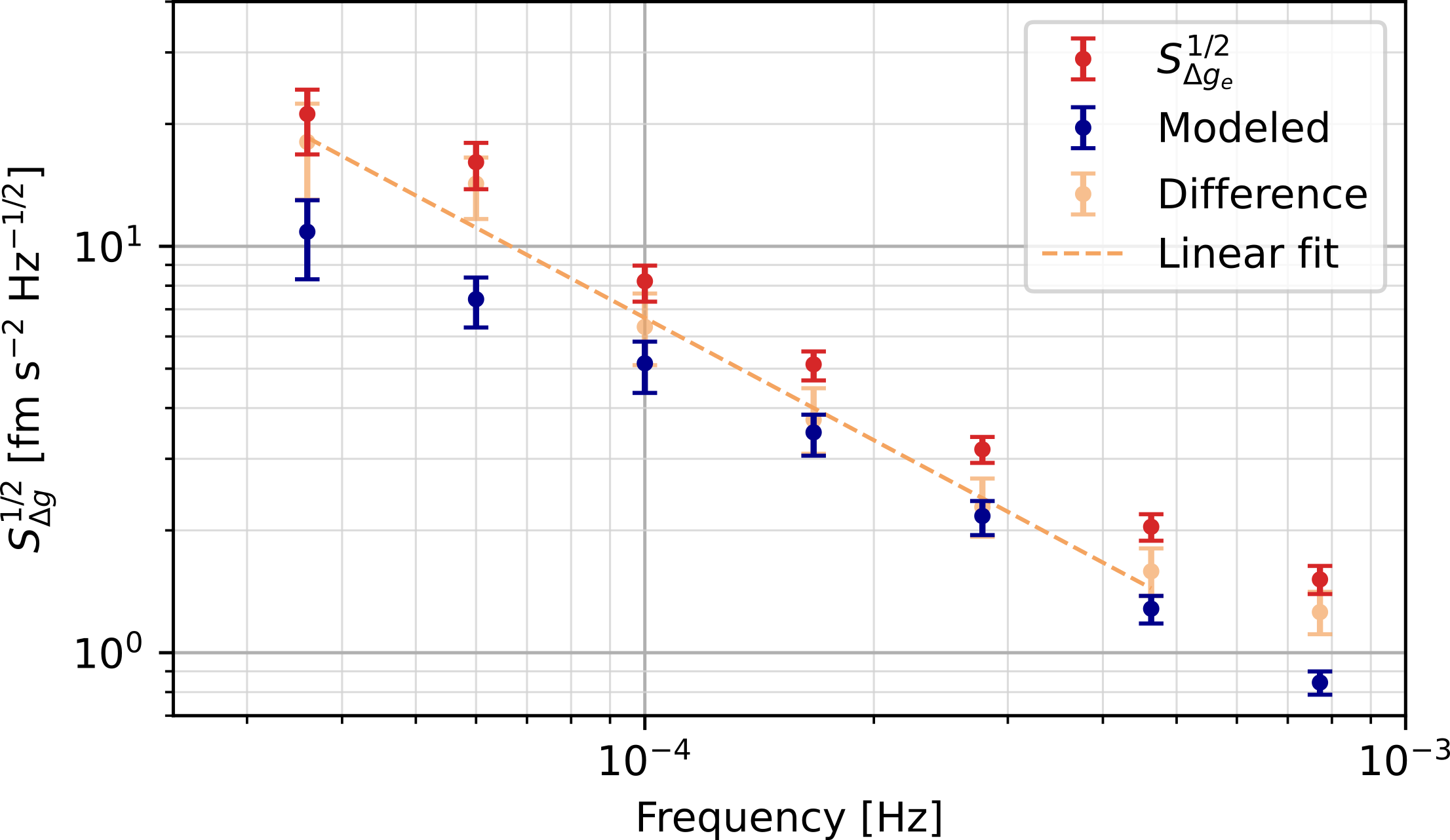}
\caption{\label{fig:findec} Total  modeled  contributions to $S_{\Delta g_e}^{1/2}$, for run~10 in the [\SI{36}{\micro\hertz},~\SI{0.77}{mHz}] frequency band. Red points, ASD $S_{\Delta g_e}^{1/2}$ of excess over Brownian; dark blue points, modeled contributions $S_{c,\text{tot}}^{1/2}$, as described in the text; light brown points, posterior for difference $\sqrt{S_{\Delta g_e}-S_{c,\text{tot}}}$. The dashed line is the best fit  to the difference with exponent $n=1.00$, $\sqrt{S_{\Delta g_e}-S_{c,\text{tot}}}=\SI{0.67}{\femto\meter\,\second^{-2}/\rtHz}\,(\SI{1}{\milli\hertz}/f)^{1.00}$.}
\end{figure}

In Fig.~\ref{fig:findec} we report the ASD of the sum $S_{c,\text{tot}}$ of all contributions to the excess over Brownian $S_{\Delta g_e}$ that we have found to be, at least at some frequency in the [\SI{36}{\micro\hertz},~\SI{0.77}{mHz}] band, statistically different from zero: tank depletion gravitational noise, magnetic fields, actuation gain fluctuations (that dominates the sum), in-band voltage fluctuations, random charging, and laser radiation pressure. 

We have built the posterior for $S_{c,\text{tot}}$ by adding, whenever available,  the samples of the $S_c$ posteriors of the various disturbances. For random charging and laser radiation pressure, for which we only had an error interval, we have assumed a Gaussian distribution with the $\pm 1\sigma$ interval coinciding with the said error one. Given the smallness of their contributions, the results are largely independent of the specific choice of such an equivalent posterior.

From the posterior for $S_{c,\text{tot}}$ and that for $S_{\Delta g_e}$, we have built for their difference, our best estimate of the residual noise after subtraction of the effect of all modeled disturbances listed above. The results are shown again in  Fig.~\ref{fig:findec} as the ASD  $\sqrt{S_{\Delta g_e}-S_{c,\text{tot}}}$.

Such an ASD can be fit, in the range $\SI{36}{\micro\hertz} \le f \le \SI{460}{\micro\hertz}$ to a power law $\widetilde{S}_0^{1/2} (1\,\text{mHz}/f)$ (see Fig.~\ref{fig:findec}),  with  $\widetilde{S}_0^{1/2}=(0.67\pm0.05)\,\text{fm\,s}^{-2}/\rtHz$ and a reduced $\chi$-square $\simeq 1$, which allows  a  direct comparison with the value of $\widetilde{S}_{\Delta g_e}^{1/2}$ (see Sec.~\ref{sec:oneoverf}) for run~10, of  $\widetilde{S}_{\Delta g_e}^{1/2}=(0.91\pm0.04)\,\text{fm s}^{-2}/\sqrt{\text{Hz}}$. The ratio of their square, $\widetilde{S}_0/\widetilde{S}_{\Delta g_e}=(0.54\pm0.09)$ gives the fraction of noise power that remains unexplained after our noise projection, the basic result of the procedure.

We have repeated our decorrelation procedure for the other runs in order to assess if some of the observed variability of $\widetilde{S}_{\Delta g_e}$ (see Fig.~\ref{fig:evolution}) might be due to corresponding variations of some of the considered disturbances, and we could not find any evidence of that.

Among the disturbances in Sec.~\ref{sec:decorrMCMC_noiseprojection}, actuation gain fluctuations are the dominating ones. Unfortunately, we cannot assess the variability of those sources, as they were all determined in dedicated experiments. To justify the observed 20\% variability of $\widetilde{S}_{\Delta g_e}$ by a corresponding variability of those, the latter  should be of order 50\%. 

On the opposite end, if the entire variability is due to the 55\% unjustified fraction, this should fluctuate by some 40\%.

Finally, we have not included, in the above analysis, the contributions that we have found to be compatible with zero: LTP distortion, spurious actuation, and thermal gradients. While the addition of the latest two would not change the results in any appreciable way, the case of the distortion is different. Actually, the  sum of the $+1\sigma$ value of $S_{c,\text{tot}}$ in Fig.~\ref{fig:findec} and the $1\sigma$ limit for $S_{c,\text{tot}}$ of Fig.~\ref{fig:XT}, falls within $\pm 1\sigma$-credible interval of $S_{\Delta g_e}$  for $f\ge f_5$. 

As said in Appendix~\ref{app:XT}, however, this limit is likely to be significantly overestimated. The argument discussed in Appendix~\ref{app:XT}, for instance, would reduce this contribution to no more than some 10\% in power at the highest frequencies.

\section{\label{sec:nonmod} The unexplained excess: summary of possible sources and implications for LISA (details in Appendix~\ref{app:nonmod})}
The previous section concludes that $(0.54\pm0.09)$ of the measured excess noise power remains unexplained by the sources for which we had a quantitative estimate. In Appendix~\ref{app:nonmod} we discuss the most likely sources of this unexplained fraction, and the measures one can possibly take to ensure that they do not compromise the LISA performance. In summary, the discussion considers the following disturbances:
\begin{enumerate}[label=(\roman*)]
\item Interaction of the quasistatic part of patch potentials with the time fluctuations of these (see Appendix~\ref{app:patchV}). This is an effect due to patch potentials that is not covered by Sec.~\ref{sec:decorrMCMC_noiseprojection}.
\item Inaccuracy in the calibration of the applied forces $g_c(t)$ (see Appendix~\ref{app:actnonlin}), which dominate the spectrum at submillihertz frequencies, including that resulting from unaccounted nonlinearities in the applied voltages time series.
\item Unmodeled gravitational noise, excluding modeled contributions from propellant tank depletion and LTP distortion, which are addressed in Appendix~\ref{app:unmodeledgrav}. Indeed any other mass motion, either because of distortion of solid parts or because of evaporation of volatile fractions, may cause gravitational force noise and may have contributed to excess noise.
\item Pressure fluctuations (see Appendix~\ref{app:pressfluct}). As the complex geometry of the TM environment may create quasistatic pressure gradients, any in-band fluctuation of such gradients would directly translate into an in-band acceleration fluctuation. 
\item High-frequency magnetic field noise (see Appendix~\ref{app:hfmag}). In addition to low-frequency effects, discussed in Sec.~\ref{sec:noiseprojection}, magnetic fields at high frequency may induce eddy currents within the test masses, and then exert Lorentz forces on them \cite{labtoLPF}. The effect is thus quadratic and would convert the low-frequency amplitude fluctuations of a high-frequency magnetic spectral line into a corresponding low-frequency force. 
\end{enumerate}

Based on the analysis of the disturbances mentioned above, we outline in Appendix~\ref{app:nonmod} several measures to control these potential noise sources. These measures include a series of ground tests focusing on the nature of the atmosphere and surface adsorbates within the VE, the magnetic characterization of the spacecraft in the audio frequency band, and the properties of the front-end electronics (FEE). Additionally, a cautious and conservative approach in designing some key features of the LISA spacecraft and GRS is recommended to minimize deviations from the LPF design.

\section{\label{sec:con}Concluding remarks}

LISA Pathfinder reached an acceleration noise performance achieving the LISA requirements with margin, and better than both its requirements and what had been estimated before launch \cite{labtoLPF}. \\
This last achievement was  mostly allowed by two facts:
\begin{enumerate}[label=(\roman*)]
\item a much better self-gravity cancellation than what had been very cautiously estimated on ground, which in turn allowed using much less electrostatic force authority than predicted \cite{armano:subfemtog}, and suppressing accordingly the actuation noise;
\item the achievement, over the course of the mission, of a lower base pressure, and thus a lower Brownian noise, than had been assumed in prelaunch estimates \cite{labtoLPF}.
\end{enumerate}

Both these facts revealed the existence of an excess noise above the Brownian noise level, with a $1/f^2$ PSD.\\
Based on this starting framework, in this paper we have:

\begin{itemize}
    \item shown that the Brownian noise evolved in agreement with the outgassing of a single gaseous species diffusing out of the immediate, complex surroundings of the TM, where such outgassing also maintains a quasistatic pressure gradient across the TM;
    \item shown that the temperature stability of the system was good enough that temperature fluctuation played a significant role in the acceleration noise only at the lowest analyzed frequency \SI{18}{\micro\hertz}, well below the LISA lower frequency of \SI{0.1}{\milli\hertz};
    \item shown that the intrinsic stability of the $1/f^2$ excess PSD (referred to in the article as $1/f$ excess for ASD) was $\pm 20\%$ in amplitude over more than 16 months, with the residual fluctuations being  independent of any traceable change in the operational  conditions that were needed to run the mission; 
    \item analyzed all sources of noise for which we had a verified model, either from correlation analysis  or from dedicated experiments, and concluded that these sources account for a fraction $0.46\pm0.09$ of the total power of the excess;
    \item finally discussed all possible explanations we could trace for the unaccounted part of the excess,  patch potentials, actuation electronics nonlinearity, gravitational noise, audio-frequency magnetic fields, and pressure fluctuations, and identified possible measures to keep them under full control during the implementation of LISA.
\end{itemize}

\section*{Acknowledgments}
This work has been made possible by the LISA Pathfinder mission, which is part of the space-science program of the European Space Agency.\\
We thank Paolo Chiggiato and his team at CERN, for very helpful discussions about the LPF outgassing environment.\\
The Italian contribution has been supported by Istituto Nazionale di Fisica Nucleare (INFN) and Agenzia Spaziale Italiana (ASI), Project No. 2017-29-H.1-2020 ``Attivit\`a per la fase A della missione LISA''. 
The UK groups wish to acknowledge support from the United Kingdom Space Agency (UKSA), the Scottish Universities Physics Alliance (SUPA), the University of Glasgow, the University of Birmingham, and Imperial College London. 
The Swiss contribution acknowledges the support of the Swiss Space Office via the PRODEX Programme of ESA, the support of the ETH Research Grant No. ETH-05 16-2 and the support of the Swiss National Science Foundation (Projects No. 162449 and No. 185051).  
The Albert Einstein Institute acknowledges the support of the German Space Agency, DLR. The work is supported by the Federal Ministry for Economic Affairs and Energy based on a resolution of the German Bundestag (No. FKZ 50OQ0501, No. FKZ 50OQ1601, and No. FKZ 50OQ1801). 
J.I.T. and J.S. acknowledge the support of the U.S. National Aeronautics and Space Administration (NASA). 
The Spanish contribution has been supported by Contracts No. AYA2010-15709 (Ministerio de Ciencia e Innovaci\'on MICINN), No.ESP2013- 47637-P, No. ESP2015-67234-P, No. ESP2017-90084-P (Ministerio de Asuntos Econ\'omicos y Transformaci\'on Digital, MINECO) and No. PID2019-106515GB-I00 and PID2022-137674NB-I00 (MCIN/AEI/10.13039/501100011033). Support from AGAUR (Generalitat de Catalunya) Contracts No. 2017-SGR-1469 and 2021-SGR-01529 is also acknowledged. It has also received partial support from the program \textit{Unidad de Excelencia Mar\'{\i}a de Maeztu} CEX2020-001058-M (Spanish Ministry of Science and Innovation). 
The French contribution has been supported by the CNES (Accord Specific de projet No. CNES 1316634/CNRS 103747), the CNRS, the Observatoire de Paris and the University Paris-Diderot. E.P. and H.I. would also like to acknowledge the financial support of the UnivEarthS Labex program at Sorbonne Paris Cit\'e (No. ANR-10-LABX-0023 and No. ANR-11-IDEX-0005-02).
N.K. is thankful for the support from a CNES Fellowship.

\newpage
\appendix
\section*{Appendix}

The following Appendixes contain some rather important information, tables, and calculations.\\
In Appendix~\ref{app:runconf}, we give details about the analyzed noise runs.\\
In Appendix~\ref{app:CPSD}, we describe our spectral analysis tools and PSD estimation methods.\\
In Appendix~\ref{app:CPSD/freqs}, we list the frequencies used in PSD estimation.\\
In Appendix~\ref{app:allpsd}, we provide the ASD of $\Delta g(t)$ for all analyzed noise runs.\\
In Appendix~\ref{app:driftcalc}, we present the calculations for instrument distortion analysis and the drift evaluation procedure.\\
In Appendix~\ref{app:BLTD}, we discuss the evolution of Brownian noise and the long-term drift.\\
In Appendix~\ref{app:obsart}, we discuss possible observational artifacts: the role of interferometer noise and under-threshold glitches.\\
In Appendix~\ref{app:decor}, we present the decorrelation of measured time series. First, we describe the analysis framework and our statistical methods, then we apply them to measurements.\\
In Appendix~\ref{app:nonmod}, we provide a rather detailed discussion about the possible sources behind the unmodeled excess noise.

\renewcommand{\arraystretch}{1.1}
\begin{table*}[htbp]
    \centering
    \begin{tabular}{>{\centering\arraybackslash}p{1.8 cm}|>{\centering\arraybackslash}p{2.2 cm}|>{\centering\arraybackslash}p{2.2 cm}|>{\centering\arraybackslash}p{2.2 cm}|>{\centering\arraybackslash}p{2.2 cm}|>{\centering\arraybackslash}p{2.2 cm}|>{\centering\arraybackslash}p{2.2
    cm}|>{\centering\arraybackslash}p{1.6
    cm}}
    
 Run & Propellant & Thruster & Actuation & TM & Voltage & Heater & ST7 \\
  & tank& branch & authority & alignment & compensation & configuration & state \\\hline\hline
     1 & 2 & \texttt{A} & \texttt{URLA} & 1 & 1 & 1 & \texttt{OFF} \\
 2 & 2 & \texttt{A} & \texttt{UURLA} & 1 & 1 & 1 & \texttt{OFF} \\
 3 & 2 & \texttt{A} & \texttt{UURLA} & 1 & 2 & 1 & \texttt{OFF} \\
 4 & 3 & \texttt{A} & \texttt{UURLA} & 2 & 3 & 1 & \texttt{OFF} \\
 5 & 3 & \texttt{A} & \texttt{UURLA} & 3 & 3 & 2 & \texttt{DIAG} \\
 6 & 1 & \texttt{A} & \texttt{UURLA} & 3 & 3 & 2 & \texttt{DIAG} \\
 7 & 3 & \texttt{B} & \texttt{UURLA} & 3 & 3 & 2  & \texttt{DIAG}\\
 8 & 1 & \texttt{B} & \texttt{UURLA} & 3 & 3 & 3 & \texttt{DIAG} \\
 9 & 1 & \texttt{B} & \texttt{UURLA} & 3 & 4 & 4 & \texttt{OFF} \\
 10 & 1 & \texttt{B} & \texttt{UURLA} & 3 & 4 & 5 & \texttt{OFF} \\
 11 & 3 & \texttt{A} & \texttt{UURLA} & 3 & 4 & 6 & \texttt{OFF} \\
 12 & 3 & \texttt{A} & \texttt{UURLA} & 3 & 4 & 1 & \texttt{OFF} \\
 13 & 3 & \texttt{A} & \texttt{UURLA} & 3 & 4 & 1 & \texttt{OFF} \\
\end{tabular}
    \caption{Experimental configuration for the 12 different runs of Table~\ref{tab:noise-only-runs}. The meaning of the numeric labels is explained in the text.
    }
    \label{tab:conf}
\end{table*}

\section{\label{app:runconf} Experimental configurations for the 13 runs}
There were minor differences in the operating conditions of the 13 runs of Table~\ref{tab:noise-only-runs}. We describe them in the following and summarize the different configurations in Table~\ref{tab:conf}.

\begin{enumerate}[label=(\roman*)]
    \item Thruster propellant was stored in three different tanks. The gravitational signal from the depletion of these tanks was different, due to the different positions. In Table~\ref{tab:conf}, we name these three tanks as 1, 2, and 3. 
    \item For redundancy reasons, LPF carried two independent branches of micro-thrusters \cite{PRDThrust}. In Table~\ref{tab:conf}, we call them A and B.
    \item The electrostatic controllers needed a setting for the maximum force/torque authority they could deliver within a linear regime \cite{ActNeda}. These settings determine the zero-actuation voltages commanded to the various electrodes. The settings were identical for all the considered runs except for the first one. Values are reported in Table~\ref{tab:act}.
    \begin{table}[ht] 
            \begin{tabular}{>{\arraybackslash}p{2.5 cm}|>{\centering\arraybackslash}p{1.4 cm}|>{\centering\arraybackslash}p{1.4 cm}}
                & ~\texttt{URLA}~ & ~\texttt{UURLA}~ \\
              \hline\hline
              ~~$F_{x,1}$ (pN)  & 0     & 0   \\
              ~~$F_{x,2}$ (pN)  & 50    & 50  \\
              ~~$N_{x,1}$ (pN\,m) & 16.37 & 4   \\
              ~~$N_{x,2}$ (pN\,m) & 16.37 & 4   \\
              ~~$F_{y,1}$ (pN)  & 3670  & 1000\\
              ~~$F_{y,2}$ (pN)  & 3670  & 1000\\
              ~~$N_{y,1}$ (pN\,m) & 13.32 & 4   \\
              ~~$N_{y,2}$ (pN\,m) & 13.32 & 4   \\
              ~~$F_{z,1}$ (pN)  & 5820  & 500 \\
              ~~$F_{z,2}$ (pN)  & 5820  & 500 \\
              ~~$N_{z,1}$ (pN\,m) & 1.5   & 1.5 \\
              ~~$N_{z,2}$ (pN\,m) & 1     & 1   \\
            \end{tabular}
  \caption{Actuation authorities, $F_{a,i}$ and $N_{a,i}$, respectively the maximum force and the maximum torque applicable on TM$i$ along axis $a$.}
    \label{tab:act}
\end{table}

\item In the attempt to reduce crosstalk between the differential interferometer readout and the motion of the remaining TM degrees of freedom \cite{PhysRevD.108.102003_LPF_TTL_2023}, we performed three different adjustments of the zero set points of all control loops. The most relevant ones were those for the angular orientations of the TM along $z$ and $y$. The three set points, named 1, 2, and 3, are listed in Table~\ref{tab:set}.
\begin{table}[ht]
    \begin{tabular}{c|c|c|c}
    ~~Angle~~&~~Set point~~&~~Set point~~&~~Set point~~\\
    ~$(\si{\micro\radian})$~&1&2&3\\
    \hline\hline
        $\phi_1$& $-59.25$ & $-56.32$ & $-61.2$ \\
        $\phi_2$& $-21.35$ & $-33.01$ & $-9.7$  \\
        $\eta_1$& $-3.5$   & $-2.14$  & $-4.9$  \\
        $\eta_2$&  $3.5$   & $10.3$   & $-3.3$  \\
    \end{tabular}
    \caption{The three different set points for the angular orientations of the TMs along $z$ ($\phi$) and $y$ ($\eta$), used during the noise runs. The index indicates which TM each angle refers to. Angles are relative to a reference frame defined by the OMS.}
    \label{tab:set}
\end{table}

\item{Random TM charging due to cosmic rays couples to dc voltage differences between the TM and the surrounding metal surfaces to produce random force and torque \cite{PhysRevLett.118.171101}. This noise may be suppressed by purposely applying some dc voltages to the various electrodes, in order to compensate the parasitic ones that are found on metal surfaces because of work function differences  \cite{PhysRevLett.118.171101}. In particular, the force along $x$ and the torque along $\phi$, may be compensated by a proper combination of voltages on the electrodes facing the $x$-faces of the TMs. These compensation voltages may be described by just voltage parameters $\Delta_x$ and $\Delta_\phi$ as explained in \cite{PhysRevLett.118.171101}. In the course of the mission we have made some adjustments of these parameters for both TMs. The various configurations are listed in Table~\ref{tab:vcomp}.
\begin{table}[ht]
    \centering
    \begin{tabular}{c|c|c|c|c}
    ~Voltage~~&~Setting~~&~Setting~~&~Setting~~&~Setting~~\\
    (mV)&1&2&3&4\\
    \hline\hline
         $\Delta_{x,1}$&0&+24&+12&+24 \\
         $\Delta_{x,2}$&0&0&0&0 \\
         $\Delta_{\phi,1}$&0&0&0&+32 \\
         $\Delta_{\phi,2}$&0&0&0&-116 \\
    \end{tabular}
    \caption{Compensation voltages for the 4 adopted settings. The index indicates the TM.}
    \label{tab:vcomp}
\end{table}

}

\item{While the noise runs in Table~\ref{tab:noise-only-runs} were all performed by using the satellite cold gas thrusters for drag-free control \cite{PRDThrust},  LPF also carried a set of alternative thrusters based on colloidal propellant technology, in the framework of NASA ST7 mission \cite{PhysRevD.98.102005}. These thrusters were used intermittently, leaving them, for purpose of diagnostics, in some activated state even when not in use, in the two epochs from June 27, 2016 to December 7, 2016, and from March 18, 2017 to April 29, 2017. In Table~\ref{tab:conf} we indicate such state as \texttt{DIAG}, while the completely off state is indicated with \texttt{OFF}.}
\end{enumerate}

\newpage
\section{\label{app:CPSD}Spectral estimation methods}

\subsection{Periodograms and their spectral properties}
We use, for the  elementary periodogram $X(k)$ of the $N$-sample series $x[n]$, sampled with sampling time $T$, the standard definition:
\begin{equation}
\label{eq:ft}
    X(k) = \sqrt{\frac{T}{N}} \sum_{n=0}^{N-1} x[n]  w[n] \, e^{-2 \pi i k n / N},
\end{equation}
where $w[n]$ are the coefficients of a  Blackman-Harris spectral window, which gives good side-lobe suppression  \cite{PhysRevD.90.042003}.

Following the  Welch  method \cite{Welch1967}, we section our data series in as many 50\% overlapping data stretches of length $N$ as they fit into the length of the data series and define the average, one-sided experimental PSD at frequency $f$:
\begin{equation}
\label{eq:expPSD}
    \Pi\left(f=\frac{k}{NT}\right) = \frac{2}{M} \sum_{l=1}^M X_{(l)}(k) X_{(l)}^*(k),
\end{equation}
where the index $l$ runs over the $M$ data stretches, and $k$ is the frequency index. 

For multiple synchronously measured data series $x_i[n]$, with $1\le i\le p$, Eq.~\eqref{eq:expPSD}  generalizes to the (one-sided) experimental cross-power spectral density (CPSD) matrix $\bPi$, with elements
\begin{equation}
\label{eq:CPSDdef}
    \Pi_{ij}\left(f=\frac{k}{NT}\right) =\frac{2}{M} \sum_{l=1}^M X_{i,(l)}(k) X_{j,(l)}^*(k),
\end{equation}
a complex and Hermitian matrix. It is also positive definite if $M\ge p$. Our founding point is that, if the $x_i(t)$ are Gaussian, zero-mean stationary stochastic processes, with a theoretical cross-spectral density matrix $\bSig$, then the matrix $\bW$, whose elements are $ W_{ij}=M \Pi_{ij}$, is distributed like a complex Wishart distribution \cite{goodman_statistical_1963,Shaman},
\begin{equation}
\label{eq:wishartSte}
    p\left(\bW\big\vert\bSig \right) = \frac{\left|\bW \right|^{M-p}}{\widetilde{\Gamma}_p(M) \left| \bSig \right|^{M}} \ \etr \left[- \bSig^{-1}\bW\right],
\end{equation}
where $\left|\cdot\right|$ is the determinant, $\etr$ the exponential trace $\etr(\cdot)=\exp\left(\mathrm{tr}\left(\cdot\right)\right)$ and $\widetilde{\Gamma}_p(M)$ is the multivariate complex Gamma function, defined by
\[
\widetilde{\Gamma}_p(M) = \pi^{\frac{1}{2} p(p-1)} \prod_{i=1}^{p} \Gamma(M-i+1)
\]

We denote this distribution as $\cw(\bSig,M)$ and indicate that $\bW$ follows this distribution as $\bW\sim\cw(\bSig,M)$. 

Note that in Eq.~\eqref{eq:wishartSte} it is required that $M\ge p$, otherwise the matrix $\bW$ is singular. This distribution is the basis of our spectral analysis method.

\subsection{PSD estimation}\label{app:PSDestimate}
For $p=1$, $\cw(\bSig,M)=\cw(S_x,M)=\Gamma(M,M/S_x)$, with $\Gamma(\alpha,\theta)$ the gamma distribution with shape parameter $\alpha$ and scale parameter $\theta$, and with $S_x$ the theoretical PSD of the sole series $x(t)$.

This distribution can be used, along with a proper prior, to build the Bayesian posterior for $S_x$. According to standard, physically sound practice, we take an uninformative flat prior in the logarithm of $S$, which is a prior proportional to $1/S$. For $p=1$ this also coincides with the Jeffreys, reparametrization-independent prior \cite{Jeffreys}.

Such choice  gives a  posterior for $S$ that is distributed as an inverse-Gamma distribution, with shape parameter $M$ and scale parameter $M\Pi$,
\begin{equation}
\label{eq:postS}
p(S\vert\Pi,M) \sim \text{inv}\Gamma\left(M, M \Pi\right)
\end{equation}
 
The expected value of this posterior, $\langle S\rangle=M\Pi/(M-1)$, is slightly biased and diverges for $M\to1$. However,  the posterior probability density function remains well behaved even in such limit case. We have checked, by a simple numerical simulation, that for all values of $M$ and for Gaussian data, the theoretical value lies in the $p$-credible interval with probability $p$.

Based on this posterior distribution, we compute the equally tailed $p$-credible interval from the $(1\pm p)/2$ quantiles of the distribution. Unless otherwise specified, error bars in our plots correspond to $p=68.3\%$, as for the $\pm 1\sigma$ interval in a normal distribution. The dots in those same plots represent the median of the distribution, which remains well behaved also for $M=1$, and has a smaller bias than the mean.  Note that, for the rest of the paper, with the ``$n\sigma$ interval'' we indicate the equal-tailed, $p$-credible interval of any probability distribution, with $p$ the probability for a normal random variable to fall in the interval $\pm n\sigma$, with $\sigma$ its standard deviation. In particular, for $n=1$, $p=0.683$, and for $n=2$, $p=0.955$. In addition, unless otherwise specified, dots in plots represent the median of the distribution, and error bars the $1\sigma$ confidence interval.

{Note that the adoption of the Jeffreys prior represents a slight modification with respect to  the choice adopted in \cite{PhysRevLett.120.061101}. There the adopted  prior  was flat in $S$, as opposed to flat in  its logarithm. Such a  choice gives  a distribution $p(S\vert\Pi,M) \sim \text{inv}\Gamma\left(M-1, M \Pi\right)$, which has a slightly larger bias, $M\Pi/(M-2)$, relative to $\Gamma\left(M-1, M \Pi\right)$. 

In the data of \cite{PhysRevLett.120.061101}, thanks to the large number of periodograms allowed by the long duration of the run,  such an  increase in bias, at the two lowest frequencies for which $M=9$, is $\simeq 15\%$,  a figure  still significantly less than width of the posterior. At higher frequencies the increase in bias becomes completely negligible. For some of the shortest runs discussed here, however, the bias resulting from a prior flat in $S$ may  become significant.}

\section{Choice of quasi-independent frequencies}
\label{app:CPSD/freqs}
As presented in Supplemental Material of \cite{PhysRevLett.120.061101}, we evaluate the CPSD at log-spaced frequencies, such that the correlation among adjacent frequencies is kept below 5\%. The number of averaged periodograms $M$ varies with frequency, so that the variance is optimally reduced as more periodograms are available. However, $M$ is the same for the first and second frequencies: even though this induces a slightly higher correlation, it allows us to analyze the lowest frequency, \SI{18}{\micro\hertz}. Frequencies are chosen so that the lowest bins always have a fair amount of periodograms, and the fourth one (\SI{0.1}{\milli\hertz}) is the lower bound of the official LISA frequency band. The frequencies are listed in Table~\ref{tab:freq}.

\begin{table}[htbp]
  \centering
  \includegraphics[]{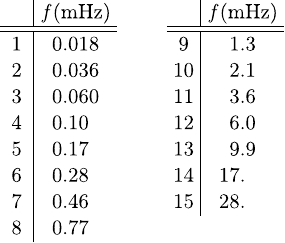}
    \caption{List of frequencies used for spectral estimation.}
    \label{tab:freq}
\end{table}

\section{\label{app:allpsd}ASD of \texorpdfstring{$\Delta g(t)$}{Dg} for all runs}
\begin{figure*}[htbp!]
  \centering
  \includegraphics[width=\textwidth]{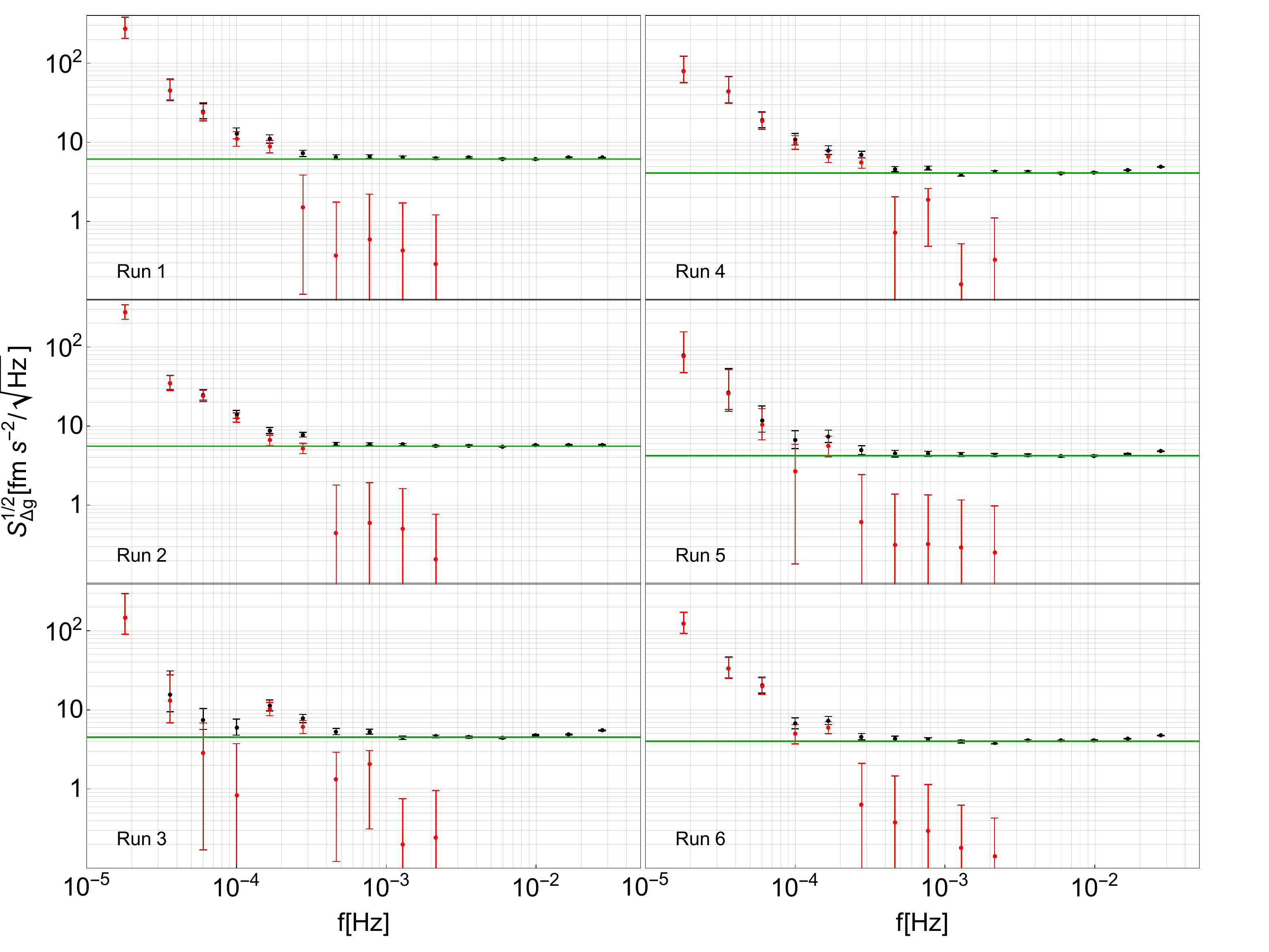}
  \caption{PSD of $\Delta g(t)$ for runs from 1 to 6. The meaning of all symbols is the same as in Fig.~\ref{fig:brownsubtr}.}
  \label{fig:batch1}
\end{figure*}

\begin{figure*}[htbp!]
  \centering
  \includegraphics[width=\textwidth]{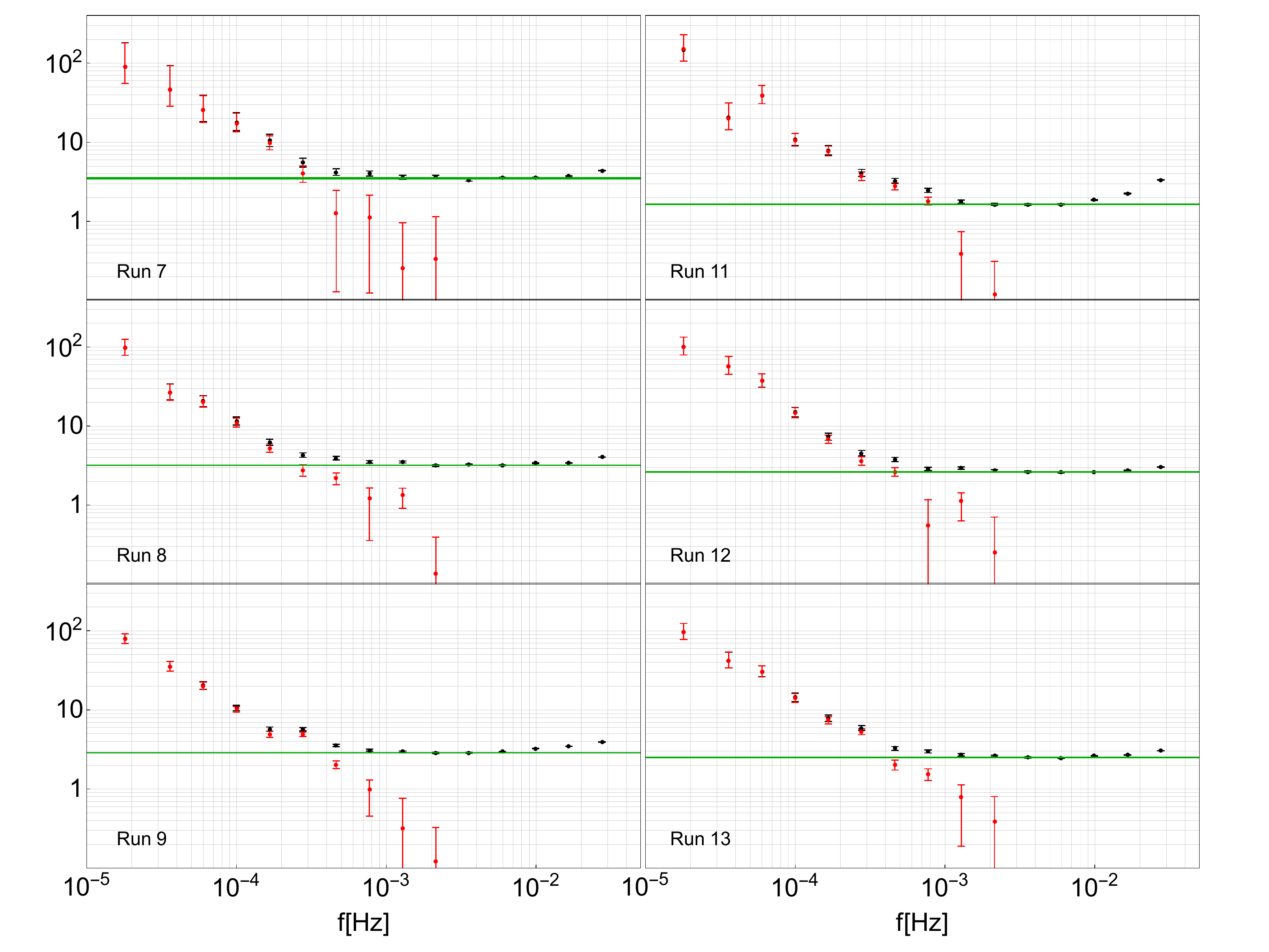}
  \caption{PSD of $\Delta g(t)$ for runs from 7 to 13, except for run~10 which has already been shown in Fig.~\ref{fig:brownsubtr}. The meaning of all symbols is the same as in Fig.~\ref{fig:brownsubtr}.}
  \label{fig:batch2}
\end{figure*}

Figures~\ref{fig:batch1} and~\ref{fig:batch2}, show the ASD for all runs of Table~\ref{tab:noise-only-runs}, except for run~10 which is shown in Fig.~\ref{fig:brownsubtr}.

\clearpage

\section{\label{app:driftcalc} Calculation of gravitational long-term variations}

\subsection{Instrument distortion}
As already said in Sec.~\ref{sec:lfbin}, the capacitive motion sensor gives a measurement of the relative motion of the GRS relative to its own test mass, $x_{\GRS,i}=x_i-X_i$. Here the suffix $i$ indicates the test mass, and $X_i$ is the coordinate of the GRS surrounding TM$i$.

As we are only interested in time variations of differences of coordinates, here the reference frame  may be taken that of the spacecraft, and each coordinate is zero at the nominal position of the body it belongs to, that is, at the position of the body in the absence of distortion. 
    
The variation of gravitational field due to such distortion-induced  motion of both  GRS and the TMs can be written as: 
\begin{equation}
    \begin{split}
      &\Delta g_{\text{Dist},0}=\\
      &-\omega_{2,2}^2\left(x_2-X_2\right)-\omega_{\text{SC},2}^2x_2-\omega_{2,1}^2\left(x_2-X_1\right)+\\
      &+\omega_{1,1}^2\left(x_1-X_1\right)+\omega_{\text{SC},1}^2x_1+\omega_{1,2}^2\left(x_1-X_2\right)
      \end{split}
    \end{equation}
Here $-\omega_{i,j}^2$ is the force gradient (per unit mass) caused by GRS $j$ on TM $i$ and $-\omega_{\text{SC},i}^2$ is that caused by all the remaining parts of the spacecraft, mostly the OMS, on TM $i$. Notice that $\omega_i^2$ in Eq.~\eqref{eq:deltag} is given by $\omega_i^2=\left(\omega_{i,i}^2+\omega_{\text{SC},i}^2+\omega_{i,j}^2\right)$.

Within this model the $\Delta g$  time series is expected to be  affected by a distortion-induced component,
    \begin{equation}
    \begin{split}
      &\Delta g_\text{Dist}=\Delta g_{\text{Dist},0}+\omega_2^2\left(x_2-x_1\right)+\left(\omega_{2}^2-\omega_1^2\right)x_1=\\
      =&\left(\omega_{2,2}^2-\omega_{2,1}^2\right)X_2-\left(\omega_{1,1}^2-\omega_{1,2}^2\right)X_1 
      \simeq \\
      &\simeq\left(\omega_{2,2}^2-\omega_{2,1}^2\right)\left(X_2-X_1\right)
      \end{split}
      \label{eq:distortion}
    \end{equation}
    
    The last simplification comes from the fact that, due the symmetry of the instrument, $\left(\omega_{2,2}^2-\omega_{2,1}^2\right)\simeq \left(\omega_{1,1}^2-\omega_{1,2}^2\right)$ within the same accuracy within which $\omega_1^2\simeq\omega_2^2$, as discussed in Sec.~\ref{sec:dyn}.
    
    The relative motion of the two GRSs, $\Delta x=X_2-X_1$, can be obtained as
    \begin{equation}
    \begin{split}
      &\Delta X=X_2-X_1=\Delta x_\text{OMS}-\Delta x_\text{GRS}=\\
      &=x_2-x_1-\left(\left(x_2-X_2\right)-\left(x_1-X_1\right)\right)  
      \label{eq:appDX}
    \end{split}
    \end{equation}
    
    Thus, in conclusion, 
    \begin{equation}
    \begin{split}
       &\Delta g_\text{Dist}=\\
       &=\left(\omega_{2,2}^2-\omega_{2,1}^2\right)\left(\Delta x_\text{OMS}-\Delta x_\text{GRS}\right)\equiv\\
       &\equiv\omega_d^2\left(\Delta x_\text{OMS}-\Delta x_\text{GRS}\right)
    \end{split}
    \end{equation}
    
    $\omega_{2,2}^2$ is dominated by both gravitational and electrical effects \cite{PhysRevD.97.122002}, while the origin of $\omega_{2,1}^2$ is just gravitational. We can get an estimate of the electrical terms from \cite{PhysRevD.97.122002}, while the gravitational components have been estimated in \cite{gravOHB}, within the work to suppress the gravitational field at the TM location \cite{gravity}.
    
    Based on  these references, we calculate  $\omega_d^2= (-3.32\pm0.05)\times 10^{-7}~\text{s}^{-2}$.

\subsection{Propellant tank depletion} 
The sign and the magnitude of $\Delta g_\text{Tanks}$ depends on the specific tanks from which the propellant is taken. Tanks 1 and 2 of Table~\ref{tab:conf} in Appendix~\ref{app:runconf}, produce, respectively, a $\Delta g_\text{Tanks}$ of $\lambda_{t,1}\simeq +39 \text{\,pm s}^{-2}\text{ kg}^{-1}$ and $\lambda_{t,2}\simeq +37 \text{\,pm s}^{-2}\text{\,kg}^{-1}$ per unit mass of contained propellant, while for tank 3 this figure is  instead $\lambda_{t,3}\simeq -43 \text{\,pm\,s}^{-2}\text{\,kg}^{-1}$.  These figures are the result of a numerical calculation on the geometry and location of the tanks and are affected by a relative error of $\simeq 5\%$.
    
The mass loss is monitored by set of flow meters, one for each of the thrusters, that measure the instantaneous flow of mass through that thruster.  As the thrusting system contains two  branches (see Appendix~\ref{app:runconf}), the measurement of the total flow of mass from the tank in use during any given run $\dot{m}$ is the sum of the readings of the six flow-meters of the thrusters belonging to the branch in use during that specific run. We estimate an absolute accuracy on  $\dot{m}$, for both branches, of the order of 10\%.

From the figures above, one can estimate $\Delta g_\text{Tank}$, during a specific run that uses tank $i$ and branch $j$, as
\begin{equation}
\begin{split}
&\Delta g_\text{Tank}(t)=\kappa_{t,i}\; \lambda_{t,i}\; \kappa_{b,j}\left(m_j(0)-\int_0^t \dot{m}_j(t)dt\right)\equiv\\
&\equiv \kappa_{t,i} \kappa_{b,j} \Delta g_{\text{Tank},0,i,j}(t)+c
\end{split}
\end{equation}

Here we have introduced
\begin{enumerate}[label=(\roman*)]
    \item a calibration factor $\kappa_{t,i}$ for the gravitational gradient of tank $i$,
    \item a calibration factor $\kappa_{b,j}$ for the reading $\dot{m}_\text{j}(t)$ of the mass outflow from tank $j$,
    \item the initial propellant mass $m_j(0)$,
    \item the nominal signal $\Delta g_{\text{Tank},0,i,j}(t)=-\lambda_{t,i}\int_0^t \dot{m}_j(t)dt$,
    \item the constant $c=\lambda_{t,i}\kappa_{t,i}\kappa_{b,j}m_j(0)$ that has no relevance for the discussion.
\end{enumerate}
In Sec.~\ref{sec:lfbin}  we discuss both the nominal case $\kappa_{t,i}=\kappa_{b,j}=1$ and some other options.

\subsection{Drift evaluation procedure}
On the long timescale of days or more we are considering here, all data series have strong autocorrelation. Thus the linear square fitting to data, which we do to estimate the residual drift,  does not allow a consistent and unbiased estimate of errors. To partially circumvent this limitation we have resorted to the following procedure.
\begin{enumerate}[label=(\roman*)]
    \item We  chose the values for $\kappa_{t,i}$, $\kappa_{b,j}$ and $\omega_d^2$. One obvious choice is $\kappa_{t,i}=\kappa_{b,j}=1$ for all values of $i$ and $j$, and  $\omega_d^2= -3.32\times 10^{-7}\text{\,s}^{-2}$ that we call the nominal calibration.
    \item For each run, we  partition all relevant data series into one day-long non-overlapping stretches. Let us call $N_k$ the resulting number of stretches for run $k$
    \item For each of the stretches in one run, we form the ``corrected'' data series $\Delta g_c(t)=\Delta g(t)-\kappa_{t,i} \kappa_{b,j} \Delta g_{\text{Tank},0,i,j}(t)-\omega_d^2\left(\Delta x_\text{OMS}-\Delta x_\text{GRS}\right)$, using the proper values for tank $i$ and branch $j$ used in that same run.
    \item Again for each of the stretches in one run, we perform a least square fitting of $\Delta g_c(t)$ with the model
    \begin{equation}
        \Delta g_c(t)=c_T T+c_t t+\text{c}
        \label{eq:dgdtT}
    \end{equation}
    \item From the  $N_k$ long sample of fitting coefficient pairs $c_{T,j},\;c_{t,j}$, with $1\le j\le N_k$, obtained by fitting the $N_k$ stretches of run $k$, we form the average partial derivatives $\left(\partial \Delta g/\partial T\right)_k\equiv(1/N)\sum_{k=1}^N c_{T,k}$ and $\left(\partial \Delta g/\partial t\right)_k\equiv(1/N)\sum_{k=1}^N c_{t,k}$ and the corresponding variances $\sigma_{T,k}^2$ and $\sigma_{t,k}^2$.
    \item As $N_k$ is quite small for many of the runs, to get a more solid estimate of the mean fluctuation of a single determination of coefficients, we make a weighted average over the variance on the entire set of run,
    \begin{equation*}
    \begin{split}
      &\sigma_T=\sqrt{\frac{\sum_{k=1}^{13}\left(N_k-1\right)\sigma_{T,k}^2}{\sum_{k=1}^{13}\left(N_k-1\right)}}\\
      &\sigma_t=\sqrt{\frac{\sum_{k=1}^{13}\left(N_k-1\right)\sigma_{t,k}^2}{\sum_{k=1}^{13}\left(N_k-1\right)}}
    \end{split}
        \end{equation*}
        \item Finally, we take  $\sigma_T/\sqrt{N_k}$ and $\sigma_t/\sqrt{N_k}$ as the uncertainty on $\left(\partial \Delta g/\partial T\right)_k$ and $\left(\partial \Delta g/\partial t\right)_k$, respectively.
\end{enumerate}
It is worth stressing that such a procedure remains largely empirical, and that coefficients from stretches within the same run may still be correlated so that the errors may be still underestimated. However, the main results of the discussion in Sec.~\ref{sec:lfbin} should depend only weakly on the details of the procedure.

\section{\label{app:BLTD} Brownian noise and long-term drift}

\subsection{\label{app:browniannoise}Brownian noise}
Power-law  evolution of pressure over time is very often  observed in vacuum systems during initial pump down \cite{jousten_handbook_2016}. When the outgassing surfaces consist predominantly of metals, like stainless steel, aluminum, or titanium, a $\propto 1/t$ behavior is very often observed, consistent with models in which water vapor is in quasiequilibrium between thermal outgassing from the metal walls and readsorption onto them. To check if this model would be consistent also with our observed  $\propto 1/t^{0.8}$ behavior, we have attempted to fit our observations to it. To this aim, we have  integrated  the differential equation (see \cite{chiggiato})
\begin{equation}
    \frac{1}{P}\frac{dP}{dt}=-\frac{\frac{1}{\tau}+\frac{P_s}{P}\frac{d\theta}{dT}\frac{dT}{dt}}{1+P_s\frac{d\theta}{dP}}
    \label{eq:dpdt}
\end{equation}
that, within such model, describes the time evolution of pressure. 
Here  
\begin{enumerate}[label=(\roman*)]
    \item $\theta\left(P,T\right)$ is the fraction of occupied adsorption sites,
    \item $P_s=N_s k_B T/V$, with $N_s$ the total number of available adsorption sites, and $V$ the vacuum enclosure volume,
    \item$\tau$ is the vacuum relaxation time of the vacuum enclosure, in the absence of outgassing and adsorption, which is set by $V$ and by the pumping speed of the vent duct and is of the order of 0.2\,s.
\end{enumerate}
Equnation~\eqref{eq:dpdt} can only be solved in combination with a choice for the functional form of  $\theta\left(P,T\right)$. A versatile form that reproduces observations in many cases, is the Temkin isotherm \cite{Redhead},
\begin{equation}
    \theta\left(P,T\right)=\frac{T}{T_2-T_1}\log\left(\frac{1+\frac{P}{P_0 e^{-T_2/T}}}{1+\frac{P}{P_0 e^{-T_1/T}}}\right),
    \label{eq:Temkin}
\end{equation}
with $P_0$, $T_1$, and $T_2$ as free parameters of the model. We have done a least squares fit of the observed values of $S_\text{Brown}/(2\kappa)$ to the prediction of Eq.~\eqref{eq:dpdt}, as a function of $P_s$, $\tau$, $P_0$, $T_1$, and $T_2$. For each choice of the parameters we have integrated Eq.~\eqref{eq:dpdt}, and calculated the $\chi$-square of the deviation of the data from the resulting  solution. To integrate the equation,  we have used the mean measured GRS temperature, mentioned before, throughout the entire integration time interval, except for the short duration epoch of the cold run, where we have used the proxy made with the LTP bay thermometers, also mentioned before.

The best-fit line is the dotted gray line in Fig.~\ref{fig:brown}. The reduced $\chi$-square of the minimum is rather poor $\simeq 2.0$, though this can be, at least in part, due to a significant underestimation of the true uncertainties in the model.

A more serious limitation with such an interpretation lies in the best-fit parameter values.  Let us first mention   that,  provided $P_0\, e^{-T_2/T}\ll P\ll P_0\, e^{-T_1/T}$, the $\chi$-square is found to depend  only on  $P_0$ and on the combination $c=(T_2-T_1)/(\tau P_s)$. Such a  simplification can also be readily derived from an inspection of Eqs.~\eqref{eq:dpdt} and~\eqref{eq:Temkin}. For these two parameters, the fit gives  a broad minimum for $\log P_0$, $\log_{10}\left(P_0/\SI{1}{kPa}\right)=(1.4\pm0.3)$ and a relatively narrow minimum for $c$, $c=(0.328\pm0.004)~\si{\kelvin\,\day^{-1}\,\micro\pascal^{-1}}$. 

The value for $P_0$, $P_0\simeq \SI{25}{\kilo\pascal}\simeq\SI{200}{Torr}$ is orders of magnitude smaller than the quoted, for instance, in \cite{Redhead} of about $6\times 10^8$\,Torr, and within the theory discussed in that same reference, the only way to get such a small number would be to assume an unreasonably low density of adsorbing sites or an equally unreasonably high molecular attempt frequency. 

Furthermore, of the four parameters that enter in the definition of $c$,  $\tau$ can be independently evaluated from the estimated conductance of the vacuum valve and the vent duct, and from the volume of the vacuum enclosure, to be $\tau \simeq \SI{0.3}{s}$. A possible choice for the temperature values is given in \cite{Redhead}, and is $T_1\simeq \SI{5.3}{\kilo\kelvin}$ and $T_2\simeq \SI{11}{\kilo\kelvin}$. The value for $T_1$ is only marginally fulfilling the condition on the pressure range. The closest value that still keeps the $\chi$-square at its minimum is $T_1= \SI{4.5}{\kilo\kelvin}$. With this choice for $\tau$, $T_1$, and $T_2$, we get $P_S=\SI{6.0\pm0.1}{\kilo\pascal}$. In turn, from this value for $P_s$, assuming, as in \cite{Redhead} a density of sites $\simeq 3 \times 10^{15}\,\text{cm}^{-2}$, we get an estimate for the internal area of the vacuum enclosure of $\simeq 180\,\text{m}^2$, a couple of orders of magnitude larger than the physical area. 

Thus to reconcile $P_0$ with the model one would need fewer absorption sites, while for $P_s$ one would need more. We believe that the chance that water readsorption has played a major role in our vacuum system is highly unlikely.

An alternative to the model of water desorption from walls, known to show as well power-law evolution over time, is diffusion-limited outgassing. In the case of hydrogen outgassing from stainless steel, for instance, a $1/\sqrt{t}$ evolution is often observed, in agreement with Fick's law prediction for one-dimensional diffusion \cite{chiggiato}. Approximate outgassing power-law evolution with an exponent even larger than 0.5 is observed in the presence of polymers and is attributed to water diffusion through them. In addition, experiments on polymer samples show that the temperature dependence of the outgassing rate is well described by a single activation exponential, as is the case of our observation \cite{chiggiato}.

Given the abundance of polymers in cable bundles, motors, and connectors within the GRS, and given in general the complex geometry of the TM surroundings with abundant solid interfaces, it looks rather likely that desorption from walls is playing a minor role, and that gas diffusion out of some of the GRS components is dominating the pressure environment.

\subsection{\label{app:discT} Long-term drift, very-low-frequency noise, and pressure gradient}

The results of Sec.~\ref{sec:lfbin} support the idea that there was a  permanent difference of pressure across one or both the TMs, whose amplitude scales with the overall pressure as measured by the Brownian noise level. This pressure gradient makes $\Delta g$ both sensitive to temperature, because of the corresponding temperature sensitivity of pressure, and drifting, because of the pressure drift due to venting to space.

We have simulated with MolFlow+ \cite{Molflow} the effect of localized outgassing on the pressure on the TM. Simulations consistently show that, due to the rather complex geometry of the TM surroundings, which creates a rather asymmetric molecular flow impedance pattern, it takes very little asymmetry of outgassing to support a pressure gradient across the TM.  

To give a scale of the phenomenon, a large fraction of any flow of molecules out of the cavity formed by the outer wall of the EH and the tungsten balance mass (see Fig.~15 of Ref.~\cite{lpf_glitch2022}), penetrates inside the EH, through a hole in the EH $x$-wall, symmetric to that for the laser beam. These molecules diffuse in the gaps surrounding the TM and, in a time of the order of milliseconds, eventually leave the EH. In their flow within the EH, these molecules exert both a mean pressure on the TM, and a pressure difference across the $x$-axis. We find that this difference is approximately 30\% of the overall mean pressure contribution around the TM. 

On the contrary, simulations show that outgassing from sources farther away from the EH, like, for instance, from the inner wall of the VE, diffuses and equalizes outside the EH and does not create any significant pressure difference across the TM.

\section{\label{app:obsart} Possible observational artifacts}

\subsection{\label{app:ifonoise} Role of interferometer noise.}

The contribution of interferometer noise $n_{\OMS}$, with ASD $S_{n_{\OMS}}^{1/2}$,  to $S_{\Delta g}^{1/2}$ is  $S_{\Delta g,n}^{1/2}=S_{n_{\OMS}}^{1/2}\left(4 \pi^2 f^2+\lvert \omega_2^2\rvert\right)$. 

Thus, for $f\le \lvert \omega_2\rvert/2\pi$,  a  branch of $S_{n_{\OMS}}^{1/2}$,  raising rapidly enough upon  decreasing frequency, may have contributed to $S_{\Delta g,e}^{1/2}$. 

In our ordinary noise measurements, there was no way to separate $S_{\Delta g,n}^{1/2}$ from the contribution of true forces. We have, however, two independent methods to put an upper limit to the former.

\begin{figure}[htbp]
  \centering
  \includegraphics[width=1\columnwidth]{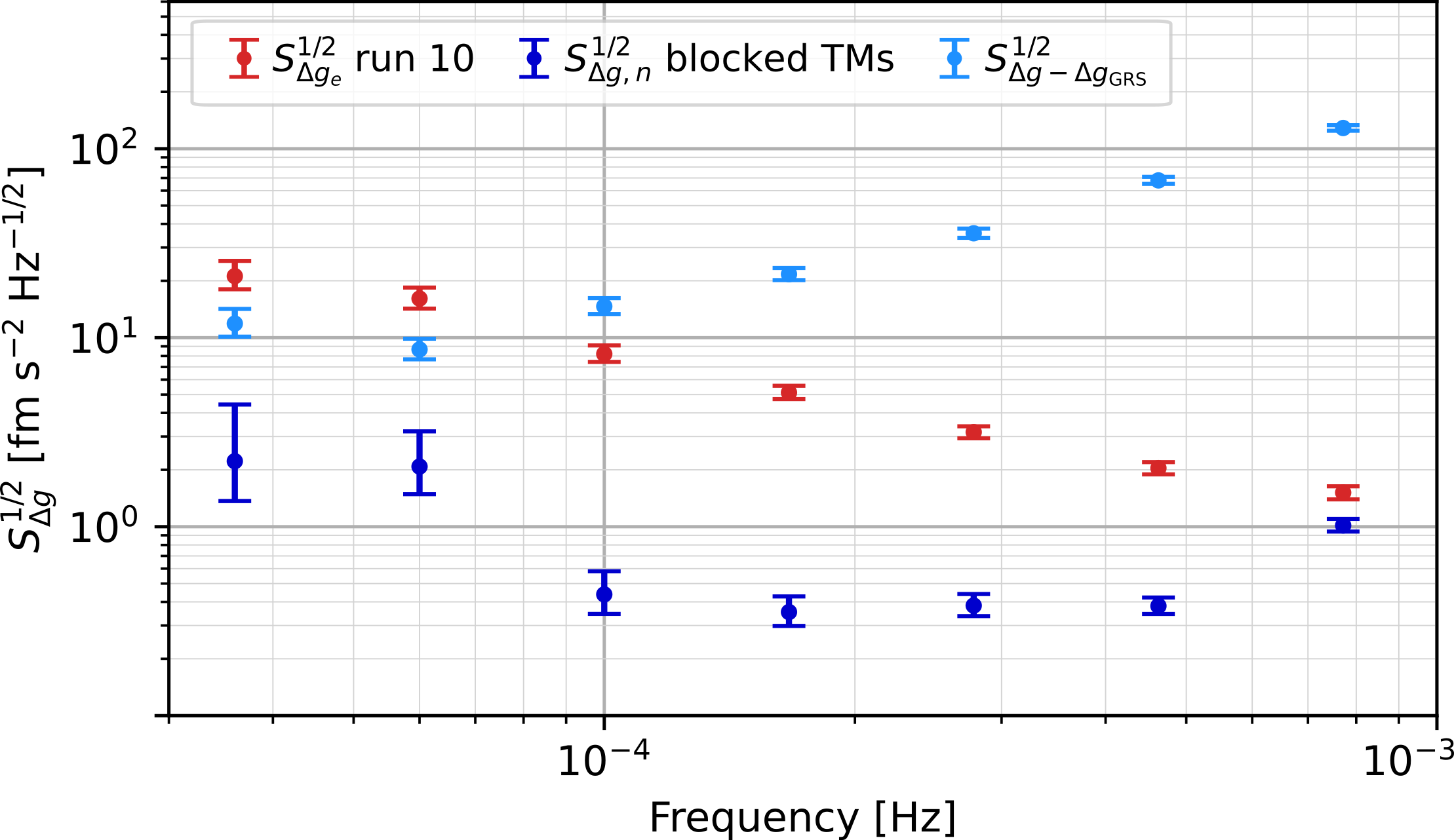}
  \caption{Role of  interferometer noise in the overall ASD of $\Delta g$: red points, $S_{\Delta g_e}^{1/2}$ for run 10 (February 2017), as in Fig.~\ref{fig:brownsubtr}; dark blue points, $S_{\Delta g,n}^{1/2}$ measured with blocked TMs; light blue points: ASD of the difference $\Delta g-\Delta g_{\GRS}$ for run 10.}
  \label{fig:ifonoise}
\end{figure}

The first method is to use the interferometer data we collected during an epoch in which the test masses were held fixed by the blocking mechanism \cite{PhysRevD.106.082001}. This gave us the chance to measure the interferometer output noise in open loop, that is, with no active control of the TM positions, which is the quantity appearing in Eq.~\eqref{eq:deltag2}. 
The results for  $S_{\Delta g,n}^{1/2}$ are reported in Fig.~\ref{fig:ifonoise} and compared to the lowest  $S_{\Delta g_e}^{1/2}$ data, that is those from run~10 of Table~\ref{tab:noise-only-runs}\footnote{The measurement epoch is the same as that used by Ref. \cite{PhysRevD.106.082001}, though we have used a longer, and possibly noisier data series, in order to be able to reach the lowest possible frequency.}.

The second method exploits the independent capacitive measurement of the relative motion of the TMs $\Delta X_{\GRS}$. As the ordinary $\Delta g$ is calculated using $\Delta X_{\OMS}$, we can calculate the analogous $\Delta g_{\GRS}$,  using $\Delta X_{\GRS}$. 
$\Delta g_{\GRS}$ is significantly more noisy than $\Delta g$ at all frequencies above \SI{80}{\micro\hertz}, while the agreement of $\Delta g$ and $\Delta g_{\GRS}$ at lower frequencies, with their independent readouts, indicates that we are observing the true force noise.

Neglecting any distortion, the difference between the two is just $\Delta g-\Delta g_{\GRS}=\lvert \omega_2^2\rvert \left(n_{\OMS}-n_{\GRS}\right)+\ddot{n}_{\OMS}-\ddot{n}_{\GRS}$, and the ASD of this difference is then  $S_{\Delta g-\Delta g_{\GRS}}^{1/2}=\left(\lvert \omega_2^2\rvert+(2 \pi f)^2\right)\sqrt{S_{n_{\OMS}}+S_{n_{\GRS}}}$. This provides then an upper limit for $\left(\lvert \omega_2^2\rvert+(2 \pi f)^2\right) S_{n_{\OMS}}^{1/2}$.  Note that $n_{\GRS}$ is an all encompassing figure and includes not just electronic noise but also the spurious pickup of degrees of freedom other than $\Delta x$.

Figure~\ref{fig:ifonoise} shows the values  $S_{\Delta g-\Delta g_{\GRS}}^{1/2}$ for run~10. It is worth to add that the values for  $(\lvert \omega_2^2\rvert+(2 \pi f)^2 )S_{\Delta X_{\GRS}}^{1/2}\simeq( \lvert \omega_2^2\rvert+(2 \pi f)^2) S_{n_{\GRS}}^{1/2} $, not shown in the figure, coincide almost exactly with those for $S_{\Delta g-\Delta g_{\GRS}}^{1/2}$, which shows that the role of $n_{\OMS}$ in these is negligible.

Both methods give an upper limit to the interferometer contribution, likely rather pessimistic. Indeed within the blocked TM measurement the contrast was poor and the interferometer performance at $f >10\,\text{mHz}$ was at least 1 order of magnitude worse than that with the free TMs. In addition, the interferometer output might still have included some residual relative motion of the TMs, due to any mechanical distortion of the instrument. On the other hand the data $S_{\Delta g-\Delta g_{\GRS}}^{1/2}$ are clearly dominated by the GRS noise, with the noise from the interferometer, and the TM motion, not contributing more than 1\% in power.

In conclusion we consider   the value measured with the TM fixed, as the relevant upper limit on interferometry noise contribution to $\Delta g$, a contribution which is  less than some 1\%--2\% in power, and probably significantly less than this figure.

\subsection{\label{app:glitchsimul}Noise from Poisson flow of glitches}

A very simple model for a stochastic process that shows a  $1/f$ ASD, is a  flow of steps $\Delta g(t)=g_i \Theta(t-t_i)$, with $\Theta(t)$ the Heaviside theta function, $g_i$ a random amplitude, and with the arrival times $t_i$ following Poisson event statistics. Such a process has indeed an ASD given by $S_{\Delta g}^{1/2}=\sqrt{2 \lambda \langle g_i^2\rangle/(2 \pi f)^2}$, with $\lambda$ the event rate. 
\footnote{Strictly speaking the calculation must be done for steps  $\Delta g(t)=g_i e^{-(t-t_i)/\tau} \Theta(t-t_i)$, and the result is valid for $f\ll 1/(2 \pi \tau)$}

The process can be generalized. Specifically, one can calculate the ASD of a flow of Poisson events $\Delta g(t)=g_i h_i(t-t_i) \Theta(t-t_i)$, with $h_i(t)$ the unit peak amplitude version  of a randomly selected glitch among those observed during ordinary LPF noise runs \cite{lpf_glitch2022}.  As we only  observed $N_g=98$ glitches, to generate an arbitrary long time series, in this process  glitch shapes  need to be repeated.

The PSD of such a process would be
\begin{equation}
    S_{\Delta g,gl}(f)^{1/2}=\sqrt{2 \lambda \langle g_i^2\rangle \frac{1}{N_g}\sum_{l=1}^{N_g}\lvert h_l(f) \rvert^2}.
    \label{eq:glitchasd}
\end{equation}

Here $h_l(t)$ is the shape of the $l$ th observed glitch, and $h_l(f)$ is its Fourier transform. For the sake of the current discussion all 98 glitches had $h_l(t)\propto e^{-t/\tau_{2,l}}-e^{-t/\tau_{1,l}}$ with $\tau_{2,l}$ and $\tau_{1,l}$ two time constants\footnote{Note that here we normalize $h_l(t)$  to have unit peak force-per-unit-mass amplitude, at variance with the normalization adopted in Ref.~\cite{lpf_glitch2022}, where $h_l(t)$ had unit impulse  per unit mass.} \cite{lpf_glitch2022}.

\begin{figure}[htbp]
    \centering
    \includegraphics[width=1\columnwidth]{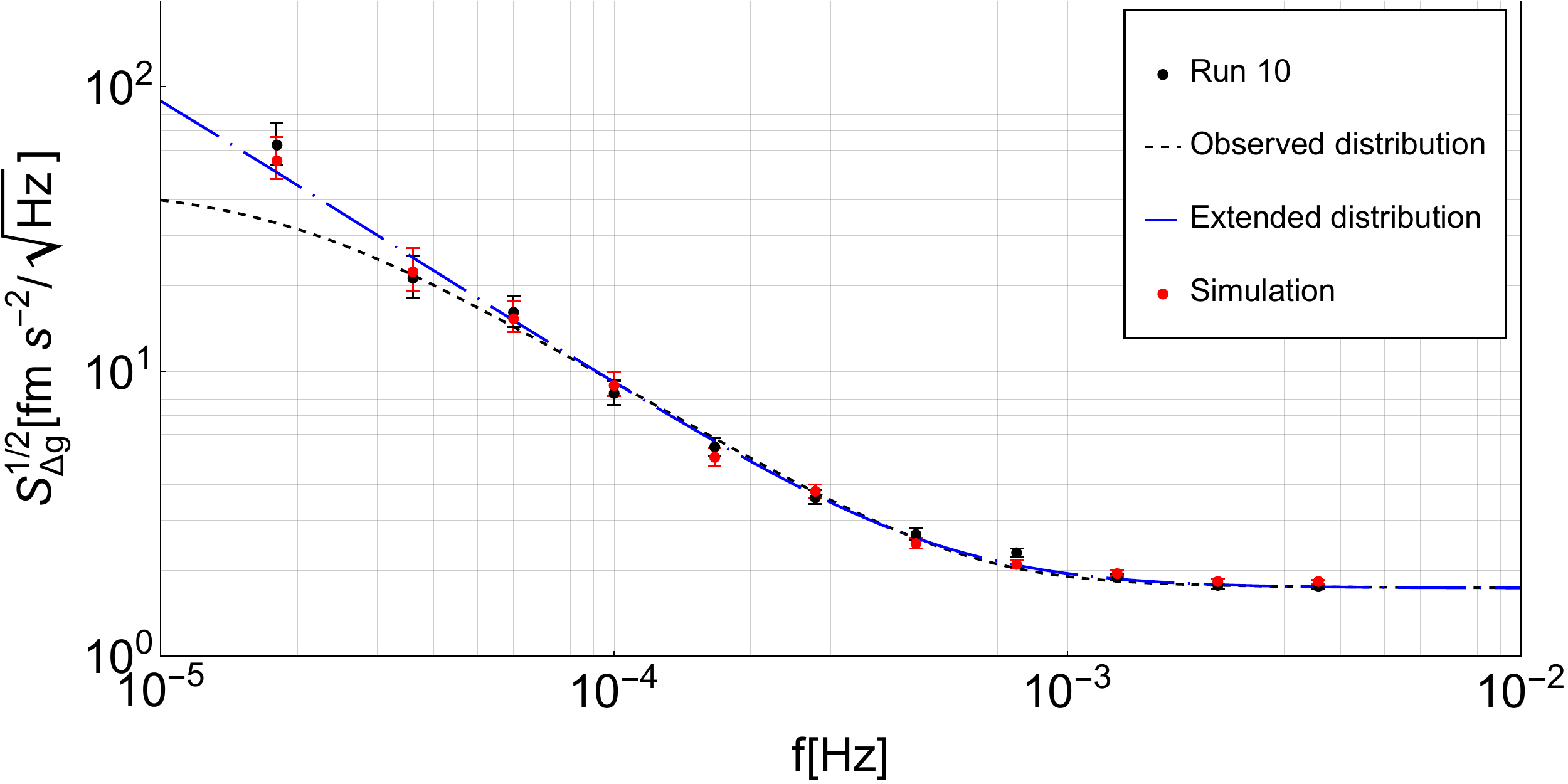}
    \caption{ASD from a Poisson flow of glitches. Dashed black line: calculated ASD of a flow of glitches taken at random from the sample of the observed ones and with $2 \lambda \langle g_i^2\rangle =9.8\times 10^{-5} \text{\,fm}^2\text{s}^{-5}$ (see text for definition of symbols); blue dot-dashed line: calculated ASD of a flow of glitches with time parameters taken at random  from the extended distribution  described in the text and  with $2 \lambda \langle g_i^2\rangle =8.0\times 10^{-5}\,\text{fm}^2\text{s}^{-5}$; red dots: ASD of a simulated time series with the same length as that of run~10,  with glitch parameters extracted from the extended simulation,  $g_i$ extracted from a uniform distribution in the interval $\pm 0.9\,\text{fm\,s}^{-2}$, and with a rate of $\lambda = 105\,\text{d}^{-1}$.}
    \label{fig:GlitchASD}
\end{figure}

Figure~\ref{fig:GlitchASD} compares the ASD calculated from Eq.~\eqref{eq:glitchasd} taking $2 \lambda \langle g_i^2\rangle =9.8\times 10^{-5} \text{\,fm}^2\text{s}^{-5}$ (the dashed black line), to  data from run~10 (the black dots). To ease the comparison, the figure reports $\sqrt{S_{\Delta g,gl}+S_\text{Brown}}$, with $S_\text{Brown}$ the maximum likelihood value for run~10.

It must be said the indistinguishable results, except for the numerical value of $2 \lambda \langle g_i^2\rangle$, are  obtained by taking just the 54 glitches for which $\tau_{2,l},\tau_{1,l}\ge 10\, \text{s}$.

The figure shows that such a simple model may well be fit to  the observed ASDs, at least for frequencies $f\gtrsim \SI{30}{\micro\hertz}$. However  below that limit the predicted ASD saturate, as none of the observed time constants ever exceeded $\simeq 8\,\text{ks}$ \cite{lpf_glitch2022}. To achieve a better result one needs then to  slightly extend the family of templates $h(t)$. 

To do that, we first note that more than 90\% of the  glitches for which $\tau_{2,l},\tau_{1,l}\ge 10\,\text{s}$, had $\tau_{2,l}=\tau_{1,l}\equiv \tau_l$, a case in which  $h_l(t)\to e (t/\tau_l)e^{-t/\tau_l}$. For these glitches the distribution of $\log_{10}(\tau_l/1\,\text{s})$ is quantitatively compatible with a uniform one in the range 1 to 4.

The simplest model extension is then to a flow of glitches of the kind $\Delta g(t)=g_i e (t/\tau_i)e^{-t/\tau_i} \Theta(t-t_i)$ with the distribution of  $\log_{10}(\tau_i/1\,\text{s})$, extended up to 5, instead of being limited to 4.

The results for such an extended model, with a fitting constant $2 \lambda \langle g_i^2\rangle =8.0\times 10^{-5}\,\text{fm}^2\text{s}^{-5}$, are reported again  in Fig.~\ref{fig:GlitchASD} (blue dot-dashed line).

We now discuss the possibility that  the glitches that form this hypothetical random flow, the ``noise glitches'', may be  the same that  occasionally become large enough to be detected as isolated signals in the data series, the ``isolated glitches''. To do that let us first discuss the  possible parameter distribution of the noise glitches. 

The model above would reproduce our observations if the values for $2 \lambda \langle g_i^2\rangle$ would only refer to  glitches with amplitude $g_i$ small enough to be undetectable against the background noise. Indeed the data series on which we have calculated noise ASD have been purged of  any detectable glitch \cite{armano:subfemtog} and thus of their contribution to the ASD. This puts a constraint on the distribution of $g_i$ for the noise glitches that depends on our ability to detect a glitch  against the background noise.

Our empirical glitch detection method has been described in \cite{lpf_glitch2022}, and its detection ability found  in substantial agreement with the prediction of a search based on a matched filter. We have repeated such an analysis for the amplitude normalization we use here, calculating the joint Fisher matrix for $g_i$, $\tau_{1,i}$, $\tau_{2,i}$ and the glitch arrival time, and  confirmed such an agreement. 

Specifically all  detected glitches have signal-to-noise ratio $\SNR\ge3.6$, with $\SNR=\lvert g_i\rvert/\sigma_g$, and $\sigma_g$ the  uncertainty on $g_i$ predicted by the Fisher matrix,  and all but four have a signal-to-noise ratio  $\SNR\ge5$. Indeed visual inspection confirms that our empirical method would detect almost certainly any glitch with $\SNR=5$, and would almost certainly not detect a glitch with $\SNR\le 3.5$. 

For a given amplitude $g_i$,  SNR  depends on $S_\text{Brown}$, but also on both $\tau_{1,i}$ and $\tau_{2,i}$. For run~10, the run with best noise, the SNR reaches  a maximum for $\tau_{1,i}=\tau_{2,i}\simeq \SI{85}{s}$. 

Putting all these elements together, and reminding the reader of the stability of the observed noise in the course of the mission, the distribution of $g_i$ and $\tau_i$ for the noise glitches should fulfill the following conditions:
\begin{enumerate}[label=(\roman*)]
    \item $g_i$ and $\tau_i$  should be independent;
    \item their distributions  should both be  independent of the run;
    \item $\log_{10}(\tau_i/1\,\text{s})$ should be uniformly distributed between 1 and 5;
    \item the distribution of $g_i$ should fulfill: $2 \lambda \langle g_i^2\rangle\simeq  8.0\times 10^{-5} \text{\,fm}^2\text{s}^{-5}$:
    \item the distribution of $g_i$ should assign  a  probability $p_t$ to $g_i+s_i\ge 5 \sigma_g$, with $s_i$ the random amplitude measurement error, such that $\lambda p_t$ is much less than the observed rate of $\simeq 1 \text{\,d}^{-1}$ of the isolated glitches. Here all quantities refer to run~10 and to $\tau_{1,i}=\tau_{2,i}\simeq 85\,\text{s}$ which is the most favorable case for glitch detection.
\end{enumerate}

Thus the question if the isolated glitches are just a sample of large amplitude tail of an overall distribution of glitch parameters translates in that if the amplitude distribution of the noise glitches, fulfilling the constraints above, merges at high amplitude into that of the isolated glitches, without the need of too many pathological assumptions to fill up the gaps among the two. 

In order to start with the minimum such gap, in the condition for $2 \lambda \langle g_i^2\rangle$ one should pick the lowest possible value for $\lambda$  corresponding to the largest for $\langle g_i^2\rangle$. Barring the nonphysical assumption of all equal $g_i$'s, the simple distribution that maximizes $\langle g_i^2\rangle$ for a given upper bound $g_{\text{max}}$, is a uniform distribution $-g_{\text{max}}\le g_i\le g_{\text{max}}$ for which $\langle g_i^2\rangle=(1/3)g_{\text{max}}^2$.

To decide the value of $g_{\text{max}}$ we take the conditions of run~10, with $\tau_{1,i}=\tau_{2,i}\simeq \SI{85}{s}$, recast $g_{\text{max}}=n_{\text{max}} \sigma_g$, and calculate the probability that a sample from a uniform distribution $-n_{\text{max}} \sigma_g\le g_i\le n_{\text{max}} \sigma_g$, plus an independent sample from a zero-mean normal distribution with variance $s_g^2$, representing the measurement error, exceeds in absolute value, the threshold $5 \sigma_g$. We pick $n_{\text{max}}$ when this probability is low enough to detect at most one detectable glitch in a run with the same duration of run~10. The procedure requires a few trial and error loops involving adjusting the value of $\lambda$ that try to minimize.

Figure~\ref{fig:GlitchASD} shows the ASD (red dots) of simulated data assuming $\lambda=102 \text{\,d}^{-1}$, $n_{\text{max}}=3$, $\sigma_g=0.1 \text{\,fm\,s}^{-2}$, and the Brownian noise level and the time duration of run~10. Inspection of the data shows no glitch we would have detected with our empirical method. 

\begin{figure}[ht]
    \centering
    \includegraphics[width=1\columnwidth]{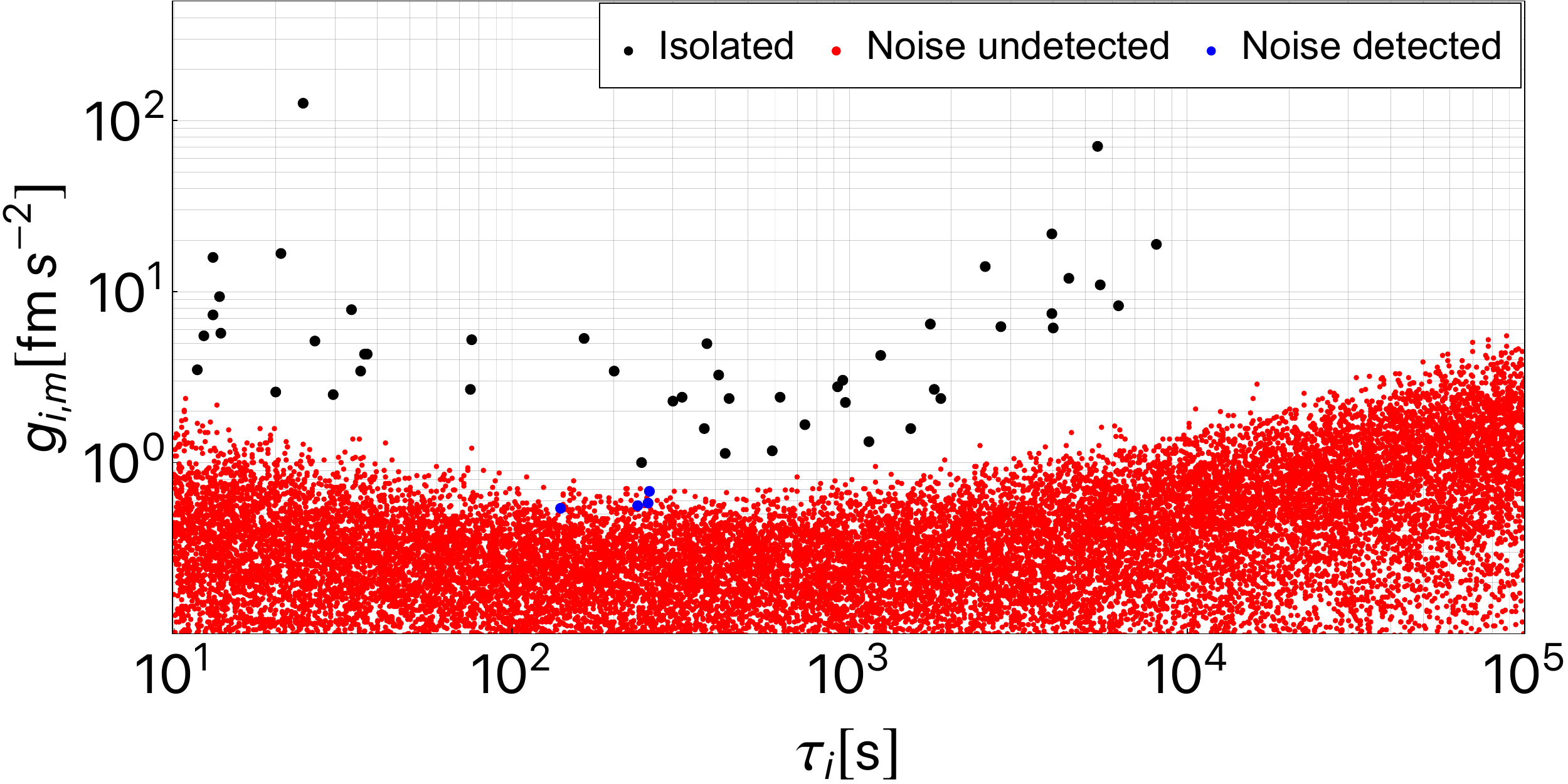}
    \caption{Measured amplitude $g_{i,m}$ vs time constant $\tau_i$ for both simulated noise glitches and observed isolated ones. Black dots, all observed isolated glitches with $\tau_{2,i}=\tau_{1,i}\equiv \tau_i$ and $\tau_i>10 \text{\,s}$. The remaining dots refer to all noise glitches from a simulation aimed at reproducing the noise ASD of all runs in which isolated glitches have been observed \cite{lpf_glitch2022}. Red dots refer to glitches with $\SNR<5$, while blue dots refer to those with $\SNR\ge5$ that would have almost certainly been detected in real data.}
    \label{fig:GlitchAmplitude}
\end{figure}

We have simulated  noise glitches to reproduce the noise for all runs in which we have observed the glitches \cite{lpf_glitch2022}, by generating amplitudes and time constants from the same distributions, independent of the run. For each of the simulated glitches, we have also generated  the   amplitude that would have been measured by the matched filter, by adding a simulated measurement error, and tested if the glitch would have been detected within the specific run to which it belonged. The results of such collective simulation are shown in Fig.~\ref{fig:GlitchAmplitude}.
The parameter space for these simulations is rather vast, and the criterion for detectability is not free of ambiguity. Nevertheless, the picture strongly suggests that
\begin{enumerate}
    \item there is an abrupt, order of magnitude large jump in the probability density between the upper edge of the noise glitches distribution and that of the isolated ones;
    \item there is a large void in the population of isolated glitches for $\tau_i \gtrsim \SI{200}{\second}$, between the amplitude of the observed glitches and that of the noise one. This void is real, glitches with amplitude in the void would be observable with high SNR.
    \item As already noted, glitches with $\tau_i\ge 10^4 \text{\,s}$, necessary to reproduce the noise, are missing in the observed sample.
\end{enumerate}

These features are unlikely to disappear as a result of the tuning of any of the model details. 

Thus it may be that the noise is due to a flow of undetectable glitches, but these appear to belong to a distribution so significantly separated from that of the observed ones, to make the conclusion that we may be witnessing two aspects of the same physical phenomenon, highly speculative.

It is also worth noting that, already at the minimal rate of $\lambda \simeq 100-200 \text{\,d}^{-1}$, the simulated time series appear  Gaussian and stationary within the statistical uncertainty, making their Poisson nature undetectable from the data. Thus again the hypothesis that $\Delta g_e(t)$ is Poisson in origin, remains a speculation.

\newpage
\section{\label{app:decor} Decorrelation of synchronous time series}

In this appendix, we give the details of the decorrelation analyses mentioned in Sec.~\ref{sec:noiseprojection} on the contribution to $\Delta g_e$  of disturbances for which we had a measurement synchronous with $\Delta g$.

\subsection{Decorrelation of  synchronous time series: Framework}
\label{app:decorrMCMC}
During noise runs we have measured, synchronously with $\Delta g$:
\begin{enumerate}[label=\textit{\alph*}.]
    \item the gravitational force loss due to fuel depletion, $\Delta g_\text{Tanks}$ defined in Sec.~\ref{sec:lfbin};
    \item the relative motion of the two GRSs, instrument distortion $\Delta X$, as defined in Appendix~\ref{app:driftcalc};
    \item temperature $T$, defined in Sec.~\ref{sec:data runs};
    \item the two temperature differences across the two electrode housings, in the $x$-direction;
    \item the three magnetic field components at four different locations;
    \item a series of spurious low-frequency voltages that have unintentionally been applied to the electrodes via the actuation circuitry nonlinearity \cite{ActNeda}.
\end{enumerate}

We treat these disturbances as small and Gaussian, and consider their effect only to first order, within the simple model
\begin{equation}
    \label{eq:decorrmodel}
    \begin{cases}
        \Delta g (t) &= \Delta g_0(t) + \sum_{i=1}^r  \int_0^\infty \alpha_i(t') y_i(t-t')\mathrm{d}t' \\
        z_i(t)&= y_i(t) + n_i(t)
    \end{cases}
\end{equation}
where $z_i(t)$ is any of the measured time series  above, 
consisting of the ``true'' physical disturbance $y_i(t)$, superimposed to its readout noise $n_i(t)$. 
$\Delta g_0$ is the ``residual'' acceleration, not correlating to $y_i$.  

Given the nature of the disturbances considered, the  \textit{susceptibilities} $\alpha_i(f)$, i.e. the Fourier transforms of the  $\alpha_i(t)$, should be, for most of them,  real and constant, but we also consider the general case to take into account the possibilities of delays and other more complex correlations.

With the model in Eq.~\eqref{eq:decorrmodel}, the elements of the cross-spectral density (CSPD) matrix  $\bSig$ of $\Delta g$  and  the $z_i$ are given by
\begin{equation}
\label{eq:sdgtot}
\begin{split}
    &\Sigma_{1,1}=S_{\Delta g}(f)=S_{\Delta g_0}(f)+\sum_{i,j=1}^r\alpha_i(f) \alpha_j^*(f) S_{y_i,y_j}(f)\\
    &\Sigma_{i+1,1}=S_{y_i,\Delta g}(f)=\sum_{j=1}^r\alpha_j^*(f) S_{y_i,y_j}(f)\\
    &\Sigma_{i+1,j+1}=S_{z_i,z_j}(f)=S_{y_i,y_j}(f)+\delta_{i,j}S_{n_i}(f)
\end{split}
\end{equation}
In Eq.~\eqref{eq:sdgtot}, $S_{y_i,y_j}(f)$ is the cross-spectral density between $y_i$ and $y_j$, while $S_{n_i}$ is the PSD of $n_i$. In all cases, except for instrument distortion and fuel depletion which we discuss separately, the readout noise $n_i$ is assumed to be independent of any of the other time series.

We deal with two broad cases. In the first case, we have limited knowledge of $\alpha_i(f)$, but we are confident that the readout noise $S_{n_i}(f)$ is negligible for the purpose of noise analysis. In the second case, we have an independent knowledge of the value of $\alpha_i(f)$, but we know or suspect that the readout noise $S_{n_i}(f)$ is significant, even dominating $S_{z_i,z_j}(f)$. These two cases are separately treated in the following. The general case of unknown susceptibilities and unknown readout noise is overdetermined and cannot be treated.

\subsection{Decorrelation: Data analysis and statistical methods}

\subsubsection{Decorrelation of disturbances with unknown susceptibilities and negligible readout noise} 
\label{app:CPSD/decorrelation}

In the case of unknown susceptibilities and negligible readout noise, we evaluate $S_{\Delta g_0}(f)$ and the susceptibilities  assuming the physically realistic  model of real and frequency-independent susceptibilities.

The starting point is that one can transform the complex-Wishart distribution in Eq.~\eqref{eq:wishartSte} to separate its dependence on  $S_{\Delta g_0}$, and the $\alpha_i$, from that on the CPSD of the disturbances $S_{y_i,y_j}$. 

To do that consider that Eq.~\eqref{eq:decorrmodel} defines a linear transformation $\Delta g_0\to \Delta g ,y_i\to y_i$ whose matrix is
\begin{equation}
\label{eq:umatrix}
    \bU=\left(\begin{matrix}
      1&\alpha_1&\dots&\alpha_r\\
      0&1&\dots&0\\
      \vdots&&\ddots&\vdots\\
      0&\dots&\dots&1
    \end{matrix}\right)
\end{equation}
Note that $\lvert\bU\rvert=1$ and that the inverse of $\bU$ is obtained with the transformation $\alpha_i\to-\alpha_i$.

As $\Delta g_0$ is independent of the $y_i$, before the transformation, the  processes have a CPSD matrix with  block representation
\begin{equation}
\label{eq:sigmaprime}
    \bSig'=\left(\begin{matrix}
        S_{\Delta g_0} &0  \\
        0 & \boldsymbol{\Sigma}_{yy}
    \end{matrix}\right)
\end{equation}
where $\boldsymbol{\Sigma}_{yy}$ is the $r\times r$ CPSD of the $y_i$.
Note that 
\begin{equation}
\label{eq:invsigmaprime}
    \bSig'^{-1}=\left(\begin{matrix}
       1/ S_{\Delta g_0} &0  \\
        0 & \boldsymbol{\Sigma}_{yy}^{-1}
    \end{matrix}\right)
\end{equation}
 and that $\bSig=\bU \bSig'\bU^T$, with $\bU^T$ the transpose of $\bU$, so that $\lvert \bSig\rvert=\lvert\bSig'\rvert=S_{\Delta g_0}\lvert\boldsymbol{\Sigma}_{yy}\rvert$.

Furthermore, in Eq.~\eqref{eq:wishartSte} one can calculate that $\etr\left[-\bSig^{-1}\bW\right]=\etr\left[-(\bU^T)^{-1}\bSig'^{-1}\bU^{-1}\bW\right]=\etr\left[-\bSig'^{-1}{\bW}' \right]$, having defined ${\bW}'=\bU^{-1}\bW(\bU^T)^{-1}$. 

We need one more step to get our separation. $\bW$ and ${\bW}'$ have  block representation 
\begin{equation}
\label{eq:wtilde}
    \bW=\left(\begin{matrix}
        W_{1,1} &\bm{W}_{1,y}  \\
        \bm{W}_{1,y}^\dagger& \bW_{yy}
    \end{matrix}\right)\qquad
    {\bW}'=\left(\begin{matrix}
        {W}'_{1,1} &{\bm{W}'}_{1,y}  \\
        {\bm{W}'}_{1,y}^\dagger & \bW_{yy}
    \end{matrix}\right)
\end{equation}
where one can calculate that 
\begin{equation}
\label{eq:w11}
\begin{split}
&{W}'_{1,1}=W_{1,1}-2\bm{\alpha}\cdot \text{Re}\bm{W}_{1,y}+\bm{\alpha}\cdot\text{Re}\bW_{yy}\cdot\bm{\alpha},
\end{split}
\end{equation}
with $\alpha$ the $r$-vector the components of which are the susceptibilities.

The consequence of Eqs.~\eqref{eq:invsigmaprime}, \eqref{eq:wtilde}, and \eqref{eq:w11} is that 
\begin{equation}
\begin{split}
    &\etr\left[-\bSig^{-1}\bW\right]=\etr\left[-\bSig'^{-1}\tilde{\bW}\right]=\\ 
    &=e^{-\frac{{W}'_{1,1}}{S_{\Delta g_0}}-\text{tr}\left[\boldsymbol{\Sigma}_{yy}^{-1}\bW_{yy}\right]}=\\
    &=e^{-\frac{W_{1,1}-2\bm{\alpha}\cdot \text{Re}\bm{W}_{1,y}+\bm{\alpha}\cdot\text{Re}\bW_{yy}\cdot\bm{\alpha}}{S_{\Delta g_0}}-\text{tr}\left[\boldsymbol{\Sigma}_{yy}^{-1}\bW_{yy}\right]}.
    \end{split}
\end{equation}

Putting all together, the distribution of $\bW$, conditional on the susceptibilities, the PSD of the residual noise, and the CPSD of the disturbances, becomes then:

\begin{equation}
\label{eq:wishartT}
\begin{aligned}
   &p\left(\bW\big\vert\alpha_i,S_{\Delta g_0},\bSig_{yy} \right) = \frac{\left|\bW \right|^{M-p}}{\widetilde{\Gamma}_p(M)}\times\\
   &\times\frac{1}{S_{\Delta g_0}^M } e^{-\frac{W_{1,1}-2\bm{\alpha}\cdot \text{Re}\bm{W}_{1,y}+\bm{\alpha}\cdot\text{Re}\bW_{yy}\cdot\bm{\alpha}}{S_{\Delta g_0}}}\times\\
   &\times\frac{1}{ \left| \bSig_{yy} \right|^{M}} \etr\left[-\boldsymbol{\Sigma}_{yy}^{-1}\bW_{yy}\right]
   \end{aligned}
\end{equation}

The distribution in Eq.~\eqref{eq:wishartT} can be used to build  a joint posterior for $S_{\Delta g_0}^{-1}$ and the $\alpha$'s, and an independent posterior for $\bSig_{yy}$.

Treating the data at the different frequencies as independent, and assuming that $S_{\Delta g_0}$ and $\bSig_{yy}$ depend on the frequency, while the $\alpha$'s do not, the joint likelihood for the Bayesian inference of $S_{\Delta g_0}$ and  $\alpha_i$  becomes:

\begin{equation}
\label{eq:decorrLLL}
\begin{split}
    &p(\alpha_i,S_{\Delta g_0}(f_1),S_{\Delta g_0}(f_2),\dots\vert \bW)\propto\\
    &\propto \prod_k  \frac{\widetilde{p}(S_{\Delta g_0}(f_k))}{S_{\Delta g_0}(f_k)^{M(f_k)}} e^{-\sum_k \frac{W_{1,1}(f_k)}{S_{\Delta g_0}(f_k)}}\times\\
    &\times e^{2\bm{\alpha}\cdot\sum_k\frac{\text{Re}\bm{W}_{1,y}(f_k)}{S_{\Delta g_0}(f_k)}-\bm{\alpha}\cdot \left(\sum_k\frac{\text{Re}\bW_{yy}(f_k)}{S_{\Delta g_0}(f_k)}\right)\cdot \bm{\alpha}},
    \end{split}
\end{equation}
having taken a flat prior  for the $\alpha$'s, and the prior $\widetilde{p}(S_{\Delta g_0}(f_k))$ for $S_{\Delta g_0}(f_k)$. As for this one, to take into account the presence of the Brownian noise, we split it as $S_{\Delta g_0}(f_k)\to S_{\Delta g_0}(f_k)+S_{\text{Brown}}$, and we take a uniform prior for  the logarithms of  $S_{\Delta g_0}(f_k)$. For  $S_{\text{Brown}}$ we use as prior the posterior obtained for its Bayesian estimate (see Sec.~\ref{sec:accPSD}). At frequencies low enough that the Brownian noise is negligible, which is where the excess is best measured, this is fully equivalent to taking the Jeffreys prior for $S_{\Delta g_0}(f_k)^{-1}$, which is indeed  a uniform prior on its logarithm.

To estimate the posterior of the parameters, we employ a parallel-tempering Monte Carlo Markov chain algorithm (\textsc{PT-MCMC} \cite{justin_ellis_2017_1037579}). We always find smooth distributions, no bimodalities and no strong cross-correlation among parameters. To implement the prior, the MCMC algorithm explores the parameter space of the  susceptibilities, and of the logarithms of  $S_{\Delta g_0}(f_k)$ and of $S_{\text{Brown}}$. 

As for $\bSig_{yy}$, Eq.~\eqref{eq:wishartT} shows that, using the intermediate prior, $\bSig_{yy}^{-1}(f_i)\sim \cw(\bW_{yy}^{-1}(f_i),M(f_i)+r-1)$. Thus the evaluation of the statistical properties of the posterior can be obtained numerically  from the relative  Wishart distribution.

Once the posteriors for $S_{\Delta g_0}(f_k)$, $S_{\text{Brown}}$, $\alpha_i$, and possibly for $\bSig_{yy}$, have been obtained, one can calculate the  posterior for $S_c=\sum_{i,j} \alpha_i\alpha_j S_{y_i,y_j}=\sum_{i,j} \alpha_i\alpha_j \left(\Sigma_{yy}\right)_{i,j}$. However, for  small values of $S_c/S_{\Delta g_0}$, the statistics of such a posterior is  biased toward high values, with the bias increasing with $r$. This is due to the positive-definite quadratic nature of $S_c$, which gets rapidly dominated by susceptibility fluctuations.

To deal with this problem we have resorted to  calibrating the bias via simulations. Our simulation consists of the following steps:
\begin{enumerate}[label=(\roman*)]
    \item We extract a sample of $\bSig_{yy}(f)$ from its calculated posterior, and form a corresponding sample for $\bSig(f)$ assuming  $S_{\Delta g_0}=0$.
    \item We generate a random vector of susceptibilities $\alpha_i$ from a normal distribution with zero mean, and standard deviation adjusted such that, at some preferred frequency,  $S_c\simeq \rho S_{\Delta g_e}$, with $\rho$ a desired value of the noise power fraction.
    \item From the susceptibility vector we form the matrix $\bU$ of Eq.~\eqref{eq:umatrix}, and with it we form the  sample of the (provisional) matrix $\bSig(f)=\bU\cdot \bSig'(f)\cdot\bU^T$. This matrix has $\Sigma_{1,1}=S_{\Delta g}(f)=S_c(f)$, a number we store as we need for our calibration procedure.
    \item We then form the final value of the matrix $\bSig$ by substituting the value of $\Sigma_{1,1}$ above, with a random sample extracted from the posterior distribution of $S_{\Delta g}(f)$ [see Eq.~\eqref{eq:postS}]. This our theoretical CPSD matrix, which has a value of $S_c(f)$, coherent with  its elements $\Sigma_{1,j}(f)$, a value for $S_{\Delta g}=S_{\Delta g_0}(f)+S_c$ coherent with its  posterior, and a value of $\bSig_{yy}(f)$ also coherent with its posterior.
    \item From the matrix $\bSig(f)$ above we generate a random  sample of the simulated $\bW$ matrix, using the Wishart distribution in Eq.~\eqref{eq:wishartSte}. On each of these simulated samples of $\bW$ we perform our entire Bayesian procedure thus getting, among others, a posterior for $S_c(f)$
\end{enumerate}

\begin{figure}[htbp]
  \centering
  \includegraphics[width=1\columnwidth]{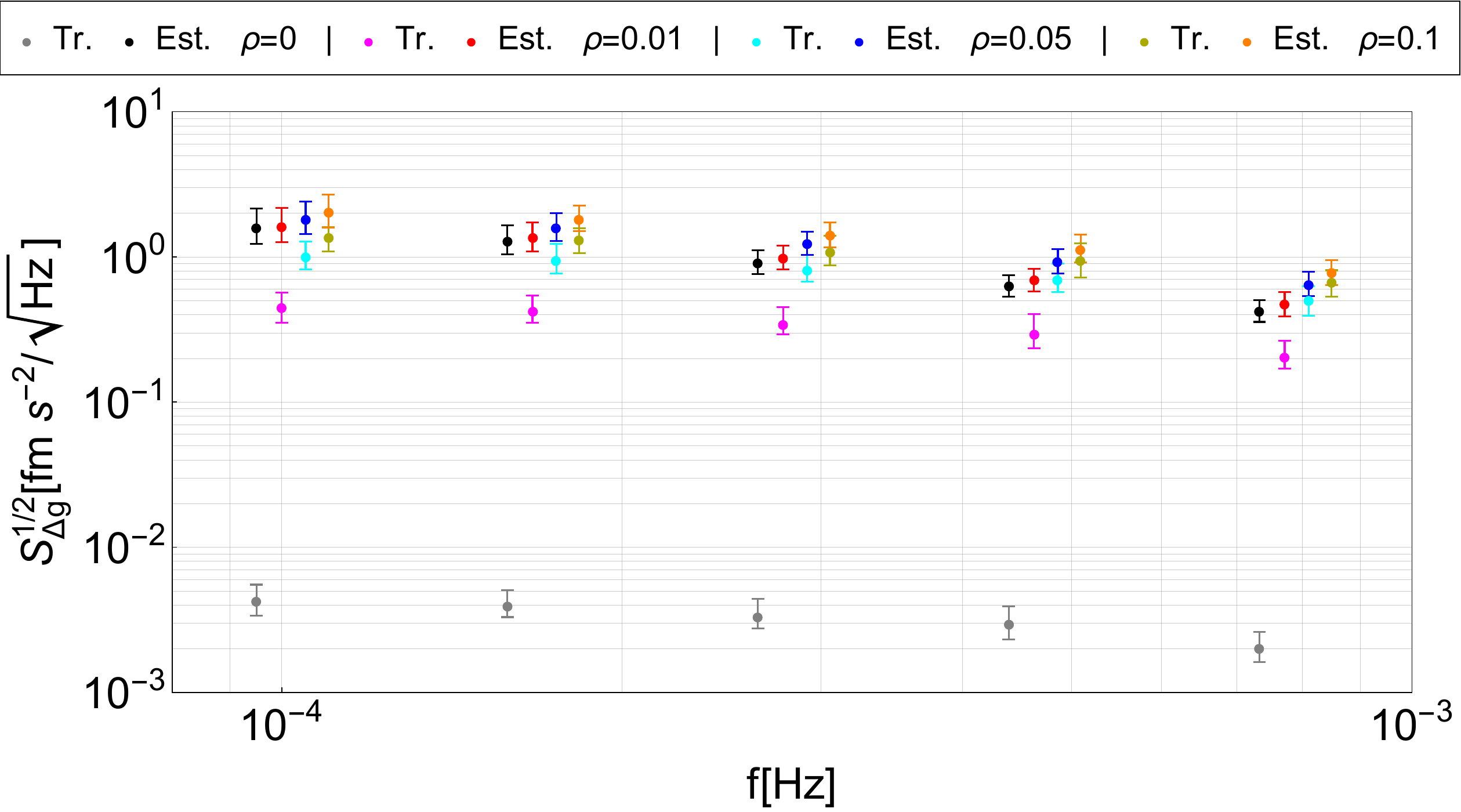}
  \caption{\label{fig:estcal}$\pm 1\sigma$ intervals for the distributions of the ``true'' (Tr.) values of $S_c(f)$ and for their global estimated posteriors (Est.) over 100 simulations for each case. The power fraction $\rho$ is an approximation of the average at the three highest frequencies. In particular $\rho=0$ is in reality calculated at $\rho\simeq 4\times10^{-6}$. Data are plotted for clarity at slightly shifted frequency coordinates, but have all been calculated at one of the frequencies $f_i$ with $4\le i \le 8$.}
\end{figure}

We repeat the procedure 100 times for different values of the power fraction $\rho$ and we get then, at each value of $\rho$ and at each of the considered frequencies, a distribution of the ``true'' values of $S_{c,t}(f)$ and a global distribution of the estimated values $S_{c,e}(f)$ for the same quantity (see Fig.~\ref{fig:estcal}). We use mean and standard deviations of these distributions, to do a linear, weighted least square fit $S_{c,t}(f)=A S_{e,t}(f)+B$, that we use to correct the observed data. The procedure gives a result close to what one would obtain by just subtracting, from the estimated value of $S_c(f)$, the median   from the simulation for $\rho=0$. Indeed in some cases, to avoid repeating time-consuming simulations at different values of $\rho$, we adopted this second method.

With simulations, we have also checked that the method is unbiased for the estimate of both $S_{\Delta g_0}$ and the susceptibilities.

As a final note, the case of the evaluation of the effect of the temperature on the first bin involves just one frequency and one disturbance. In such case the likelihood in Eq.~\eqref{eq:decorrLLL} can be integrated analytically, to give a marginal probability density for $S_{\Delta g,0}\sim \text{inv}\Gamma\left(M-1,M /(\bPi^{-1})_{1,1}\right)$, for the general case of a complex $\alpha$, and $S_{\Delta g,0}\sim \text{inv}\Gamma\left(M-1/2,M /(\text{Re}\bPi^{-1})_{1,1}\right)$  if $\alpha$ is real. We have used these formulas to evaluate the effect of the temperature.

\subsubsection{\label{app:CPSDnoisy} Decorrelation of noisy series with known susceptibilities}

This is the case when we have a relatively narrow posterior distribution $f(\alpha_i)$ for the susceptibility $\alpha_i$, but $S_{n_i}$ may be large, actually may even dominate $S_{z_i,z_i}$ in Eq.~\eqref{eq:sdgtot}, and refers to the joint analysis of fluctuations of the average temperature and of LTP distortion in Appendix~\ref{app:XT}, and to that of the tank depletion gravitational signal in Appendix~\ref{app:G}.

Though the two cases differ substantially, they both share the step of estimating the Bayesian posterior for $\bSig$, the theoretical CPSD.


To that purpose, our starting point is again the distribution in Eq.~\eqref{eq:wishartSte} that can be directly used to estimate the Bayesian posterior for $\bSig$, or its inverse $\bQ=\bSig^{-1}$, from the observation of $\bW$, once a prior distribution for either $\bSig$ or $\bQ$ has been assumed.

Natural, noninformative choices are either a uniform prior on all elements of $\bQ$, or the Jeffreys prior \cite{Jeffreys}, which,  for a $p\times p$-dimensional $\bQ$ matrix, is $\propto \lvert \bQ\rvert^{-p}$ \cite{lundberg,villegas}. For the uniform prior, the posterior is $\bQ\sim\cw(\bW^{-1},M+p)$, while  for the Jeffreys prior  $\bQ\sim\cw(\bW^{-1},M)$ 

Both choices have limitations. Of those of the uniform prior when $p=1$ we have discussed in Appendix~\ref{app:PSDestimate}. On the other hand, the Jeffreys prior is affected by  bias and by a significant  inconsistency when $p>1$. 

The bias consists of the fact that while the mean value of $\bQ$ is unbiased, the mean value of $\bSig$ is equal to $\bPi M/(M-p)$. Thus the bias depends on $p$ and may become large at low values of $M$. 

The inconsistency  stays in the fact that, when $p>1$, the marginal posterior of any of the diagonal elements of $\bSig$, is not the same as the posterior in Eq.~\eqref{eq:postS}, which  one would calculate from the Jeffreys prior with  $p=1$.

The bias can be reduced and made independent of $p$, and the inconsistency solved, if one takes the ``intermediate'' prior $\propto \lvert \bQ\rvert^{-1}$, that still coincides with  the logarithmic prior for the case $p=1$.  

Note that,  also with this choice, $\bQ$ is Wishart distributed,  $\bQ\sim\cw(\bW^{-1},M+p-1)$, which guarantees, without imposing any further prior constraint, that $\bQ$ and its inverse are positive-definite matrices, a fundamental constraint for CPSD. We use this posterior to estimate both the entire matrix, in the case of temperature/LTP distortion, or its   $S_{z_i,\Delta g}(f)$ elements, in the case of the tank depletion noise.

As we have a closed formula, we generate random samples both from such posterior for $\bQ$ and from the posterior for the $\alpha_i$, and then calculate the samples of the quantities needed in the two cases, as discussed in their  relative sections.

\subsection{Decorrelation of synchronous time series: Results for series with negligible readout noise}
\label{sec:decnonoise}

For the time series we analyze here,  we have limited prior knowledge about the susceptibilities, except that they are real and frequency independent. This is because delays between the time series and the forces they exert on the TMs are negligible. 

We then do a simultaneous Bayesian fit at all frequencies as explained in Appendix~\ref{app:CPSD/decorrelation}. This way we get a posterior for the residual $S_{\Delta g_0}(f)$, for the susceptibilities $\alpha_i$, and for the joint contribution  to $S_{\Delta g}$, $S_c(f)=\sum_{i,j=1}^r\alpha_i \alpha_j S_{y_i,y_j}(f) $ [see Eq.~\eqref{eq:sdgtot}]. 

$S_c(f)$ may be subject to a large positive bias in case of large $r$. For that reason we calibrate the method with extensive simulations, as explained again in Appendix~\ref{app:CPSD/decorrelation}.

\subsubsection{\label{sec:spact}Spurious actuation due to in-band noise from digitized electrostatic actuation}
   
   The capacitive actuation design employed digitally synthesized audio-frequency sinusoidal voltages \cite{ActNeda}. As the force is quadratic in the voltage, this applies a force at low frequency, proportional to the mean square value of the sinusoid, while avoiding mixing down low-frequency voltage noise from the final amplifiers. 
    
    The truncation error mentioned earlier affects the above mean square value and, if uncorrected, gives a wrong estimate of the applied force.
    
    The nominal sinusoidal voltage signal has zero time average and then zero {dc} value. The truncated signal acquires a nonzero average, that is, quasi-{dc} voltages $\delta V_i(t)$ of order $\sim\si{\micro\volt}$ amplitude that we were able to calculate  from the   commanded voltages with negligible numerical noise.

    These voltages are able to apply  electrostatic forces onto the test masses by coupling to  its charge and to the parasitic voltages biasing the electrodes \cite{prl108.5.2012}.  To linear terms in the $\delta V_i(t)$'s this force is given by:

    \begin{equation}
    \label{eq:Luigi}
    \Delta g_D(t)=\sum_{i\in[1,24]} \alpha_i \delta V_i(t) 
    \end{equation}
    where the sum is over the 24 electrodes that surround the two TMs, and where the susceptibility $\alpha_i$ is a linear combination of charge and patch potentials and cannot be predicted except, possibly,  in order of magnitude.

In reality, due to cosmic rays, the TM charge is subject to an approximately linear variation in time, $q(t)\sim q_0 + \dot{q}t$ \cite{PhysRevD.107.062007_Charging2023}, with $\dot{q}\sim+23~\si{e/s}$. 
On the charge itself the relative effect may be large, as we have often operated by putting a proper negative charge on both TMs at the beginning of a noise run, to find an equal but opposite charge on them at the end of the run.  

Hence, a complete analysis should, in principle, include 48 charge parameters. To circumvent this problem, we have taken advantage of the fact that the differential force contribution due to the coupling of the $\delta V_i(t)$ to the drifting charge is $\left(\dot{q}_2 t \Delta_{x,2}-\dot{q}_1 t\Delta_{x,1}\right)$. Here $\Delta_{x,i}$ is linear combination, with $\pm1$ coefficients,  of the $\delta V_i(t)$ of the $x$-electrodes of GRS$i$, as defined in \cite{prl108.5.2012}, and $\dot{q}_i$ is the  charging rate of TM$i$ \cite{prl108.5.2012}.

We have then formed the two time series $t \Delta_{x,1}(t)$ and $t\Delta_{x,2}(t)$ and added them to our analysis. This obviously breaks the hypothesis of noise stationarity, but should still help highlight the existence of significant correlation.

To keep $r$ not too large, we have done two separate analyses. In the first we have  included the  24 voltage series only. Note that with $r=24$ the analysis could only be performed for $f\ge f_4$ (\SI{0.1}{\milli\hertz}) as lower frequencies would not have a sufficient number of periodograms (see Appendix~\ref{app:CPSD/decorrelation}). 

\begin{figure}[htbp]
  \centering
  \includegraphics[width=1\columnwidth]{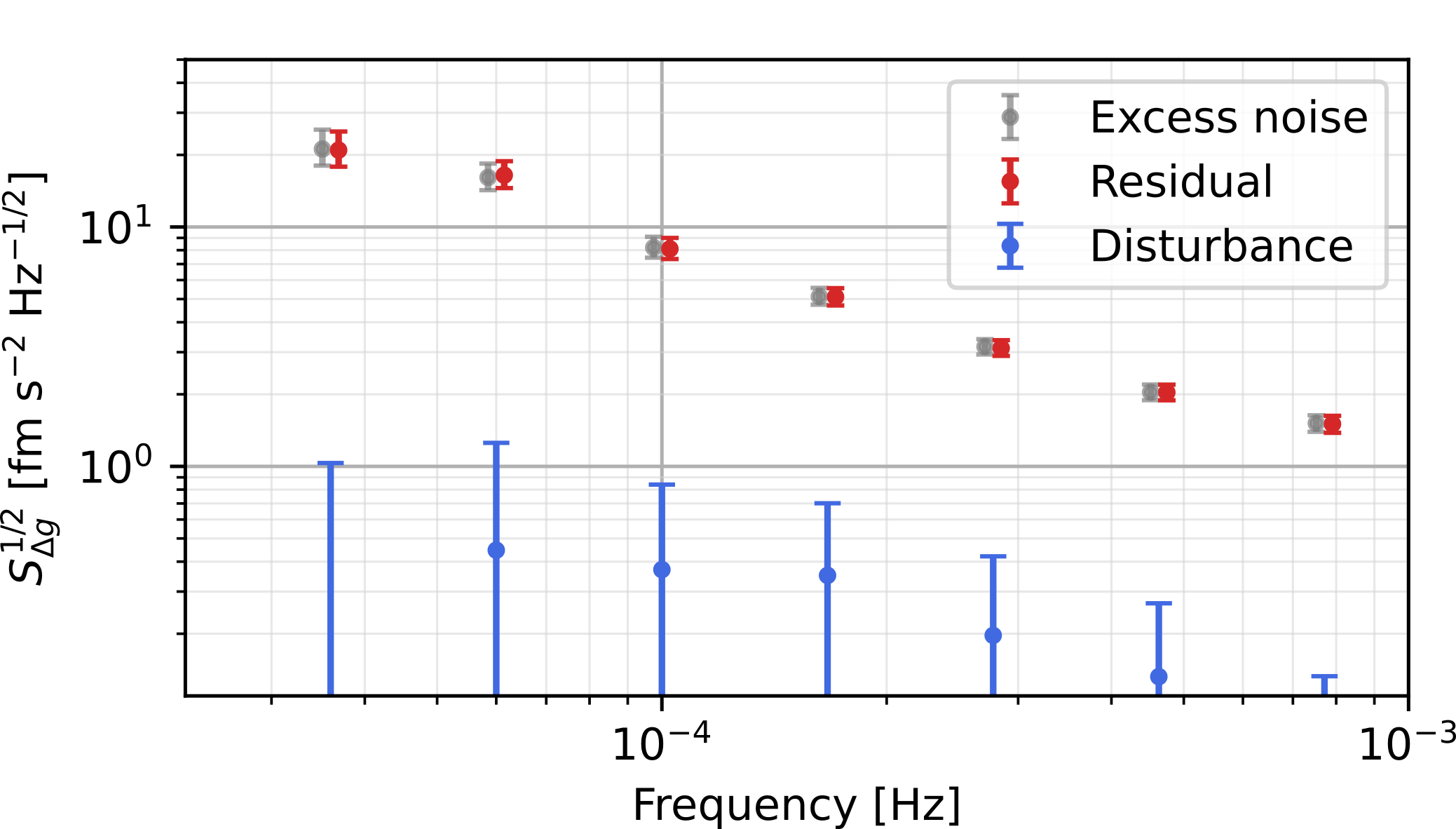}
  \caption{\label{fig:decorrelation q} Decorrelation of the coupling of spurious actuation to TM charging for run~10, over  the [\SI{0.036}{mHz},~\SI{0.77}{mHz}] frequency band. Frequencies are slightly shifted for clarity.
  Gray points, ASD of total excess noise over Brownian, $S_{\Delta g_e}^{1/2}$, as in Fig.~\ref{fig:brownsubtr}; red points, Bayesian posterior $\pm 1\sigma$ interval for the  residual ASD after decorrelation of the spurious actuation along the {$x$-axis}; blue points, estimated ASD $S_c^{1/2}$ of the contribution of spurious actuation along the $x$-axis.}
\end{figure}

In the second we have considered the two series $t \Delta_{x,1}(t)$ and $t\Delta_{x,2}(t)$, but we have also added the two series $ \Delta_{x,1}(t)$ and $\Delta_{x,2}(t)$, to consider the complete effect of the charge, and to reduce the bias due to direct correlation between $\Delta g$ and the $\delta V_i(t)$ involved in the $\Delta_{x,i}$. This allows us to include just four signal time series, hence being able to analyze a wider set of frequencies. We show in Fig.~\ref{fig:decorrelation q} the results of the second analysis.

The results of Fig.~\ref{fig:decorrelation q} clearly show that this effect is compatible with zero within the resolution of the measurement. The  $1 \sigma$ error, on the other hand, is  compatible with a contribution slightly less than $\simeq 1\%$  of total power at $f\ge f_5$, and  much less below that. A similar  result is obtained from the  analysis of the 24 voltage series.


The posteriors of the susceptibilities are all compatible with zero, except, perhaps, for one. More specifically we have analyzed, for all the 24 $\alpha_i$, the likelihood, that we call  $\mathcal{L}_0$, assigned by the posterior to the less likely of the two tails $\alpha_i <0$ and $\alpha_i >0$, as a very low likelihood of one of the two tails would indicate that $\alpha_i$ has a well-defined sign. All likelihoods are found to be larger than 5\%,  except for one, the voltage applied to one of the $x$-electrodes of TM1, which is $\mathcal{L}_0\simeq 1\%$. For this last series the susceptibility is $\alpha =(10\pm5) \text{ fm s}^{-2}/\text{mV}$. To give an order of magnitude, if the coupling was due to a uniform stray voltage on the said electrode, this should be $(0.07\pm0.03)\,\text{V}$, a figure in the range of observed patch potentials  \cite{prl108.5.2012}. 

This slightly significant susceptibility is reflected, in the analysis with a linearly drifting charge, in an equivalently significant susceptibility to $\Delta_{x,1}(t)$, that contains the series above, while the susceptibilities to $t\Delta_{x,1}(t)$ and $t\Delta_{x,2}(t)$ are, within their large uncertainty, both compatible with zero and with the observed values of the charging rates \cite{PhysRevLett.118.171101}.

\subsubsection{Magnetic fields}
Below $\sim\SI{1}{mHz}$, the noise part of our magnetometer signals, as shown by their almost complete cross-correlation and by the absence of any measurable fluctuating gradient \cite{magnetic-mnras},  was dominated by the interplanetary magnetic field and had negligible readout noise. 
Because of their negligible gradient, the interplanetary field fluctuations interact with the TMs only as they induce, via the  TM residual diamagnetism, a fluctuating magnetic moment that couples to any static magnetic field gradient at the TM location. The force due to this interaction is given by
    \begin{equation}
    \label{eq:magcontrib}
    \Delta g(t) = \frac{\chi L^3}{M \mu_0}  \left(\bm{\nabla}B_{x,\text{dc,2}}-\bm{\nabla}B_{x,\text{dc,1}}\right) \cdot \bm{B} (t),
    \end{equation}
    where $\bm{\nabla} B_{x,\text{dc},i}$ is the magnetic gradient of the static $B_x$ component averaged over the volume of TM$_i$, $\mu_0$  is the vacuum magnetic permeability, and $\chi=(-3.37 \pm 0.15)\times 10^{-5}$ is the magnetic susceptibility of the test masses \cite{magneticprl,magneticLPF-paper-2024}.
    
    Note that the static magnetic gradient may be different from the value that can be extrapolated, at the test mass location, from the differences of the magnetometer readings, which typically is $\simeq \SI{0.5}{\micro\tesla/\meter}$.  
    Indeed a gradient up to  $\SI{10}{\micro\tesla/\meter}$ is expected to be created by a series of thermistors, containing ferromagnetic materials, placed on the outer surface of the EH.

    These thermistors are too far away from the magnetometers to give any significant measurable signal. Thus we had no in-flight information on the gradient that these thermistors were creating  at the location of both TMs.   A dedicated, in-flight experiment using oscillating magnetic fields  \cite{magneticLPF-paper-2024} was  able to give an estimate  just for  $\partial_x B_{x,\text{dc},1}$, that is, just  for one of the three required components of the static magnetic gradient and at the location of just one of the TMs.
    
\begin{figure}[htbp!]
  \centering
  \includegraphics[width=1\columnwidth]{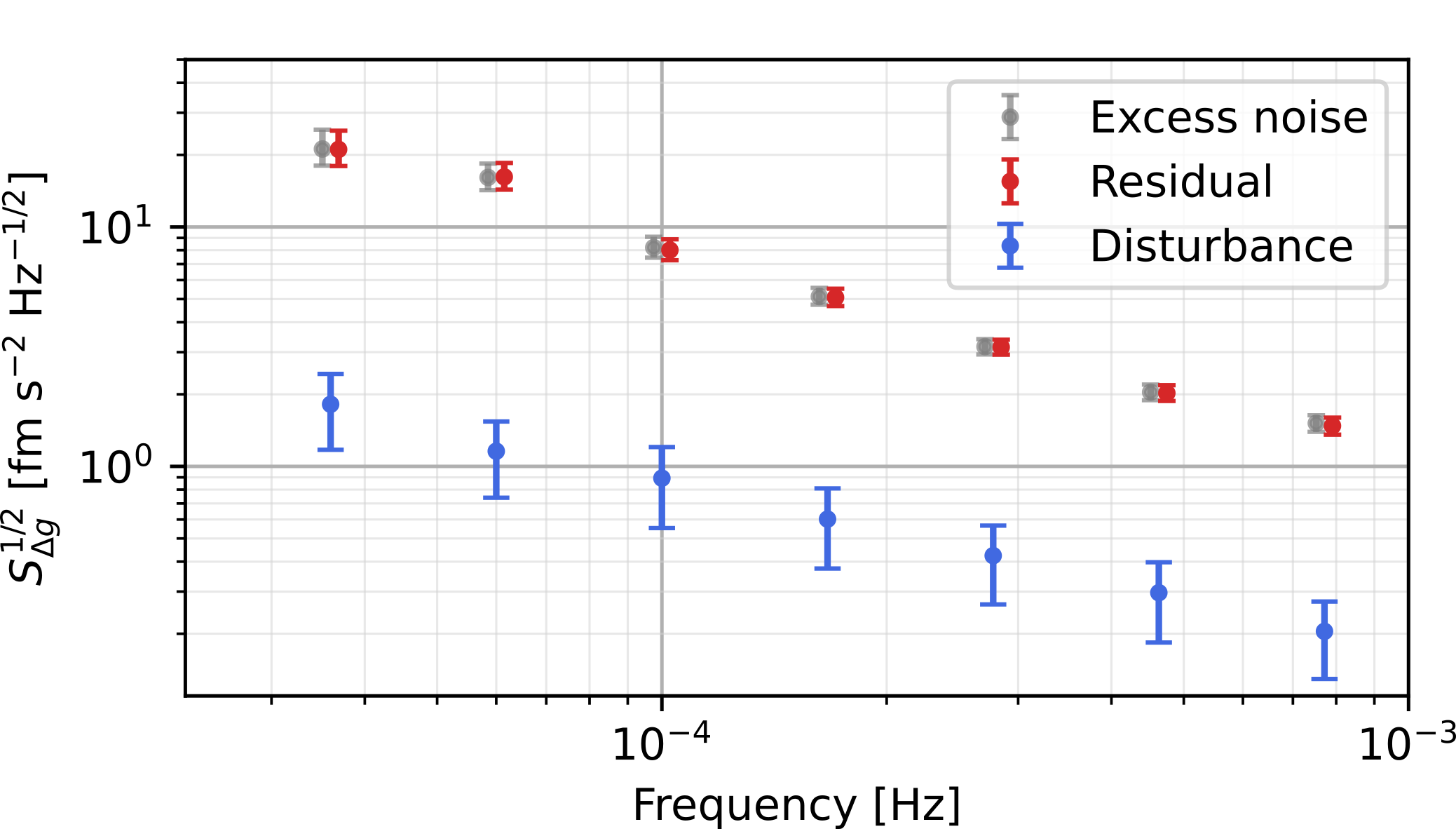}
  \caption{\label{fig:decorrelation B} Decorrelation of low-frequency magnetic fields for run~10, over  the [\SI{36}{\micro\hertz},~\SI{0.77}{mHz}] frequency band.  Frequencies are slightly shifted for clarity.
  Gray points, ASD of total excess noise over Brownian, $S_{\Delta g_e}^{1/2}$, as in Fig.~\ref{fig:brownsubtr}; red points, Bayesian posterior $\pm 1\sigma$ interval for the residual ASD after decorrelation of the three magnetic field time series; blue points, estimated ASD $S_c^{1/2}$ of the contribution of magnetic fields to $\Delta g_e$. }
\end{figure}

    This lack of knowledge of the static magnetic gradient difference at the TM locations left us with  only a possible order of magnitude estimate of the corresponding susceptibilities. Hence, we have performed our decorrelation analysis with no prior assumptions on the susceptibilities. The results are shown in Fig.~\ref{fig:decorrelation B}.

At all frequencies, $S_c(f)$ is significant at $1 \sigma$, while the lower $2\sigma$ quantile of the posterior is negative.\\
Some significance is also supported by the susceptibilities. In particular, while $\mathcal{L}_0 \ge 0.16$ for the $x$ and $z$ components of the field,  $\mathcal{L}_0\simeq 0.001$ for the $y$ component. The corresponding susceptibility is $\alpha_y=(-8\pm2)\,\si{\femto\meter\,\second^{-2}/\micro\tesla}$, that, if due to a gradient acting on just one of the two TMs, would correspond to $\lvert\delta\partial_y B_{x,\text{dc}}\rvert=(6\pm 2)\,\si{\micro\tesla/\meter}$, close to what one would expect because of the magnetic thermistors \cite{magneticLPF-paper-2024}. The $\delta$ is meant to highlight that the measured quantity is the difference of magnetic gradients at the two TM locations.

We note that our estimate of the magnetic force noise ASD,  based on decorrelation,   is consistent with that reported in \cite{magneticprl}, which is based instead on a calculation from measured or estimated values of all the involved quantities. In particular, in addition to the in-flight measurements of the magnetic field fluctuations that we also use, Ref.~\cite{magneticprl} uses the aforementioned measured value for $\partial_x B_{x,\text{dc},1}$, while the other components of $\bm{\nabla}B_{x,\text{dc,}i}$ are estimated based on a model for the statistical distribution of the magnetic dipoles associated with above-mentioned thermistors.

\subsubsection{Thermal gradients}
   In addition to single thermistor readout, that we used to form the average temperature signal, we also had two differential readouts, one for each GRS,  each reading a pair of the thermistors located on the opposite faces of the EH of the corresponding GRS. 
   
   As the drift, which was the source of extra noise on the single readouts, is largely common mode among the thermistors , these channels were basically immune to the nonlinearity noise that plagued the average temperature and  fall then in the category of negligible noise time series.

\begin{figure}[htbp]
  \centering
  \includegraphics[width=1\columnwidth]{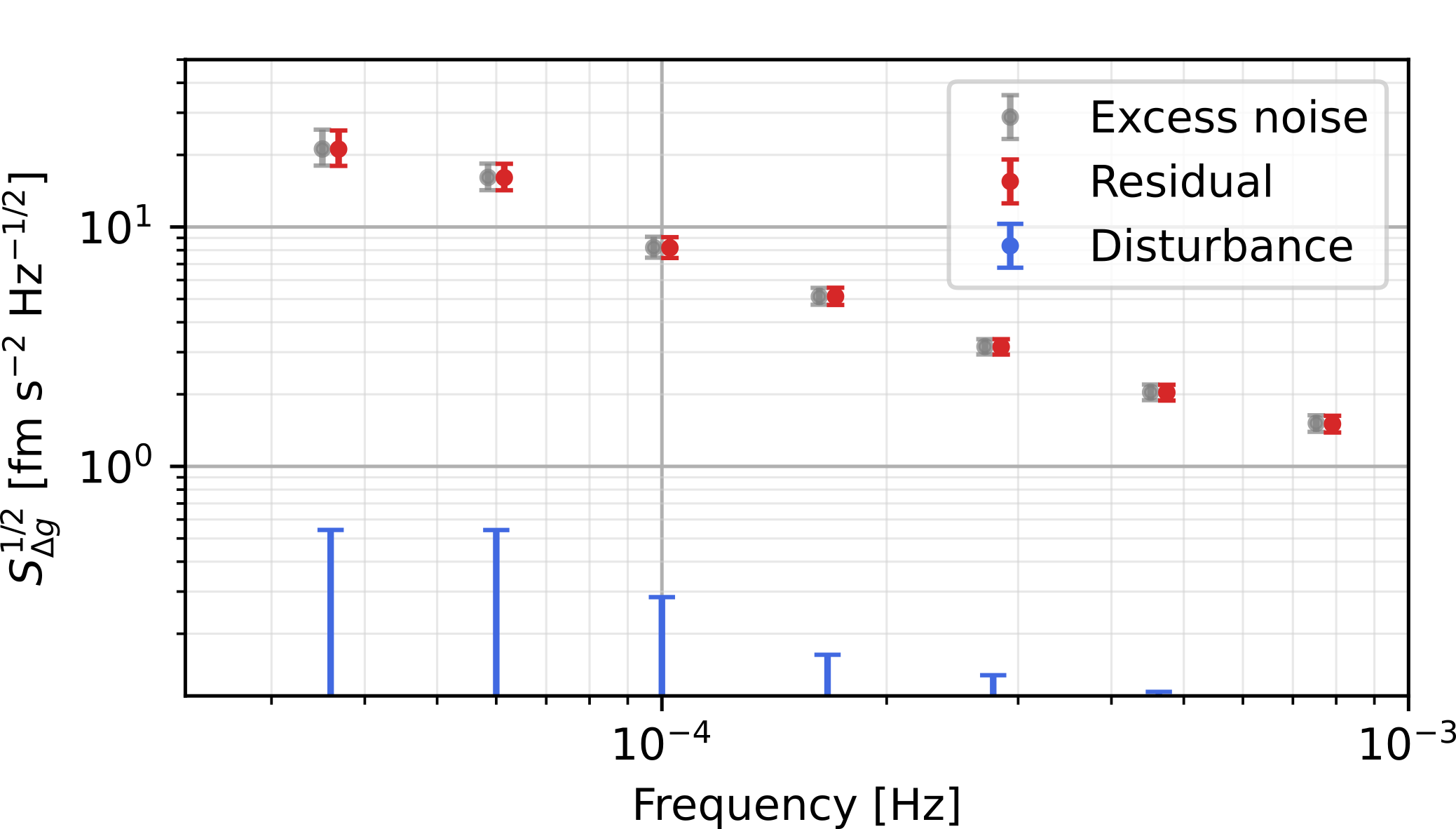}
  \caption{\label{fig:decorrelation dT} Decorrelation of thermal gradients for run~10, over  the [\SI{36}{\micro\hertz},~\SI{0.77}{mHz}] frequency band.  Frequencies are slightly shifted for clarity.
  Gray points, ASD of total excess noise over Brownian, $S_{\Delta g_e}^{1/2}$, as in Fig.~\ref{fig:brownsubtr}; red points, Bayesian posterior $\pm 1\sigma$ interval for the  residual ASD after decorrelation of the two thermal gradients time series; blue points, estimated ASD $S_c^{1/2}$ of the contribution of thermal gradients to $\Delta g_e$.}
\end{figure}

   Thermal gradients are the source of various forces, dominating ones being radiometer effect and asymmetric outgassing \cite{Thermal2007}. While  for the radiometer effect the susceptibility can be estimated to within some 30\%, $\alpha_R\simeq2\times 10^3\,\si{\femto\meter\,\second^{-2}/\kelvin}$, for asymmetric outgassing the uncertainty is much larger \cite{Thermal2007}, not better than an order of magnitude. The results of the decorrelation procedure are shown in Fig.~\ref{fig:decorrelation dT}.

 This contribution is clearly undetectable. Note that the susceptibilities, $(1\pm5)\times10^3 \text{ fm s}^{-2}/\text{K}$, and  $(1\pm7)\times10^3 \text{ fm s}^{-2}/\text{K}$, are zero within errors, errors that comfortably include the value expected of the radiometric effect. This lack of contribution is due substantially to the good stability of the thermal gradient, with an ASD in the considered range of $\simeq \SI{40}{\micro\kelvin}/\sqrt{\text{Hz}}\sqrt{0.1\,\text{mHz}\,/\,f}$.

\subsection{\label{app:decres} Decorrelation of  synchronous time series: Results for noisy series}

\subsubsection{\label{app:G} Tank depletion}  
For the case of the gravitational signal from tank depletion, the PSD of the readout noise of the propellant flow meter is unknown. The susceptibility is  $\alpha_{\text{Tanks}}=\kappa_{t1} \kappa_{bB}$ (see Sec.~\ref{sec:lfbin}) a real, frequency-independent figure known to within a 10\% uncertainty.
 
We estimate the contribution of this source of noise to the total excess noise as $\alpha_{\text{Tanks}}S_{\Delta g,\Delta g_{\text{Tanks}}}$, as explained in Appendix~\ref{app:CPSDnoisy}, with $\Delta\tilde{g}_{\text{Tanks}}$ the true gravitational noise, free of any readout noise contamination.

\begin{figure}[htbp]
  \centering
  \includegraphics[width=1\columnwidth]{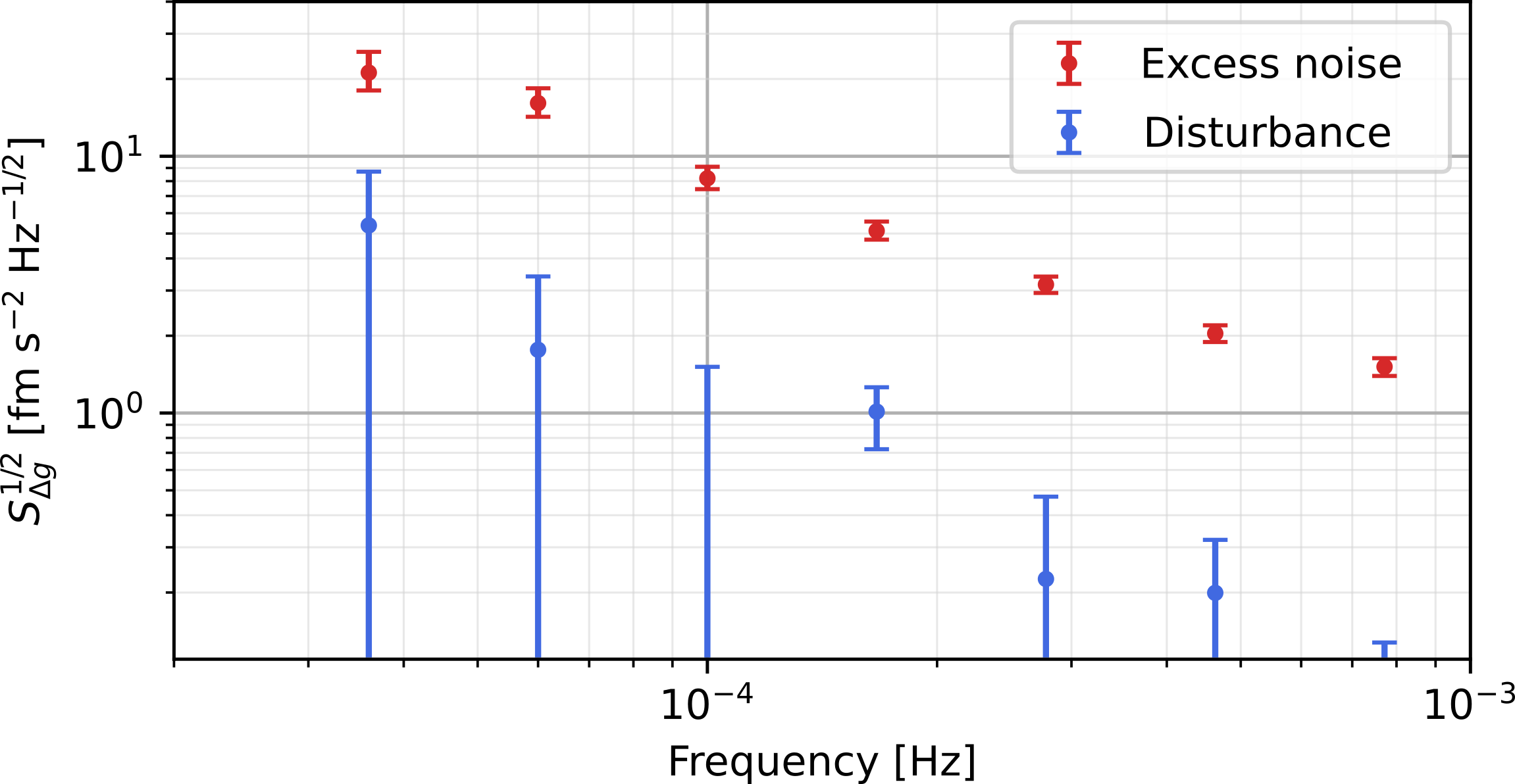}
 \caption{Contribution of gravitational noise due to tank depletion to the ASD of $\Delta g_e$. Red data, ASD of the excess over Brownian $\Delta {g_e}$ from Sec.~\ref{sec:accPSD}; blue data, $\pm 1\sigma$ posterior interval for the gravitational noise contribution from CPSD and susceptibility posteriors.}
  \label{fig:G}
\end{figure}

We take the posterior for $S_{\Delta g,\Delta g_{\text{Tanks}}}$ from the proper Wishart distribution (see Appendix~\ref{app:CPSDnoisy}), and that for $\alpha_{\text{Tanks}}$ to be a normal distribution with mean and standard deviation, respectively, 0.92 and 0.10 (see Sec.~\ref{sec:lfbin}). Results are shown in Fig.~\ref{fig:G}.

Note that $\sum_{i,j=1}^r\alpha_i(f) \alpha_j^*(f) S_{y_i,y_j}(f)$ is a positive real number, while there is no guarantee that a posterior sample for $\sum_{i=1}^r\alpha_i(f) S_{y_i,\Delta g}(f)$ is even real. 

We have dealt with this problem in two ways. We have first checked that the posterior distribution of the imaginary part of $\sum_{i=1}^r\alpha_i(f) S_{y_i,\Delta g}(f)$ was statistically compatible with zero. 

We have then taken the posterior just for the real part of  $\sum_{i=1}^r\alpha_i(f) S_{y_i,\Delta g}(f)$, checking that, whenever the distribution extended to negative values, the zero was within the $\pm 1 \sigma$-credible interval.

Except for the lowest frequency, the $+1\sigma$ limit is well below 10\% in power. At the opposite end, except for one frequency, zero falls always within the $\pm 1 \sigma$ interval.

\subsubsection{\label{app:XT} Temperature effects and LTP distortion}
The instrument distortion $\Delta X(t)$ and temperature $T(t)$ give  rather significant contributions, $\Delta g_{\text{Dist}}(t)= \omega_d^2\Delta X(t)$ and $\Delta g_T(t)=\partial \Delta g/\partial T\times T(t)$, to the quasi-dc long-term evolution of $\Delta g$ (see Sec.~\ref{sec:lfbin}). In this section, we address their possible roles in contributing to the in-band excess noise, for $f\ge f_2$. As we found a significant correlation between these series \cite{FEEnoise}, we consider them together.\\
Unfortunately, these two series are affected by a significant readout noise. Here briefly follows our knowledge of the properties of such noise.

\emph{Instrument distortion} --- The experimental PSD of $\Delta X$ peaks at about 0.1\,mHz and slowly decays above that \cite{FEEnoise}. The experimental CPSD between $\Delta X$ and $\Delta g$ parallels somewhat this behavior, its real part becoming significantly different from zero and positive above about  0.3\,mHz, slowly decaying above that. In the frequency range in which the real part is significant, the imaginary part is also significantly different from zero, this time with a frequency-dependent sign.
This behavior of the CPSD is hardly compatible with the gravitational signal from mechanical distortion, which is virtually instantaneous, and then free of imaginary CPSD,  and supports instead the existence of a dominating readout noise of electrical origin. \\
Actually, the linear instrument distortion $\Delta X$ is the combination of four signals. Each of these signals consists of an independent differential capacitance measurement \cite{PRD_96_062004_capacitive}.
We find that only two of these signals bear some significant correlation with $\Delta g$, the correlation being in different frequency ranges for the two signals.\\ 
Both these signals refer to TM2, to whom the largest actuation is applied, and cross-correlation is likely due to electronic crosstalk between the actuation command signals and the TM motion sensing ones. At low frequencies the $\Delta g(t)$ time series is dominated by the actuation contribution $g_c(t)$ in Eq.~\eqref{eq:deltag}, which brings an indirect correlation between $\Delta g$ and TM motion sensing within the GRS.

Analogously to the linear distortion $\Delta X$, the GRS allows one to measure the angular distortion $\Delta \Phi\equiv\left(\Delta \phi_\text{OMS}-\Delta \phi_\text{GRS}\right)$, which is built with different combinations of the same four capacitance measurements. Once corrected with the analogous interferometer signal $\Delta\phi_{\OMS}$, this signal measures the much suppressed angular distortion, while being insensitive to the linear one. Hence, it carries the same readout noise that affects $\Delta X$, though in a different combination. Including $\Delta\Phi$ in our model should lead to a better constraint of the contribution of these sources to $\Delta g$.

\emph{Temperature} --- The average temperature at the location of the test masses was measured by averaging the readings of various thermistors located on the external $x$-faces of the two electrode housings. \\   
The time series of these thermistors were dominated, above about $\sim\SI{30}{\micro\hertz}$, by excess electronic noise \cite{temperature}. This fact was made particularly evident by the loss of mutual coherence among the time series of the different thermistors above that frequency, a coherence that below $\sim\SI{30}{\micro\hertz}$ was nearly complete. \\
The noise was caused by a subtle interaction between drift and a nonlinearity of the analog-to-digital converter. For two of the eight available thermistors, such noise was so large that we had to discard the corresponding time series. 

\vspace{2mm}
\emph{Susceptibilities} --- Regarding susceptibilities, we take those from the long-term behavior discussed in Sec.~\ref{sec:lfbin}. In addition, we measured the dependence of $\Delta X$ on $T$ by using the ``low distortion'' runs, that is by excluding runs~7, 11, and 12  (see Sec.~\ref{sec:lfbin}). We find consistent results both by measuring the slope $\gamma=\partial \Delta X/\partial T$ in a linear fit over one-day-long data stretches, and from the CPSD between the two data series at \SI{18}{\micro\hertz}. {A set of thermal experiments \cite{thermal-paper-rita} has shown the possibility of delay effects between $T$ and the other time series, reasonably represented by a simple pole filter with a cutoff frequency in the \si{\micro\hertz} range. We summarize all this information by taking the following distributions for the susceptibilities.} We recall here that $\omega_d^2$, $\alpha_T(f)$, and $\gamma(f)$ are, respectively, the susceptibilities of $(\Delta g,\Delta X)$, $(\Delta g, T)$, and $(\Delta X,T)$. 
\begin{enumerate}[label=(\roman*)]
\item $\omega_d^2$ is frequency independent and  normally distributed with $\omega_d^2=(-0.33 \pm 0.03)\times 10^{-6}\,\si{s^{-2}}$.
\item $\alpha_T(f)$, the susceptibility of $\Delta g$ to $T$, is given by $\alpha_T(f)=\alpha_{T,0}\left(1+i f/f_0\right)^{-1}$, with $\alpha_{T,0}$ normally distributed with $\alpha_{T,0}=(0.4\pm0.2)\,\si{\pico\meter\,\second^{-2}\,\kelvin^{-1}}$, and $\log f_0$ uniformly distributed between of $\log(\SI{1}{\micro\hertz})$ and $\log(\SI{100}{\micro\hertz})$.
\item $\gamma(f)$, the susceptibility of $\Delta X$ to $T$,  is given by $\gamma(f)=\gamma_0 \left(1+i f/f_1\right)^{-1}$, with $\gamma_0$ normally distributed with $\gamma_0=(1.04\pm0.07)\,\si{\micro\meter\,\kelvin^{-1}}$, and  $f_1$  with the same distribution  as for $f_0$. 
\end{enumerate}

\begin{figure}[htbp]
 \centering
 \includegraphics[width=1\columnwidth]{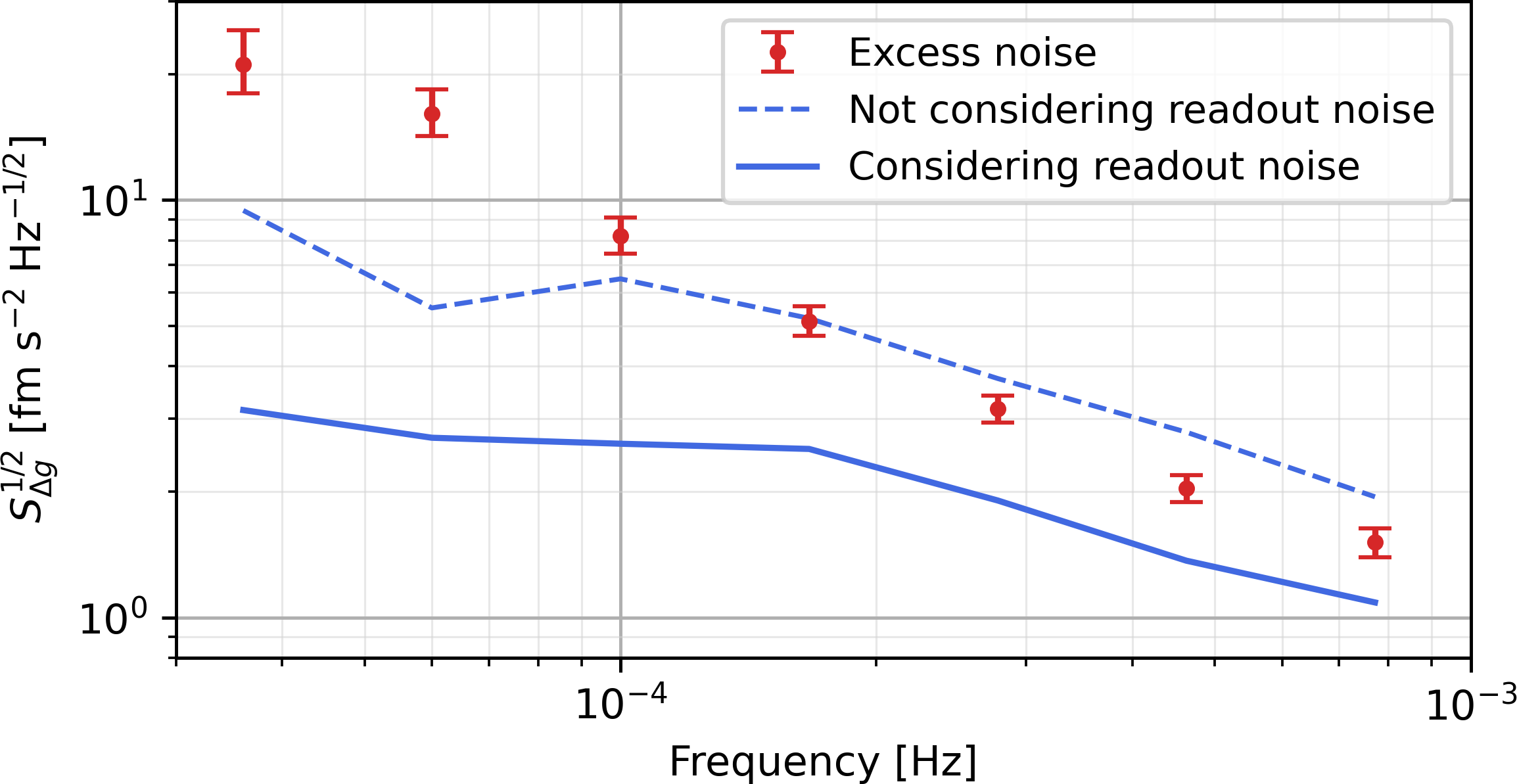}
 \caption{Upper bound ($1\sigma$) of the contribution of LTP distortion and of pressure-mediated temperature effect to $S_{\Delta g_e}^{1/2}$. Red data, ASD of the excess over Brownian $\Delta g_e$ from Sec.~\ref{sec:accPSD}; dashed blue line, $1\sigma$ upper quantile of the estimate assuming time series free of readout noise, ``naive'' model; solid blue line, $1\sigma$ upper quantile considering the presence of readout noise, as described in the text. }
\label{fig:XT}
\end{figure}

\vspace{2mm}
We address the problem with two different approaches.\\
The first approach is ``naive'', as it assumes that the time series are not affected by any readout noise. It is represented as a dashed line in Fig.~\ref{fig:XT}. The bound is completely dominated by the instrument distortion term, the effect of temperature alone being orders of magnitude smaller. However, it overcomes the entire excess noise above 0.1\,mHz, resulting in a nonphysical limit. Hence, this result carries no information.

{The second approach, on the contrary, is more refined as it takes into account that time series are affected by readout noises, possibly cross-correlating. The $1 \sigma $ upper bound is reported in Fig.~\ref{fig:XT}, solid line.}

\vspace{2mm}
\emph{First approach} --- The first obvious, naive way to put an upper limit to the combined contribution of  $\Delta g_{\text{Dist}}$ and $\Delta g_T$ to the PSD of $\Delta g$, is to: assume that these are not affected by any readout noise; take the posterior for the joint CPSD of $\Delta X$, $\Delta T$, and $\Delta g$ from the proper distribution, as explained in Appendix~\ref{app:CPSDnoisy}; take the posterior for the susceptibilities as explained above. From all this, the total contribution to $S_{\Delta g}$ of $\Delta X$ and  $\Delta T$ would be
\begin{equation}
\omega_d^4 S_{\Delta X}(f)+\left(2 \text{Re}\left[ \omega_d^2\alpha_T(f)\gamma^*(f)\right]+\lvert \alpha_T(f)\rvert^2\right)S_T(f),
\end{equation}

\emph{Second approach} --- A more informative bound can be obtained explicitly including the presence of the readout noise affecting the time series. We include the following considerations:
\begin{enumerate}
    \item The measured $\Delta X$ and $T$ are affected by unmodeled and uncorrelated readout noises. Moreover, the noise affecting $\Delta X$ correlates to $\Delta g_0$ in Eq.~\eqref{eq:sdgtot}. 
    \item $\Delta \Phi$ is also affected by readout noise, which correlates with $\Delta X$.
    \item The angular distortion $\Delta \Phi$ has a negligible contribution to the linear acceleration noise $\Delta g$.
\end{enumerate}

We define the matrices $\bSig_{\Delta X}$, $\bSig_{T}$, and $\bSig_0$ so that the CPSD matrix of $(\Delta g, \Delta X, T, \Delta\Phi)$ can be written as
\begin{equation}
\label{eq:matricesDXT_bSig}
    \bSig=\bSig_0 + S_{\Delta X} \bSig_{\Delta X} + S_T \bSig_T + S_{\Delta \Phi}\bSig_{\Delta \Phi}  
\end{equation}
Here, $\bSig_0$ represents the CPSD between $\Delta g_0$ and the readout noise part of $\Delta X$, $T$, and  $\Delta \Phi$. According to the previous assumptions, the matrices read

\begin{equation}
\label{eq:matricesDXT}
\begin{split}
&\bSig_T=\left(\begin{matrix}
|\alpha_T|^2 + 2\text{Re}\left(\alpha_T \omega_d^2 \gamma^*\right) & \alpha_T\gamma^* & \omega_d^2\gamma + \alpha_T & 0\\
\alpha_T^*\gamma & 0 & \gamma & 0 \\
\omega _d^2 \gamma^* + \alpha_T^* & \gamma^* & 1 & 0 \\
0 & 0 & 0 & 0
\end{matrix}\right)\\
&\bSig_{\Delta X}=\left(\begin{matrix}
\omega_d^4 & \omega_d^2 & 0 & 0 \\
\omega_d^2 & 1 & 0 & 0\\
0 & 0 & 0 & 0\\
0 & 0 & 0 & 0
\end{matrix}\right)
\end{split}
\end{equation}

Within this model, we estimate the contribution to $S_{\Delta g}$. First, we take a sample from the posterior of $\bSig$ (see Appendix~\ref{app:CPSDnoisy}) and one from that of the susceptibilities, from which we build $\bSig_0$ using Eqs.~\eqref{eq:matricesDXT_bSig} and~\eqref{eq:matricesDXT}. Then, we search the values of $S_{\Delta X}$, $S_T$, and $S_{\Delta \Phi}$  that give the maximum noise contribution $\left(S_{\Delta X} \bSig_{\Delta X}+S_T \bSig_T\right)_{1,1}$, subject to the condition that $\bSig_0$ is a positive definite matrix, as a CPSD should always be. 
We repeat the calculation over a thousand different posterior samples, deriving a posterior also for this upper limit.

\vspace{2mm}
This upper limit is once again dominated by the effect of $\Delta X$ and constrained by its correlation with $\Delta \Phi$; the role of temperature is completely negligible. Despite being tighter than the ``naive'' one, this upper limit is still set by the large correlation between $\Delta g$ and the readout noise of $\Delta X$. Note that, as the temperature is irrelevant, the effect would amount to some nonthermal distortion of the LTP, such as long-term creep due to stress release. One would expect that, below the system mechanical resonances, all greater than $\SI{10}{Hz}$, the ASD of this kind of effects to be some, possibly rather steep power law of frequency \cite{LevinCreep}. But even for a $\propto f^{-1/2}$ dependence, the value of the limit at $f_2$ would set an upper limit at the highest frequencies at least a factor 4 in power smaller than that indicated by the solid line in Fig.~\ref{fig:XT}. Thus this limit again probably significantly overestimates the effect. 

Finally, note that this technique does not give an explicit lower limit, as the hypothesis that time series are constituted by readout noise only is fully compatible with the observations (solution $S_X=0$, $S_T=0$).

\section{\label{app:nonmod} The unexplained excess: Possible sources and implications for LISA}

In this Appendix, we discuss the most likely sources of the unexplained fraction of the excess noise, and the measures one can possibly take to ensure that they do not compromise LISA performance.

\subsection{\label{app:patchV} Patch potential fluctuations}
Patch potentials \cite{RevModPhys.64.237} may cause force noise in many ways \cite{prl108.5.2012}: through the interaction of their time fluctuations with any static potential applied to the TM, including that due to the TM charge; by creating a quasistatic potential difference that may interact with electronic noise and charge fluctuations; through the interaction of their quasistatic part with their own fluctuations. 

The first two effects have already been discussed in Sec.~\ref{sec:decorrMCMC_noiseprojection}. An analysis of the third effect, the self-interaction of the patch fields \cite{patch2024}, also based on the results of the experiments with charge bias \cite{actuation-paper-bill} in LPF, has appeared recently. The analysis indicates contamination as the most likely source of patches and shows that a model of diffusion-driven fluctuations of contaminant density, that has been considered to explain self-heating in ion traps \cite{IonTraps}, may indeed predict force noise with $\propto 1/f^2$ PSD, with an amplitude in the range of the observed excess, for model parameter values that are not unreasonable.

Though such scenario remains unproven, its possibility would suggest a few precautions to be followed in the development of LISA. These have been discussed in \cite{patch2024}, and we report them here for convenience.

First,  torsion pendulum experiments with LISA-like TM have achieved sensitivities  \cite{prl108.5.2012, russano_measuring_2018}  that may allow a direct detection of, or a significant upper limit to the noise we are discussing here. This would require though to reduce the gaps around the TM  to around 1\,mm or smaller. Needless to say, a direct measurement would bring this potential source of noise under full control.

In the absence of a direct measurement, a measure of precaution is to investigate the nature and the extent of the adsorbate that may have been present on the surface of LPF TM and EH during its operations.  The objective is that  of keeping TM contamination in LISA close to or better than that in LPF. 

This requires a campaign of surface characterization on samples that have undergone a similar preparation history to that of LPF test masses. A systematic experimental study, with the  Kelvin probe technique, of the quasistatic distribution of patch potentials, would be an important part of such a characterization campaign.

It is reasonable to assume that if no new contaminants are introduced in LISA, that had not been present in LPF, and if the amount of contamination can be kept below that of LPF, the noise performance of LPF that fulfills LISA requirements may confidently be reproduced.

\subsection{\label{app:actnonlin}Actuation force calibration and additional voltage noise}
An obvious source of excess noise could be an inaccuracy in the calibration of the applied forces $g_c(t)$, which dominate the spectrum at submillihertz frequencies; similarly, unaccounted nonlinearities in the applied voltages time series could lead to inaccuracies in the applied forces time series. 
Experiments were carried out in flight to calibrate the actuation system, against differences between commanded and applied forces/torques. In particular \cite{PhysRevD.97.122002,actuation-paper-bill}, sinusoidal ``calibration tones'' of amplitude $\SI{100}{\femto\newton}$ and frequency $\SI{10}{\milli\hertz}$ were injected out of loop, effectively inducing a controlled force on TM2. These experiments yielded fluctuations in the calibration coefficient $<0.5\%$, which could not account for the observed excess noise. 

However, these experiments also led to the discovery of the truncation error, already introduced in Sec.~\ref{sec:dyn} \cite{ActNeda}, which required a deterministic correction on the actuation voltage amplitudes of the order of \SI{10}{\micro\volt} rms, out of a nominal bit resolution of \SI{153}{\micro\volt}. This correction was relevant to the success of the calibration tone experiments: the injection of external sinusoidal forces resulted in an apparently time-varying actuation gain factor, as well as a series of spurious lines at several harmonics of the injection frequency. These lines could be effectively suppressed --- within statistical errors --- only by taking into account such correction. 

Driven by these results, we performed simulations to understand if additional nonlinearities, smaller than the truncation error rms value, could result in increased noise levels in noise-only runs, but at the same time go unnoticed in calibration tone experiments. We showed that, if the effective amplitudes of applied voltages $V_x$ and $V_\phi$ (see \cite{PhysRevD.97.122002,actuation-paper-bill} for definitions) were affected by additional small nonlinearities of the order of 1--2\,\si{\micro\volt} rms -- i.e., of the order of 1/100th of the nominal bit resolution -- neither the calibration tone experiment, nor any other in-flight experiment could rule out the presence of such nonlinearities. 

\begin{figure}
  \centering
  \includegraphics[width=0.95\columnwidth]{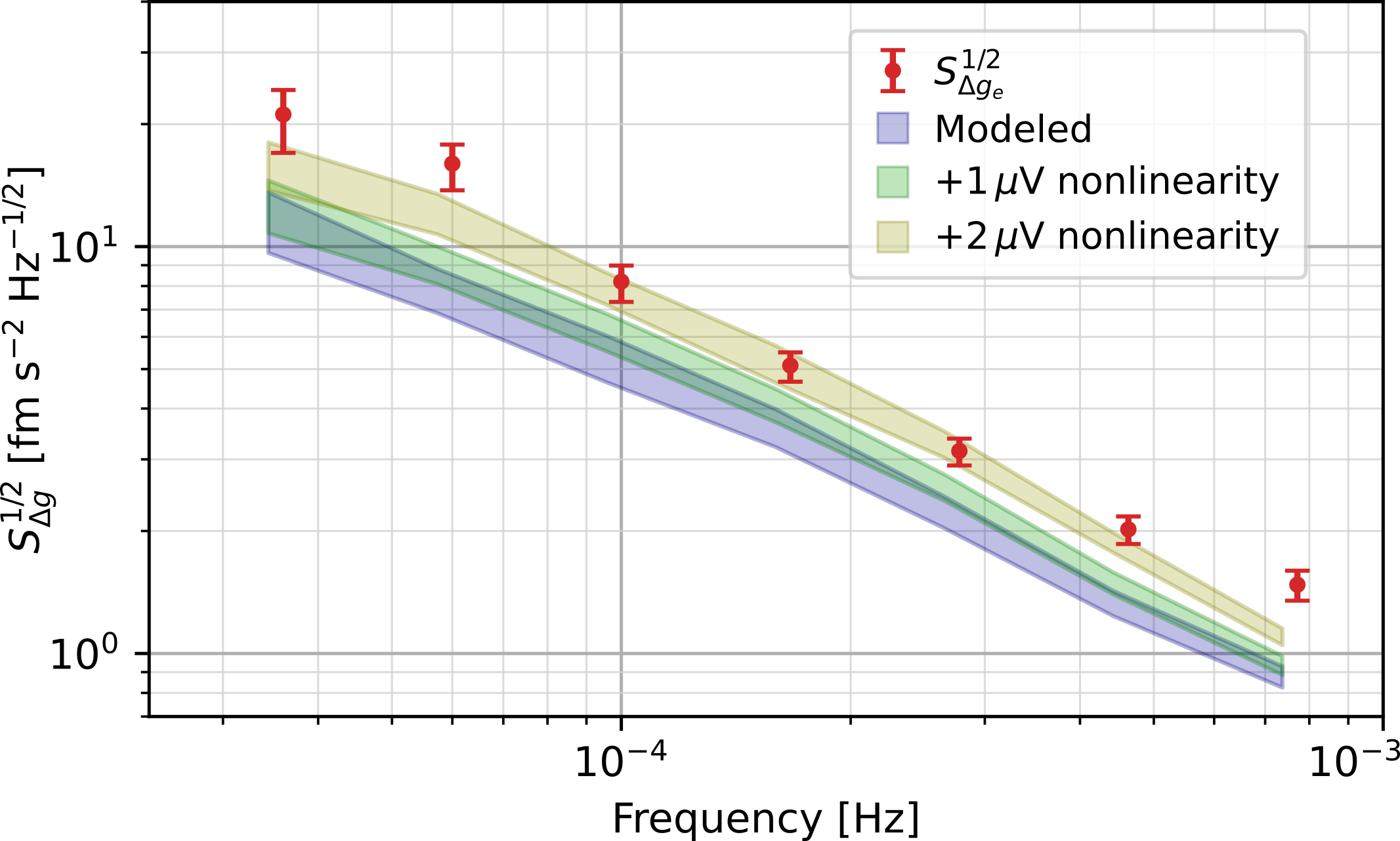}
\caption{\label{fig:NoiseProj_1muV_2muV_66} Total modeled noise, including potential inaccuracies in the applied actuation voltage amplitudes. Red points and blue span, respectively, excess noise $S_{\Delta g,e}^{1/2}$ and total modeled noise, as in Fig.~\ref{fig:findec}; green and yellow spans, modeled noise, summed to voltage amplitude nonlinearities with \SI{1}{\micro\volt} and \SI{2}{\micro\volt}, respectively. The additional PSD due to such inaccuracies is the joint posterior distribution of ten independent realizations with the given rms, as described in the text.}
\end{figure}

These nonlinearities would, however, have a non-negligible impact on noise-only runs. With forces applied during February 2017, run~10 of Table~\ref{tab:noise-only-runs}, the presence of a \SI{1}{\micro\volt} or a \SI{2}{\micro\volt} rms nonlinearity would indeed have a relevant impact on the total modeled noise as shown in Fig.~\ref{fig:NoiseProj_1muV_2muV_66}. For each level of rms deviation (\SI{1}{\micro\volt}-\SI{2}{\micro\volt}), we simulated ten realizations and showed their joint posterior distribution. In Fig.~\ref{fig:NoiseProj_1muV_2muV_66}, the red points and blue span represent, respectively, the excess $S_{\Delta g,e}^{1/2}$ and the total modeled noise, as in Fig.~\ref{fig:findec}. Adding deviations of \SI{1}{\micro\volt} and \SI{2}{\micro\volt} rms, respectively, the total modeled noise becomes the one represented by the green and yellow spans. 

Potentially, an inaccuracy of \SI{2}{\micro\volt} rms, or even less, could result in non-negligible force noise, explaining a relevant fraction of the detected excess noise. At the same time, it would go undetected in in-flight calibration experiments. Preliminary measurements on LPF-prototype FEE models show that the presence of residual nonlinearities of the magnitude above can not be excluded, i.e., measurements are compatible with the presence of a nonlinearity of \SI{1}{\micro\volt} rms. However, we are currently planning deeper and more systematic tests, to better characterize nonlinearities and assess their impact.

Another mechanism through which voltage disturbances could lead to increased in-band force noise and, at the same time, go undetected in dedicated measurement campaigns is through down-conversion of high-frequency spurious disturbances. If voltage anomalies should be present at frequencies outside the measurement band, they would down-convert into the band because of the quadratic nature of the electrostatic force. In addition, due to the lack of significant associated torque, the voltage anomaly should have involved one or more pairs of electrodes facing the same face of the same TM. 

To fix the amplitude scale of such a disturbance, the force due to the voltage $V$, when applied to both electrodes in one such pair, exerts a force $\lvert\Delta g\rvert=(1/M)\lvert\partial C_x/\partial x\rvert V^2\simeq 0.15\,\text{pm\,s}^{-2}\,(V/100\,\text{mV})^2$. Thus a line with a mean amplitude of 100\,mV, and a noisy relative amplitude fluctuation with ASD $\simeq 5\times 10^{-3}/\rtHz\,(1\, \text{mHz}/f) $ would produce the observed noise. We note that the actuation circuitry connected to the electrodes includes a passive two-stage low-pass filter with a bandpass near 2\,kHz, thus this putative high-frequency disturbance would have needed to be accordingly higher at the amplifier outputs for frequencies larger than 2\,kHz.

No such line has been detected either during the ground testing of the FEE flight hardware, or during laboratory testing of its various prototypes, but we cannot exclude, for instance, some damage due to the launch stresses. 

{As a final note, we want to emphasize that detecting any of these effects in the LISA GRS FEE is feasible through ground testing, either via electronic measurements or by using torsion pendulum measurements of the induced forces.}

\subsection{\label{app:unmodeledgrav} Unmodeled gravitational noise}
We have discussed the gravitational noise due to propellant tank depletion and that due to LTP distortion. We have accounted for the first and estimated that it is unlikely that the second may have caused more than 10\% of the unaccounted noise.

In addition to those effects, any other mass motion, either because of distortion of solid parts or because of evaporation of volatile fractions, may cause gravitational force noise and may have contributed to excess noise. We discuss here a few possibilities, first for the case of distortion and then for that of evaporation.

The tungsten balance mass is the most intense source of gravitational field gradient at the TM location. The LTP distortion, already  discussed, moves the  balance mass, relative to the TM, together with the entire GRS. However, in addition to that, any internal GRS distortion may  also have  moved the balance mass relative  to the TM. 

With a gravitational gradient  $\simeq\partial g_x/\partial x_s\simeq 5\times10^{-7}\,\text{s}^{-2}$, a random jitter $\delta x$ of the balance mass position with  ASD $S_{\delta x}^{1/2}\simeq \SI{1.5}{\nano\meter/\rtHz}\,\left(\SI{1}{mHz}/f\right)$, would account for the entire unexplained excess. Such a jitter, if thermally induced, given the construction materials and the geometry of the GRS, would amount to thermal fluctuations with an ASD of  $S_{T}^{1/2}\simeq \SI{4}{\milli\kelvin/\rtHz}\,\left(\SI{1}{mHz}/f\right)$, definitely larger than the measured one \cite{temperature}.

As for nonthermal deformation, an obvious example  would be long-term, noisy  mechanical secondary creep due to stress release, like that due to the unlock of the TM on orbit. Crudely approximating this creep as a Poisson sequence of steps with rms amplitude $\delta x$, and average relaxation  rate $\delta \dot{x}$, one would need $\delta x \delta \dot{x}\simeq \SI{0.2}{\micro\meter^2/\year}$. As realistically $\delta \dot{x}\ll \,\si{\micro\meter^2/\year}$, this would require steps of $\SI{0.2}{\micro\meter}$, happening then at a rate of less than 5 per year, very different from the observed time series.

Thermal or nonthermal distortion resulting in the motion of massive components farther away from the TMs may also exert significant forces. Calculations  \cite{gravASU} show, for instance,  that the spacecraft alone, without the LTP, exerts a static difference of force on the TM of $\Delta g_{sc}\simeq 5\times 10^{-9}\,\si{\meter\,\second^{-2}}$, very  similar to that of a homogeneous square toroid with an inner diameter of 1\,m, an outer one of 2\,m, and a mass of $\simeq \SI{300}{kg}$, a crude approximation to the spacecraft shape. 

One can calculate that a homogeneous  relative distortion $\delta$ of any of the toroid dimensions causes a variation of differential force  $\delta \times \Delta g_{sc}$. It would thus take $\delta$ to be a random process with ASD $S_\delta^{1/2}\simeq 1.4\times 10^{-7}\,\si{\hertz^{-1/2}}\,(\SI{1}{mHz}/f)$ to justify the unexplained noise.

We have no way of assessing if such mechanical distortions of the spacecraft (about $\SI{0.1}{\micro\meter}$ root-mean-square variation of the corresponding dimension over 1~day) did take place during operations. A correlation analysis of $\Delta g$ with the solar power hitting the spacecraft gave no significant results. A simulation of thermal induced distortion and of the resulting gravitational noise performed during mission development \cite{gravASU}, pointed to 1 order of magnitude smaller figures. This simulation was done though on a simplified model and was significant only for $f\gtrsim\SI{2}{mHz}$.

All that said, detailed thermomechanical simulations are standard practice in aerospace engineering. In addition, for LPF rather complete tools have been set up to interface the thermomechanical model of the system with the gravitational one. We calculate that a thorough dynamical simulation of the LISA spacecraft's gravitational field should have the accuracy and the sensitivity to keep gravitational  noise due to mechanical distortion under control.

Outgassing of volatile fractions from spacecraft has been observed, for instance, in the Rosetta mission,  to continue years after launch \cite{Rosetta}. In particular, after a  water desorption-dominated phase lasting 100--200\,days, a longer, possibly diffusion-dominated phase was observed with a very slowly, if at all, decaying evaporation rate, at least for the following 500\,days. 

Given the nonsymmetric geometry of the LPF spacecraft, even a  spatially homogeneous noisy outgassing would have caused  some $\Delta g$ noise. Definitely  more so if the outgassing had also some  patchy pattern, that would have further reduced the spatial symmetry.

To fix the scale of the effect, we have considered  the spacecraft  structure, which is made out of sandwich panels of carbon fiber reinforced plastic with an aluminum honeycomb core. We have considered the case of $N$ outgassing patches of size small enough to approximate sums with integrals, having  noisy  evaporation rates all with same ASD, incoherently adding up to a total evaporation rate with ASD $S_{\dot{M}}^{1/2}$. 

We calculate that $S_{\Delta g}^{1/2}\simeq 0.5 \left(\mathcal{G}/L_0^2\right) S_{\dot{M}}^{1/2}/(2 \pi f)$, with $L_0=37.6\text{ cm}$ the distance between the centers of the two TMs, and $\mathcal{G}$ the gravitational constant. To match our observation of $S_{\Delta g}^{1/2}\simeq \SI{0.7}{\femto\meter\,\second^{-2}/\rtHz}\,(1\,\text{mHz}/f)$, we need a frequency-independent value  $S_{\dot{M}}^{1/2}\simeq \SI{0.02}{\milli\gram\,\second^{-1}/\rtHz}$.

Just for the sake of discussion, we note that such a  ``white'' evaporation rate behavior would naturally be obtained if the outgassing on LPF consisted of a Poisson succession, with mean evaporation rate $\langle \dot{M}\rangle$, of discrete outgassing events from any of the patches, with a rms value $\delta m$. This process  would indeed give $S_{\dot{M}}^{1/2}=(\delta m \, \langle \dot{M}\,\rangle)^{1/2}$.

Reference~\cite{Rosetta} estimates the mass loss of Rosetta in several hundred grams per year. LPF was a lighter satellite, by almost a factor~3, but with more plastic and closer to the Sun. Thus a direct projection of the outgassing properties may be rather speculative. 

Nevertheless, a crude estimate based on the outgassing properties of LPF materials, gives a lower limit for $\langle \dot{M}\rangle$ at $\langle \dot{M}\rangle\simeq \SI{0.1}{kg\,y^{-1}}=\SI{3}{\micro\gram\,\second^{-1}}$, not far from what one would extrapolate from Rosetta based just on the mass ratio of the two satellites. Using this figure for $\langle \dot{M}\rangle$ in the Poisson noise ASD formula, one gets $\delta m\simeq \SI{0.2}{mg}$ and $\lambda=\langle\dot{M}\rangle/\delta m\simeq \SI{0.01}{\second^{-1}}$.

To our knowledge, there are no studies on the spatial distribution of outgassing in spacecraft, nor on its dynamics. Thus this crude scenario of noisy outgassing remains rather speculative. We note, however, that nothing in the figures above or in the current knowledge allows us to rule it out. We conclude that a cautious approach for LISA would be to stay as close as possible to the LPF design in terms of quantity and distribution of materials with significant volatile components.
\subsection{\label{app:pressfluct}Pressure fluctuation}
As said in Sec.~\ref{sec:lfbin}, the complex geometry of the TM environment may create quasistatic pressure gradients. Any in-band fluctuation of such gradients would directly translate into an in-band acceleration fluctuation. 

To account for the measured excess $S_{\Delta g_e}^{1/2}$ of about $0.7\,\si{\femto\meter\,\second^{-2}/\rtHz}\,(\SI{1}{mHz}/f)$, the ASD of these fluctuations should be  $S_{\Delta p}^{1/2} \sim 0.6\times10^{-12}\,\si{\pascal/\rtHz}\,(\SI{1}{mHz}/f)$. 


We note that the process at the basis of this hypothetical fluctuating pressure should  necessarily  be different from that generating the static gradient discussed in Sec.~\ref{sec:lfbin}. Indeed (see Sec.~\ref{sec:brown}) the latter steadily decreased during the mission, following the general decay of the pressure,  whereas the former (Sec.~\ref{sec:oneoverf}) did not. 

Hence, if the  process responsible for the $1/f$ noise has been due to the outgassing of some species for the TM environment, it must have been substantially stable over the entire mission  and independent of the  outgassing rate of the main fraction. This observation parallels that on the possible outgassing origin of force glitches in LPF \cite{lpf_glitch2022} that also had properties that have been stable over the course of the entire mission. 

It does not require much outgassing to produce the $1/f$ noise. If, for instance, such hypothetical gas phase were hydrogen diffusing out of the various elements of the TM environment, and if the noisy outgassing took place as the series of undetectable glitches discussed in Sec.~\ref{sec:glitches} and in the Appendix~\ref{app:glitchsimul}, it would only take a mean outgassing rate of $\simeq 6 \text{ pg/d}$, that is some 3 ng of total emission over the course of the entire mission, to explain the noise. We have no  specific piece of experimental evidence to support this hypothesis, neither could we trace any relevant study on pressure fluctuations in vacuum systems. Neither have we, however, any evidence proving the model false or unlikely, neither from our own experiments nor from the literature. It is also unlikely that we could devise a laboratory experiment able to detect such tiny pressure fluctuations. 

For this reason, we conclude again that the LISA GRS outgassing environment should be kept as close as possible to that of LPF to retire the risk of unwanted large pressure gradient fluctuations.

\subsection{\label{app:hfmag} High-frequency magnetic field noise}
In addition to low-frequency effects, discussed in Sec.~\ref{sec:noiseprojection}, magnetic fields at high frequency may induce eddy currents within the test masses, and then exert Lorentz forces on them \cite{labtoLPF}. The effect is thus quadratic and would convert the low-frequency amplitude fluctuations of a high-frequency magnetic spectral line into a corresponding low-frequency force. 

To give a scale of the effect,  a recent finite-element  electromagnetic  calculation by the LISA project \cite{eddy_CarloValerio} has shown that the effect of a dipole of $\SI{1}{\milli\ampere\,\meter^2}$ located at a distance $d=\SI{20}{cm}$ from the test mass and oscillating at the frequency of 100\,Hz, would cause a force of $\Delta g \simeq \SI{4}{\femto\meter/\second^2}$. The effect reaches its peak at \SI{100}{\hertz}, while at lower frequency the induced current decreases, and above that, the screening effect of the metallic electrode housing attenuates the oscillating field. The effect of a dipole source decreases with $d^{-7}$, so that at the closest distances of about $\SI{0.4}{m}$ between the test mass and any active device on the LPF spacecraft the effect might be $\sim 100$ times smaller.

The spacecraft prime contractor, during LPF development, performed a test campaign on ground  against audio-frequency magnetic lines \cite{acmagnetics}. A few, barely significant lines have  been identified, with peak amplitudes $\ll \SI{1}{\nano\tesla}$ at the position of the test masses. In the point dipole model at a distance of $d\simeq \SI{0.4}{\meter}$, a $1\,\text{nT}$ line would be generated by a dipole of $\simeq\SI{0.3}{\milli\ampere\,\meter^2}$ at most, and would exert a static force $\Delta g \simeq \SI{4e-3}{fm\,s^{-2}}$. To reproduce the observed excess noise with a noisy amplitude modulation of one of these lines $\SI{1}{\nano\tesla}$, one would need a relative amplitude modulation with ASD $\simeq 2.5\times 10^2 \text{\,Hz}^{-1/2} \left(1\,\text{mHz}/f\right)$.

One can calculate that the rms fluctuation of a process $x$ with ASD given by $S_x=S_0^{1/2}(f_0/f)$, over a data stretch of duration $T$, is  $\langle x_\text{rms}\rangle=S_0^{1/2}\pi f_0 \sqrt{T/3} $. With $S_0^{1/2}\simeq 2.5\times 10^2 \text{\,Hz}^{-1/2}$, it would take about 5\,s to make the relative rms fluctuation of the line amplitude 100\%. Such a large fluctuation would have been noticed during testing. 

The required amplitude fluctuation for such lines would become smaller in the presence of more than one line. However, consider that 1\,nT is a real conservative upper limit for the observed lines and that the effect is quadratic in the amplitude of the lines, so that the combined effect of many lines would be dominated by the few brightest among them. 

Despite these considerations, that would rule out this source of noise, it must be stressed that unfortunately we had no magnetometer on board sensitive to the audio band. Thus, as the operating conditions may have been different from those during testing,  we cannot exclude that additional, more intense, amplitude-modulated lines had been generated once on orbit.

For instance, a single line generated by a dipole of $\simeq\SI{2}{\milli\ampere\,\meter^2}$ located at a distance $d=\SI{20}{cm}$ from the test mass,  with a more reasonable relative amplitude fluctuation  ASD $\simeq \SI{0.2}{\hertz^{-1/2}} \left(1\,\text{mHz}/f\right)$, would explain the excess noise. This is a relatively large magnetic dipole, for instance 32\,mA in a fully uncompensated square loop of 25\,cm size, the existence of which we are not aware.  Nevertheless we certainly recommend that in LISA thorough testing is performed on ground, and that onboard diagnostic magnetometers with sensitivity up to, at least, \SI{1}{\kilo\hertz} are seriously considered.

\newpage

\bibliography{A_bibliography_tidy.bib}
\end{document}